\documentclass[%
 reprint,
 amsmath,amssymb,
 aps,
]{revtex4-2}

\usepackage{braket}% bra/ket notation
\usepackage{color}% colors for comments
\usepackage{dsfont}% double slash, for identity 1
\usepackage{slashed}% Feynman slash notation
\usepackage{graphicx}% Include figure files
\usepackage{dcolumn}% Align table columns on decimal point
\usepackage[caption=false]{subfig}
\usepackage{bm}% bold math
\newif\ifdocomment
\docommentfalse
\docommenttrue
\ifdocomment

\newcommand{\acomment}[1]{\textcolor{red}{[#1]}}
\newcommand{\ccomment}[1]{\textcolor{blue}{[#1]}}
\else
\newcommand{\acomment}[1]{\PackageError{acomment}{unresolved comment}{}}
\newcommand{\ccomment}[1]{\PackageError{ccomment}{unresolved comment}{}}
\fi

\newcommand{\amuv}{\ensuremath{a_\mu^{\rm v}}}

\newcommand{\conniso}{{\rm conn.,isospin}}
\newcommand{\disciso}{{\rm disc.,isospin}}

\begin{document}

\preprint{APS/123-QED}

\title{Consistency of hadronic vacuum polarization between lattice QCD and the R-ratio}

\author{Christoph Lehner}
 \email{christoph.lehner@ur.de}
\affiliation{
Universit\"at Regensburg, Fakult\"at für Physik, 93040, Regensburg, Germany
}
\affiliation{
Physics Department, Brookhaven National Laboratory, Upton, New York 11973, USA
}

\author{Aaron S. Meyer}%
 \email{ameyer@quark.phy.bnl.gov}
\affiliation{
Physics Department, Brookhaven National Laboratory, Upton, New York 11973, USA
}%

\date{\today}% It is always \today, today,

\begin{abstract}
There are emerging tensions for theory results of the hadronic vacuum polarization contribution to the muon anomalous magnetic moment both within recent lattice QCD calculations and between some lattice QCD calculations and R-ratio results.
In this paper we work towards scrutinizing critical aspects of these calculations.  We focus in particular on a precise calculation of Euclidean position-space windows defined by RBC/UKQCD that are ideal quantities for cross-checks within the lattice community and with R-ratio results.  We perform a lattice QCD calculation using physical up, down, strange, and charm sea quark gauge ensembles generated in the staggered formalism by the MILC collaboration.  We study the continuum limit using inverse lattice spacings from $a^{-1}\approx 1.6$~GeV to $3.5$~GeV, identical to recent studies by FNAL/HPQCD/MILC and Aubin {\it et al.}~and similar to the recent study of BMW.  Our calculation exhibits a tension for the particularly interesting window result of $a_\mu^{\rm ud, conn.,isospin, W}$ from $0.4$~fm to $1.0$~fm with previous results obtained with a different discretization of the vector current on the same gauge configurations.  Our results may indicate a difficulty
related to estimating uncertainties of the continuum extrapolation that deserves further attention.  In this work we also provide results for $a_\mu^{\rm ud,conn.,isospin}$, $a_\mu^{\rm s,conn.,isospin}$, $a_\mu^{\rm SIB,conn.}$ for the total contribution and a large set of windows.  For the total contribution, we find $a_\mu^{\rm HVP~LO}=714(27)(13) 10^{-10}$, $a_\mu^{\rm ud,conn.,isospin}=657(26)(12) 10^{-10}$, $a_\mu^{\rm s,conn.,isospin}=52.83(22)(65) 10^{-10}$, and $a_\mu^{\rm SIB,conn.}=9.0(0.8)(1.2) 10^{-10}$,  where the first uncertainty is statistical and the second systematic.  We also comment on finite-volume corrections for the strong-isospin-breaking corrections.
\end{abstract}

\maketitle

\section{Introduction}
The established theory result for the muon anomalous magnetic moment, $a_\mu$,
exhibits a $3.3\sigma$ \cite{Davier:2019can} to a $3.8\sigma$ \cite{Keshavarzi:2019abf} tension with the 
results of the BNL experiment \cite{Bennett:2006fi}.  Within this year, we
expect the Fermilab $g-2$ experiment~\cite{Grange:2015fou} 
to release first results towards their target to reduce the uncertainties of the BNL experiment by a factor of 4.  In the near future, we also look forward to results from the methodologically independent experimental program at J-PARC \cite{Abe:2019thb}.  These results are highly anticipated and are accompanied
by a concerted effort of the theory community to improve upon and scrutinize the existing standard model results, most importantly for the hadronic vacuum polarization (HVP) \cite{Hagiwara:2011af,Davier:2010nc,Davier:2017zfy,Jegerlehner:2017lbd,Keshavarzi:2018mgv,Davier:2019can,Burger:2013jya,Blum:2015you,Blum:2016xpd,Chakraborty:2016mwy,Clark:2017wom,Lehner:2017kuc,Chakraborty:2017tqp,Boyle:2017gzv,DellaMorte:2017dyu,Borsanyi:2017zdw,Blum:2018mom,Giusti:2018mdh,Izubuchi:2018tdd,Davies:2019efs,Shintani:2019wai,Gerardin:2019rua,Borsanyi:2020mff,Lehner:2019wvv,Giusti:2019xct,Giusti:2019hkz} and hadronic light-by-light (HLbL) \cite{Prades:2009tw,Blum:2019dmy} contributions, which currently limit the precision of the theory result.
The Muon $g-2$ Theory Initiative, a multi-year community effort \cite{tiworkshop:firstplenary,tiworkshop:kek2018,tiworkshop:uconn2018,tiworkshop:secondplenary,tiworkshop:thirdplenary},
is now in the final stages of writing a whitepaper summarizing the current theory status \cite{tiwp}.

For the HLbL contribution, new analytic approaches \cite{Colangelo:2015ama,Colangelo:2017fiz,Colangelo:2017qdm,Colangelo:2019uex,Colangelo:2019lpu} as well as the first
ab-initio lattice QCD calculation \cite{Blum:2019dmy} building on multi-year methodology development \cite{Blum:2014oka,Blum:2015gfa,Blum:2016lnc,Blum:2017cer,Asmussen:2016lse,Green:2015sra} so far show consistent results and rule out the HLbL contribution as an explanation for the current tension between theory and experiment.

For the HVP contribution, however, tensions exist within lattice QCD calculations \cite{Aubin:2019usy} as well as between lattice QCD calculations and R-ratio results \cite{Aubin:2019usy,Borsanyi:2020mff}.  At this point, the lattice calculations exhibiting a tension with R-ratio results share some aspects.  They are performed at physical pion mass, with staggered sea quarks and a conserved valence vector current, and use inverse lattice spacings in the range from $a^{-1}\approx 1.6$~GeV to $a^{-1}\approx 3.5$~GeV.  At the same time, a joint study of lattice QCD and R-ratio results performed by the RBC/UKQCD collaboration \cite{Blum:2018mom} using domain-wall sea quarks at physical pion mass with $a^{-1}=1.7$~GeV to $2.4$~GeV showed no significant tension.  

Concretely, there are two tensions for the
 isospin-symmetric quark-connected light-quark contribution (Fig.~\ref{fig:overviewmain}).
The first is for the Euclidean position-space window
 $a_\mu^{\rm  ud,conn.,isospin,W}$ for times $t_0=0.4$~fm,
 $t_1=1.0$~fm, and $\Delta=0.15$~fm as defined by
 RBC/UKQCD~\cite{Blum:2018mom} between
 Aubin {\it et al.}~\cite{Aubin:2019usy} on one side and
 RBC/UKQCD \cite{Blum:2018mom} and a combined R-ratio/Lattice
 result~\cite{Blum:2018mom,Aubin:2019usy,Borsanyi:2020mff}
 on the other side.
The second is a tension between the total
 $a_\mu^{\rm  ud,conn.,isospin}$ with high values
 for BMW \cite{Borsanyi:2020mff} and lower values for
 FNAL/HPQCD/MILC \cite{Davies:2019efs}
 and ETMC~\cite{Giusti:2019hkz}.
In this work, we focus on scrutinizing the first tension.

To this end, we use the same lattice QCD ensembles as Aubin {\it et al.}~\cite{Aubin:2019usy}
but use a site-local current instead of a conserved current.  Within this framework, we then consider different approaches towards the continuum limit for windows in the staggered formalism and provide an analysis with minimal input from effective theories.  In our analysis, we find a substantially lower value for $a_\mu^{\rm  ud,conn.,isospin,W}$ compared to Ref.~\cite{Aubin:2019usy}.  This is particularly noteworthy since the same sea-quark sector is used and may indicate difficulties with properly estimating uncertainties associated with the continuum limit.

Within our numerical framework, we can also access the connected strong-isospin breaking contribution as well as the strange-quark connected contribution.  We provide results also for these contributions including a wide range of different windows.  We hope that these results will prove useful to further understand the current tensions.

This manuscript is organized as follows:
Section~\ref{sec:methodology} discusses the main methods
 used in the analysis, including the window method and
 a correlator smoothing technique to reduce the unwanted
 parity partner contributions.
Section~\ref{sec:numerical} gives information about
 the ensembles used for this study and the
 computational setup for our data generation.
In Section~\ref{sec:results}, we describe our analysis and uncertainty estimates.
In this section we in particular also comment on finite-volume corrections to the strong-isospin breaking contributions.
In Section~\ref{sec:conclusion}, we summarize and
 give some concluding remarks.
Appendix~\ref{sec:appresults} provides additional
tables of results for cross-comparisons with other analyses.

\section{Methodology}
\label{sec:methodology}
\subsection{General setup}
\begin{figure}
  \subfloat[Diagram c]{~\includegraphics[height=1.5cm]{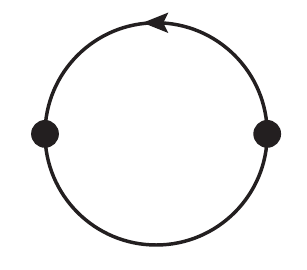}~}
  ~~~~~~~~
  \subfloat[Diagram d]{\includegraphics[height=1.5cm]{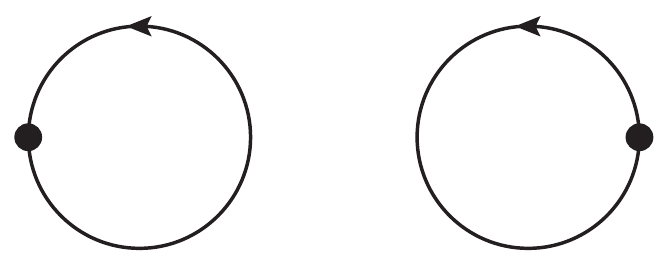}} 
    \caption{
     Feynman diagrams for the isospin symmetric contribution to the HVP.
     The dots represent the vector currents coupling to external photons.
     These diagrams represent gluon contributions to all orders.
    }
    \label{fig:diagramcd}
\end{figure}

\begin{figure}
  \centering
  \subfloat[Diagram M]{~~~~~~\includegraphics[height=1.5cm]{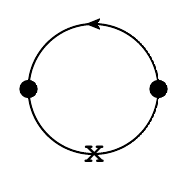}~~~~~~}
  ~~~
  \subfloat[Diagram O]{\includegraphics[height=1.5cm]{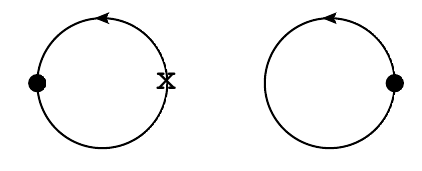}}\\
  \subfloat[Diagram R]{~~~~~~\includegraphics[height=3cm]{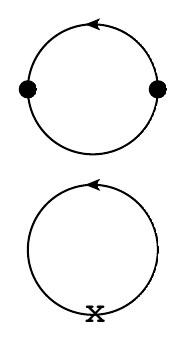}~~~~~~}
  \vspace{-0.15cm}
   ~~~
  \subfloat[Diagram $R_d$]{\includegraphics[height=2.85cm]{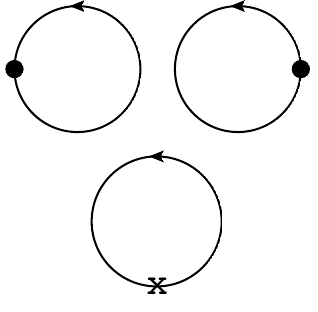}}
  \caption{Connected and disconnected strong-isospin breaking (SIB) diagrams.  The cross denotes the insertion of a scalar operator.  Also here each diagram represents gluon contributions to all orders.}
  \label{fig:mdiag}
\end{figure}

In this work we perform a calculation of the HVP contribution
to $a_\mu$ using the Euclidean time-momentum representation
\cite{Bernecker:2011gh}
\begin{equation}% sum only over positive t, this also leads to factor 8 instead of 4 for w_t
 a^{\rm HVP}_\mu = \sum_{t=0}^\infty w_t C(t)
 \label{eq:amuhvp}
\end{equation}
with sum over Euclidean time $t$
and
\begin{equation}
 C(t) = \frac{1}{12\pi^2} \int_0^{\infty}
 d(\sqrt{s}) R(s) s e^{-\sqrt{s}t}
 \label{eq:rratio}
\end{equation}
with R-ratio
$R(s)=(3s/4\pi\alpha^2)\,\sigma(s,e^+e^-\to{\rm had.})$.
We can also relate $C(t)$ to vacuum expectation values
of vector currents $V_\mu$ that we compute in lattice QCD+QED
as
\begin{align}\label{eq:correlationfn}
C(t) = \frac{1}{3} \sum_{i,\vec{x}}
 \langle V_i(\vec{x},t) V_i(\vec{0},0) \rangle \,,
\end{align}
where the sum is over spatial indices $i$ and all points $\vec{x}$ in the
spatial volume and
\begin{align}
 V_\mu &= \frac{2}{3}i (\bar{u} \gamma_\mu u + \bar{c} \gamma_\mu c) - \frac{1}{3}i (\bar{d} \gamma_\mu d + \bar{s} \gamma_\mu s + \bar{b} \gamma_\mu b)
\end{align}
with quark flavors $u,d,s,c,b$.  
In the absence of QED but the presence of a quark-mass splitting between up
and down quarks with individual quark masses
\begin{align}\label{eqn:isomassdef}
 m_u &= m_l - \Delta m \,, \notag\\
 m_d &= m_l + \Delta m \,,
\end{align}
the total up, down, and strange contributions can be written as
\begin{align}
 a_\mu^{\rm uds} &= a_\mu^{{\rm ud,}\conniso}
 + a_\mu^{{\rm s,}\conniso}
 + a_\mu^{{\rm uds,}\disciso} \notag\\
 &\quad + a_\mu^{\rm SIB,conn.}  + a_\mu^{\rm SIB,disc.} \,.
\end{align}
In this work, we focus on the connected contributions $a_\mu^{{\rm ud,}\conniso}$,
$a_\mu^{{\rm s,}\conniso}$, and $a_\mu^{\rm SIB,conn.}$, which we express as
\begin{align}\label{eqn:amuudconnisospin}
    a_\mu^{{\rm ud,}\conniso}
 &= 5a_\mu^{v}(m_l) \,, \\
 a_\mu^{{\rm s,}\conniso}
 &= a_\mu^{v}(m_s) \,,
\end{align}
where $m_v$ denotes the mass of the valence quark and
\begin{align}
a_\mu^{v}(m_v)
 &= \frac{1}{9}c(m_v)
\end{align}
in terms of the diagrams of Fig.~\ref{fig:diagramcd}.
The connected strong-isospin breaking (SIB) contribution can be written as
\begin{align} % -M is mass derivative (-scalar op), M=-c'
\label{eqn:amusib}
     a_\mu^{\rm SIB,conn.} &= \frac23 \Delta m M(m_v=m_l) \notag\\
     &= 3 \lambda_0 \frac{\kappa - 1}{\kappa + 1} \lim_{\lambda\to \lambda_0} \frac{\partial }{\partial \lambda}a_\mu^{v}(\lambda=m_v / m_s) 
\end{align}
where diagram M and O of Fig.~\ref{fig:mdiag} are related to diagrams c and d of Fig.~\ref{fig:diagramcd} by
\begin{align}\label{eqn:Mderivc}
 \frac{\partial}{\partial m_v}c(m_v) &= -2 M(m_v) \,, \\\label{eqn:Oderivd}
 \frac{\partial}{\partial m_v}d(m_v) &= -2 O(m_v) \,,
\end{align}
and
\begin{align}
 \lambda_0 &\equiv \frac{m_l}{m_s} \,, & \kappa \equiv \frac{m_u}{m_d} \,.
\end{align}
Both $\kappa$ and $\lambda_0$ are obtained from
 FLAG~2019~\cite{Aoki:2019cca}, where $\lambda_0$ is
 taken from 2+1+1 flavor simulations and $\kappa$
 is from 2+1 flavor simulations.
Only one 2+1+1 flavor result for $\kappa$ is available,
 so we choose to use the 2+1 flavor simulation so as
 not to tie our results to a single external measurement.
The values for these quantities are
\begin{align}
 \kappa &= 0.485(19) \,, &
 \lambda_0^{-1} &= 27.23(10).
\end{align}
%http://flag.unibe.ch/2019/Media?action=AttachFile&do=get&target=FLAG_qmass.pdf

The diagrams $R$ and $R_d$ do not contribute in the definition of the isospin symmetric
point given in Eq.~\eqref{eqn:isomassdef}.

%% dMrho/dMpi and dGrho/dMpi computed from:
%% https://arxiv.org/pdf/1907.01237.pdf
%% ensembles A30.32 and B25.32, with thermal subtraction

The weighting kernel in Eq.~(\ref{eq:amuhvp})
 is determined as
 \cite{Bernecker:2011gh,Lehner:2015bga}
\begin{align}
 w^f_t &= 8 \alpha^2 m_\mu^2 \int_0^\infty ds K(s,m_\mu) f(t,\sqrt{s}) \,, \\
 K(s,m_\mu) &= \frac{s Z(s,m_\mu)^3(1-s Z(s,m_\mu))}{1 + m_\mu^2 s Z(s,m_\mu)^2} \,, \\
 Z(s,m_\mu) &= \frac{\sqrt{s^2 + 4 m_\mu^2 s} - s}{2 m_\mu^2 s}\,,
 \label{eq:bmweight}
\end{align}
where we will use two alternative choices for the function $f$,
  \begin{align}
       f_p(t,q) &= \frac{\cos(qt) - 1}{q^2} + \frac12 t^2 \,, \\
 f_{\hat{p}}(t,q) &= \frac{\cos(qt) - 1}{(2\sin(q/2))^2} + \frac12 t^2 \,.
\end{align}
We refer to the choice $f=f_p$ as the $p$ prescription and to the choice of $f=f_{\hat{p}}$ as the $\hat{p}$ description.  Both are well-motivated within a lattice
calculation and differ only due to discretization errors.  We will provide results for
both and scrutinize the difference when considering uncertainties associated with the continuum limit.

\subsection{Window Method}

It is instructive to isolate specific ranges of
 Euclidean time in order to better understand
 their contributions to \amuv.
This can be accomplished by constructing windows
 that suppress contributions outside of the window
 region~\cite{Blum:2018mom}.
Rather than using Heaviside step functions to isolate
 these ranges,
 which would have significant dependence on the
 lattice cutoff near the boundary of the window,
 a smoothed step is
 considered~\cite{Bernecker:2011gh,Lehner:2017kuc}:
\begin{equation}
 \Theta(t,\mu,\Delta) = [1+{\rm tanh}[(t-\mu)/\Delta]] /2 .
 \label{eq:window}
\end{equation}
This step function suppresses all values below $\mu$
 and has a width parameterized by $\Delta$.
From these step functions, windows into specific
 regions of \amuv {} Euclidean time can be studied
 by instead summing the integral relation
\begin{align}\label{eq:hvpwindow}
 &a_\mu^{\rm v,W}(t_0,t_1,\Delta) \notag\\
 &= \sum_{t=0}^\infty w_t C(t)[ \Theta(t,t_1,\Delta) - \Theta(t,t_0,\Delta) ]\,.
\end{align}
We will quote results both for the total contribution,
corresponding to $t_0 \to -\infty$ and $t_1\to \infty$,
as well as specific windows.  
It should be noted that
windows of \amuv {} that isolate specific Euclidean distance scales
can be related to specific windows of time-like $s$
in the experimental data used in the R-ratio \cite{Blum:2018mom}.

\subsection{Parity Improvement}
\label{sec:ipa}

When performing computations with staggered quarks,
 parity projections are not possible and
 correlation functions receive contributions
 from parity partner states.
These parity partners have different spin and taste
 quantum numbers and constitute unwanted contributions
 to the correlation function.
The unwanted contributions come as oscillating terms
 with a prefactor proportional to $(-1)^{t/a}$.
To suppress these contributions to the correlation functions,
 we also study the
Improved Parity Averaging (IPA) procedure
 which averages
 neighboring timeslices~\cite{Bailey:2008wp},
\begin{equation}
 C^{\rm IPA}(t) = \frac{e^{-m_\rho t}}{4} \left[
    \frac{C(t-1  )}{e^{-m_\rho  (t-1)   }}
 + 2\frac{C(t)}{e^{-m_\rho (t)}}
 +  \frac{C(t+1)}{e^{-m_\rho (t+1)}}
 \right] \,.
 \label{eq:smoothing}
\end{equation}
%see smoothed effective mass Eq.(37) of Ref.~\cite{Bailey:2008wp}
The correlation function times are weighted by
 exponential factors that reflect the falloff
 of the correlation function in order
 to better enforce the cancellation
 of oscillating parity partner contributions.
The exponent used is the $\rho$ meson mass,
 obtained from PDG~\cite{Tanabashi:2018oca},
 which is expected to give the best cancellation
 in the $\rho$ resonance peak.
The $\rho$ resonance region accounts for the
 majority of the contribution to \amuv,
 and so cancellation in this region would be
 most beneficial.  In the continuum limit
 the choice of $m_\rho$ is irrelevant, however,
 the IPA prescription using the rho mass is not well-motivated  for very short
 or long distances or for heavier quark-masses.

\section{Numerical setup}
\label{sec:numerical}

The computation in this work is performed
 with the Highly-Improved Staggered Quark
 (HISQ) action for both valence and sea quarks.
The ensembles were generated by the MILC
 collaboration~\cite{Bazavov:2014wgs},
 and details about these ensembles are given in 
 Table~\ref{table:ensembledata}.
\begin{table}[h]
\begin{tabular}{c|cccccccc}
 Ens & $L^3\times T$ & $w_0/a$ & $Z_V(\bar{s}s)$ & $M_{\pi_5}$ & $M_{\pi}L$\\
 \hline
 48c & $48^3\times64$  & $1.41490(60)$ & $0.99220(40)$ & $132.73(70)$ & 3.9 \\
 64c & $64^3\times96$  & $1.95180(70)$ & $0.99400(50)$ & $128.34(68)$ & 3.7 \\
 96c & $96^3\times192$ & $3.0170(23)$  & $0.9941(11)$  & $134.95(72)$ & 3.7 \\
\end{tabular}
\caption{
 List of ensemble parameters for ensembles used in this study.
 Table values reproduced from Table I of Ref.~\cite{Davies:2019efs}.
 The gradient flow scale in the continuum, $w_0=0.1715(9)~{\rm fm}$,
  taken from Eq.(18) of Ref.~\cite{Dowdall:2013rya}.
 \label{table:ensembledata}
}
\end{table}
For the staggered quark action,
 the vector current operator
 is written
\begin{equation}
 V_i(x_4) = \sum_{\vec{x}} (-1)^{x_i/a} \epsilon(x) \bar\chi(\vec{x},x_4) \chi(\vec{x},x_4)
 \label{eq:vectorcurrentstaggered}
\end{equation}
 where $\chi(\vec{x},x_4)$ is
 the staggered one-component spinor and
 $\epsilon(x)$ is the usual staggered sign phase
\begin{equation}
 \epsilon(x) = (-1)^{\sum_\mu x_\mu/a}.
\end{equation}
This vector current bilinear has the advantage of
 being local to a single site.
Vector currents of other tastes may be formed
 by distributing the quark and antiquark over the
 unit hypercube, but these bilinear combinations
 require extra inversions and so were not explored.

Sources are inverted on random noise vectors that solve the Green's function equation
\begin{equation}
 \sum_y \slashed{D}_{xy}^{ab} G^{bc}_{y;t}
 = \eta^{ac}_{x} \delta_{x_0,t}
 \label{eq:greensfn}
\end{equation}
 with $\eta$ satisfying the condition
\begin{equation}
 \langle \eta^{ab}_x (\eta^{bc}_{y})^\dagger \rangle
 = \frac{1}{8} \delta^{ac} \delta_{xy}
 \prod_{i=1}^{3} (1-(-1)^{x_i/a}) \,.
 \label{eq:noisecondition}
\end{equation}
The phase factor on the RHS of Eq.~(\ref{eq:noisecondition})
 results from projecting out sites where $x_i/a$ is
 odd for at least one $i$.
When the propagator obtained from Eq.~(\ref{eq:greensfn})
 is contracted with its Hermitian conjugate at the source,
 the construction produces an operator that couples
 to many staggered spin-taste meson
 irreducible representations.
The vector current of Eq.~(\ref{eq:vectorcurrentstaggered})
 is contracted explicitly at the sink, projecting
 out the unwanted spin-taste irreps at the source
 and reproducing the correlation function of
 Eq.~(\ref{eq:correlationfn}) up to a factor of 8.
In the propagator solutions for the Dirac equation,
 the Naik epsilon term set to zero.

Results are computed on three ensembles
 with 2+1+1 flavors of sea quarks and
 up to 7 choices of valence quark mass
 per ensemble.
The parameters for the 3 ensembles used in this study are 
 given in Table~\ref{table:ensembledata}.
The sea quark masses are given in
 Table~\ref{table:ensemblequarkmasses}
 along with retuned quark masses for the strange quarks.
The valence quark masses used in this study are rational
 fractions $\lambda$ times the tuned strange quark masses
 from Table~\ref{table:ensemblequarkmasses}.
The list of rational fractions is given in
 Table~\ref{table:nconfigpermass}
 along with the number of time sources per configurations
 and the number of configurations used for each
 ensemble and mass combination.
\begin{table}[h]
\begin{tabular}{c|ccc|ccccc}
 Ens & $m_{\ell}$ & $m_{s}$ & $m_{c}$ & $m_{s,\text{tuned}}$ \\
 \hline
 48c & 0.00184 & 0.0507 & 0.628 & $0.05252(10)$ \\
 64c & 0.00120 & 0.0363 & 0.432 & $0.03636( 9)$ \\
 96c & 0.0008  & 0.022  & 0.260 & $0.02186( 6)$ \\
\end{tabular}
\caption{
 Sea and valence quark masses used for each ensemble.
 Sea quark masses are listed in Ref.~\cite{Davies:2019efs}.
 Valence quark masses are taken as fractional ratios of the tuned strange quark masses
 given in Table V of Ref.~\cite{Bazavov:2014wgs}.
 \label{table:ensemblequarkmasses}
}
\end{table}
\begin{table}[h]
\begin{tabular}{c|c|ccccccc}
 Ens & $t_{\text{src}}/$conf & \multicolumn{7}{c}{$N_{\text{conf}}$[$\lambda$]} \\
 \hline
 & &   $1/1$ & $3/4$ & $2/3$ & $1/2$ & $1/3$ & $1/6$ & $1/12$ \\
 \hline
 48c & 16            & 50 &  50 &  50 & 100 & 100 & 800 & 800 \\
 64c & $24(48^\ast)$ & 32 &  32 &  32 &  64 &  64 & 192 & $20/1540^\ast$ \\
 96c & 24            & 32 & $-$ & $-$ & $-$ &  32 & $-$ & $-$ \\
\end{tabular}
\caption{
 Number of configurations and time sources used for each ensemble
 and valence quark mass combination.
 The valence quark masses are quoted as ratios of the valence quark mass to the tuned
  strange quark mass, $\lambda\equiv m_{\text{valence}}/m_{s,\text{tuned}}$,
  obtained from Table~\ref{table:ensemblequarkmasses}.
 The 64c $\lambda=1/12$ mass point was inverted on double the number of time sources
  compared to the other 64c mass points, for a total of 48 time sources.
 This ensemble/mass point combination was computed using the
  truncated solver method \cite{Collins:2007mh} in an AMA setup \cite{Shintani:2014vja} with 20 configurations solved
  with full precision and 1540 configurations solved with a residual of $10^{-4}$.
 The 20 full-precision solves are used to correct the bias introduced by this procedure,
  which was tuned to have negligible impact on the results.
 \label{table:nconfigpermass}
}
\end{table}

\section{Results}
\label{sec:results}

\subsection{Bounding method}
\begin{figure}
 \centering
 \includegraphics[width=6.5cm,page=69]{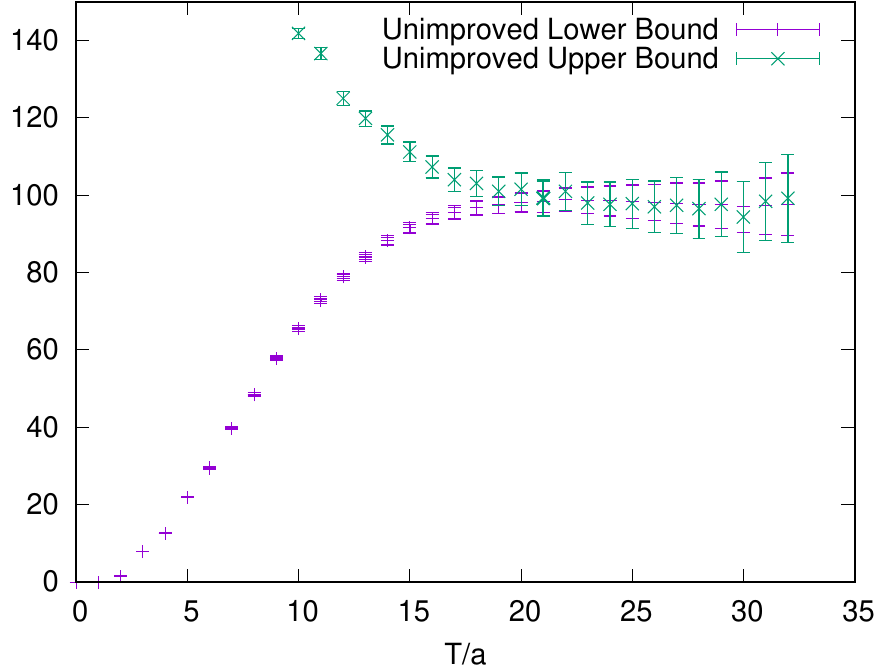}
 \includegraphics[width=6.5cm,page=69]{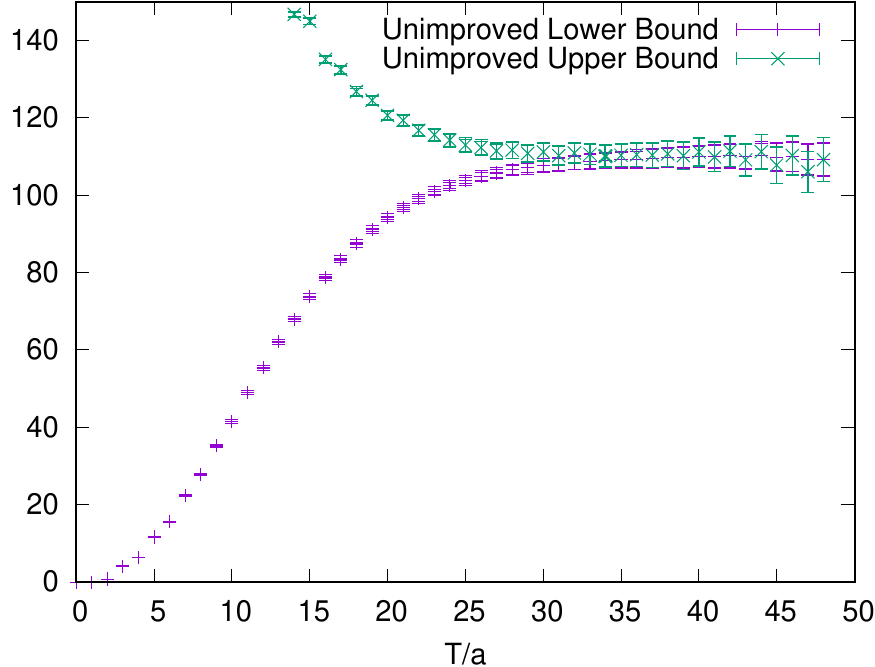}
 \includegraphics[width=6.5cm,page=69]{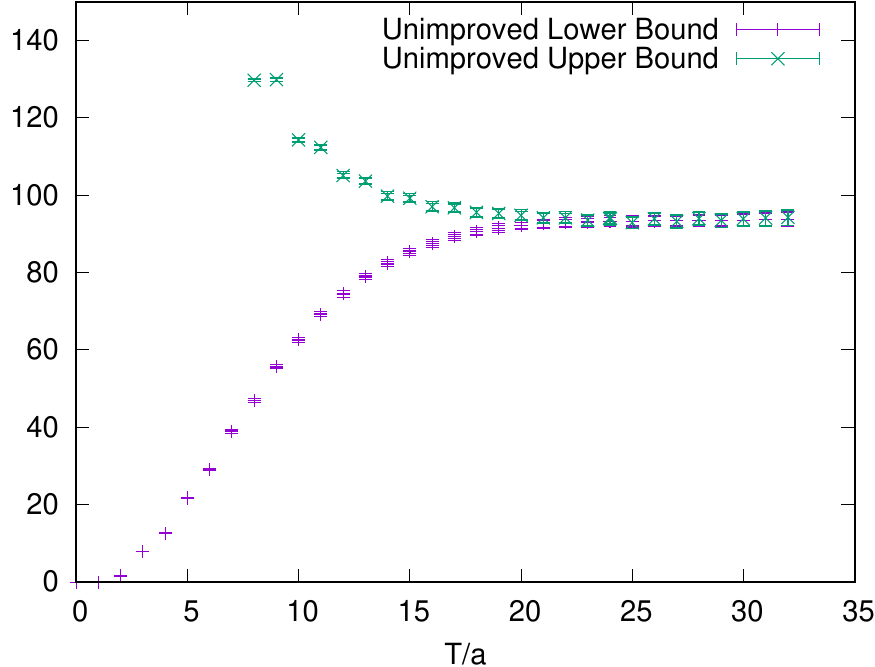}
 \includegraphics[width=6.5cm,page=69]{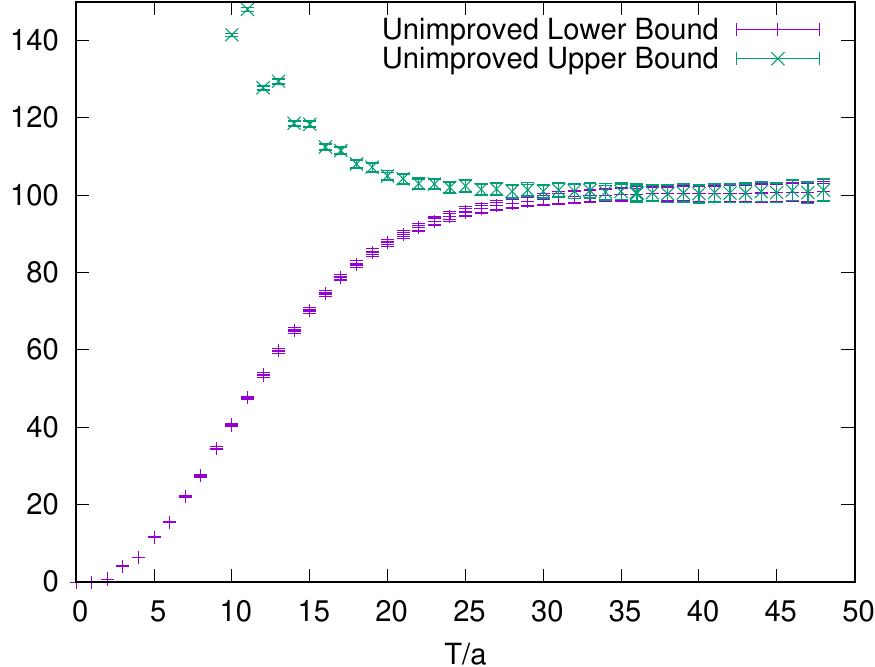}
 \caption{
 We show $10^{10}a_\mu^{v,\rm BND}(T)$ with $a_\mu^{v,\rm BND}(T) = \sum_{t=0}^\infty w_t C(t,T,\tilde{E})$ for both the upper bound $\tilde{E}$ equals to the free two-pion ground state energy as well as the lower bound $\tilde{E}=\infty$.  From top to bottom, the results are for $\lambda=1/12$ on the 48c ensemble, $\lambda=1/12$ on the 64c ensemble, $\lambda=1/6$ on the 48c ensemble, and $\lambda=1/6$ on the 64c ensemble.
 }
 \label{fig:bound}
\end{figure}

For the two lightest masses $\lambda=1/12$ and $\lambda=1/6$,
we also employ the bounding method \cite{LehnerTalkLGT16,Borsanyi:2017zdw,Blum:2018mom}
to create strict upper and lower bounds for $a^v_\mu$.  We show
 results for the total $a_\mu^v$ in Fig.~\ref{fig:bound}.  
In the bounding method, one replaces the correlator $C(t)$ by
\begin{align}
 \tilde{C}(t; T, \tilde{E}) =
\begin{cases}
C(t) & t < T \,, \\
C(T) e^{-(t-T) \tilde{E} } & t \geq T
 \end{cases}
\end{align}
which then defines a strict upper or lower bound of $C(t)$ for each $t$
given an appropriate choice of $\tilde{E}$.  For the upper bound we use $\tilde{E}$ equals to the free two-pion ground state energy and for the lower bound we use $\tilde{E}=\infty$
\cite{Blum:2018mom}.  We select the data points of $T=T_0$, where upper and lower bounds agree.  For $\lambda=1/12$, we use
$T_0/a=21$ for the 48c ensemble and $T_0/a=34$ for the 64c ensemble.  For $\lambda=1/6$, we use
$T_0/a=24$ for the 48c ensemble and $T_0/a=36$ for the 64c ensemble.

\subsection{Continuum Extrapolation}
The first part of the analysis consists of taking
 the continuum limit of each individual mass point.
This extrapolation is applied before considering
 extrapolations in quark mass or volume.
 
Figs.~\ref{fig:continuumextrap_fullwindow1o1}~and~%
 \ref{fig:continuumextrap_fullwindow1o3}
 show the extrapolation of the
 $m_v/m_s=1/1$~and~$1/3$ data to the continuum for both
 the unimproved and improved data.
These mass points have data for all three
 ensembles and are used to study the
 continuum extrapolation behavior of the windows.
A linear extrapolation is performed on the 64c and 96c
 ensembles to obtain the continuum limit.
A systematic uncertainty is obtained by taking
the difference between the 48c-64c extrapolation
and the 64c-96c extrapolation for both these
mass points.
We use the average of these systematic uncertainties for $\lambda=1/1$
and $\lambda=1/3$ as an additional systematic uncertainty that we apply
also to all other mass points, where there is no third ensemble.

\def\fwid{6.5cm}
\begin{figure}
 \centering
 \includegraphics[width=\fwid,page=2]{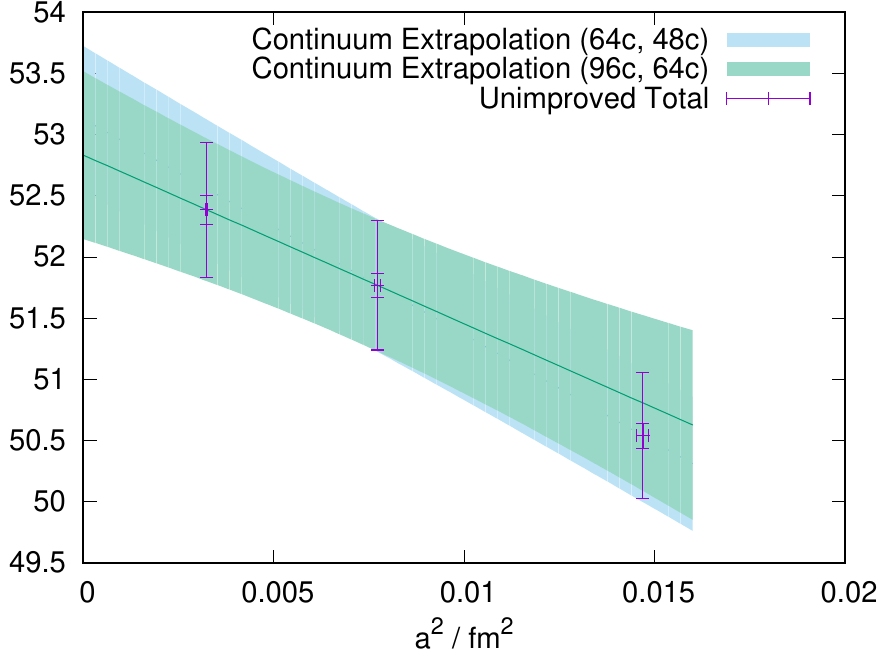}
 \includegraphics[width=\fwid,page=3]{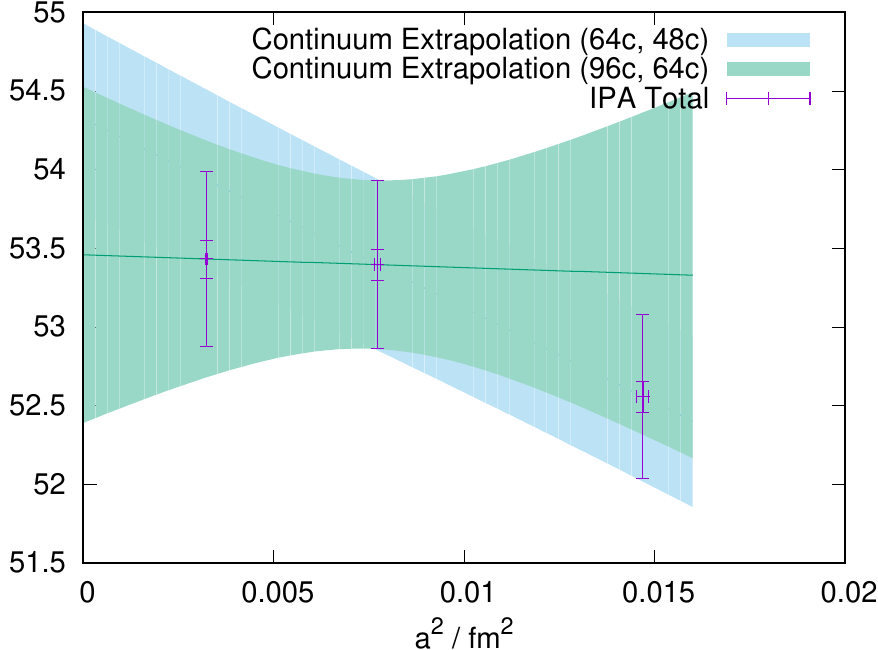}
 \includegraphics[width=\fwid,page=89]{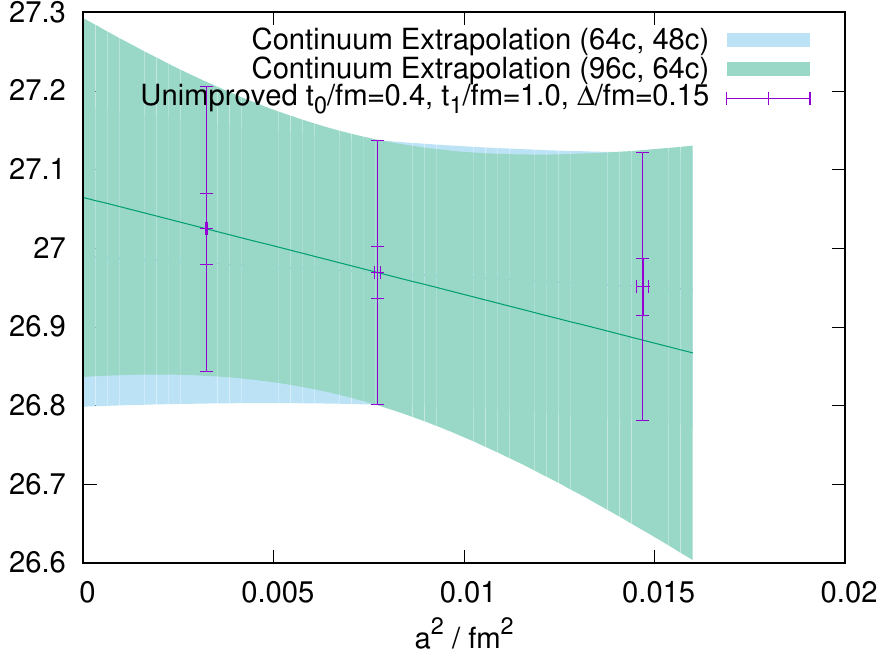}
 \includegraphics[width=\fwid,page=90]{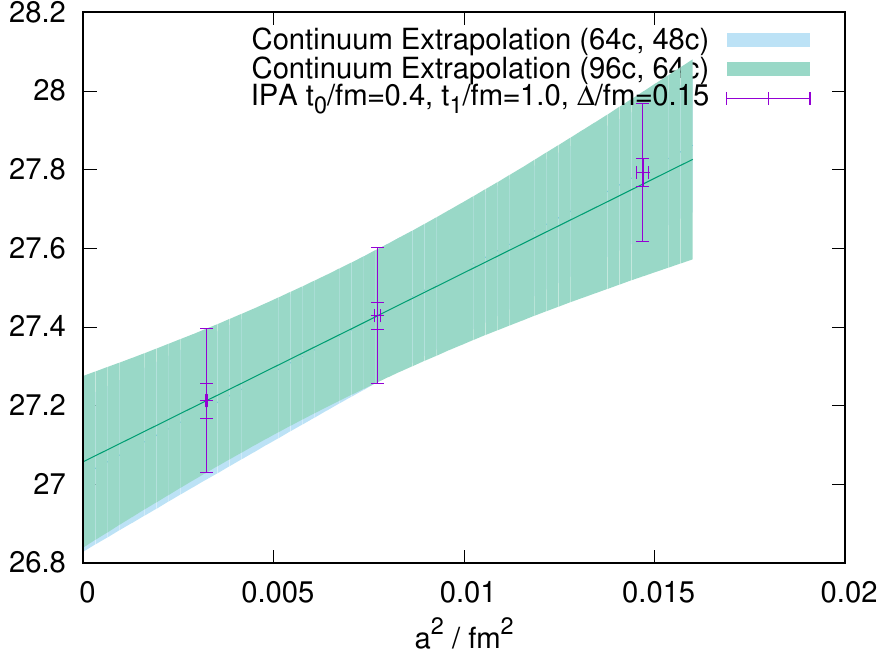}
 \caption{
  Continuum extrapolation of the $m_v/m_s=1/1$ data
  for both the total \amuv {} contribution
  and for the window with $(t_0,t_1)=(0.4,1.0)~{\rm fm}$.
  We show the unimproved
  data and as well as data with the parity improvement
  described in Section~\ref{sec:ipa}.
  The data include both statistical and systematic
  uncertainties.
  The extrapolations with either 48c and 64c or with
  64c and 96c ensembles are shown as shaded bands.
 }
 \label{fig:continuumextrap_fullwindow1o1}
\end{figure}

\def\fwid{6.5cm}
\begin{figure}
    \centering
    \includegraphics[width=\fwid,page=2]{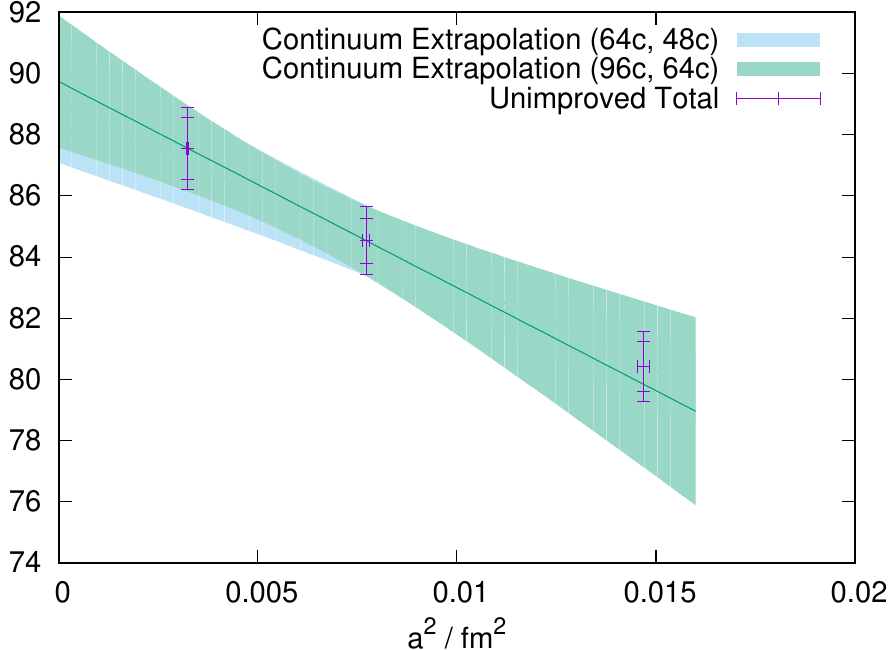}
    \includegraphics[width=\fwid,page=3]{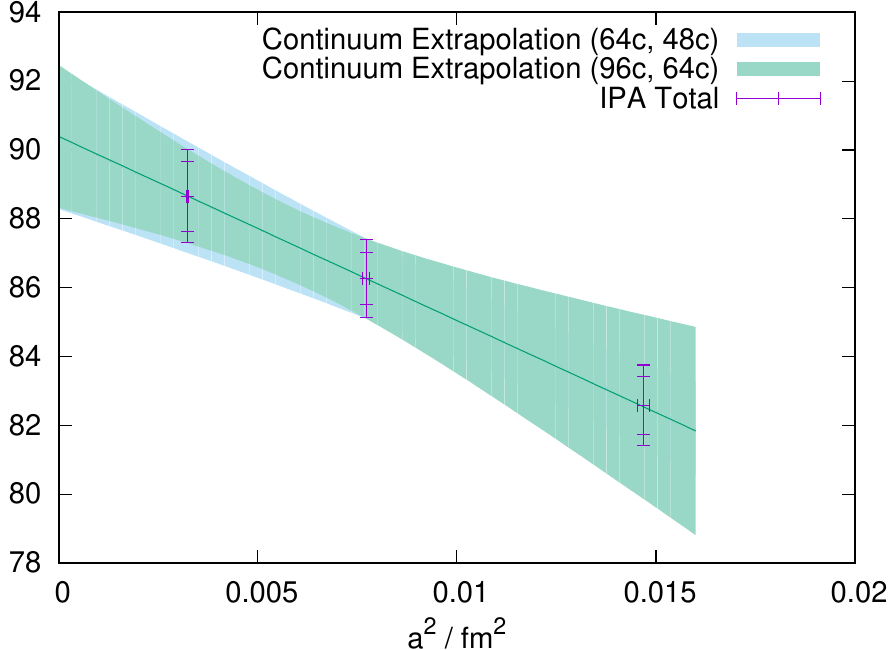}
    \includegraphics[width=\fwid,page=89]{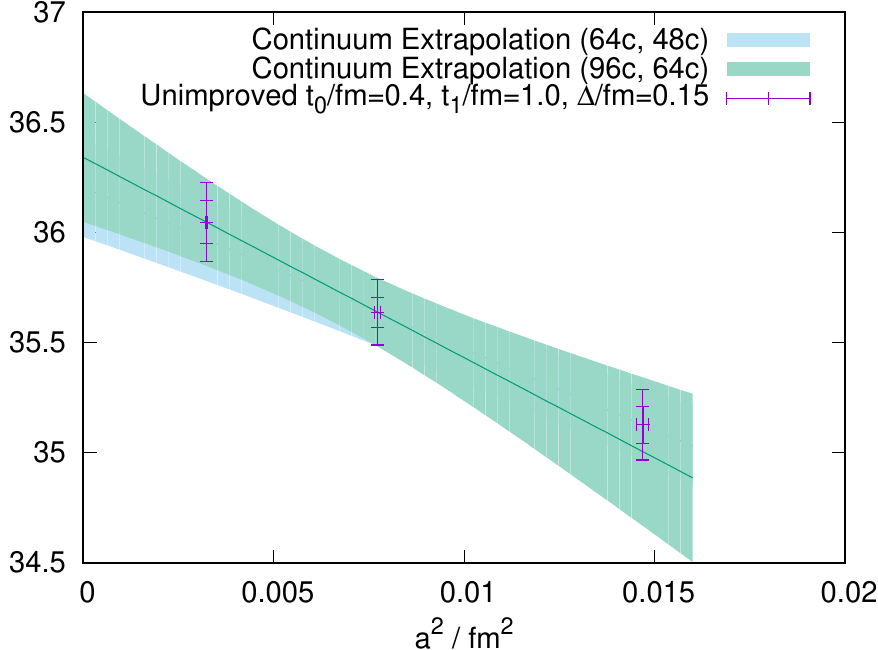}
    \includegraphics[width=\fwid,page=90]{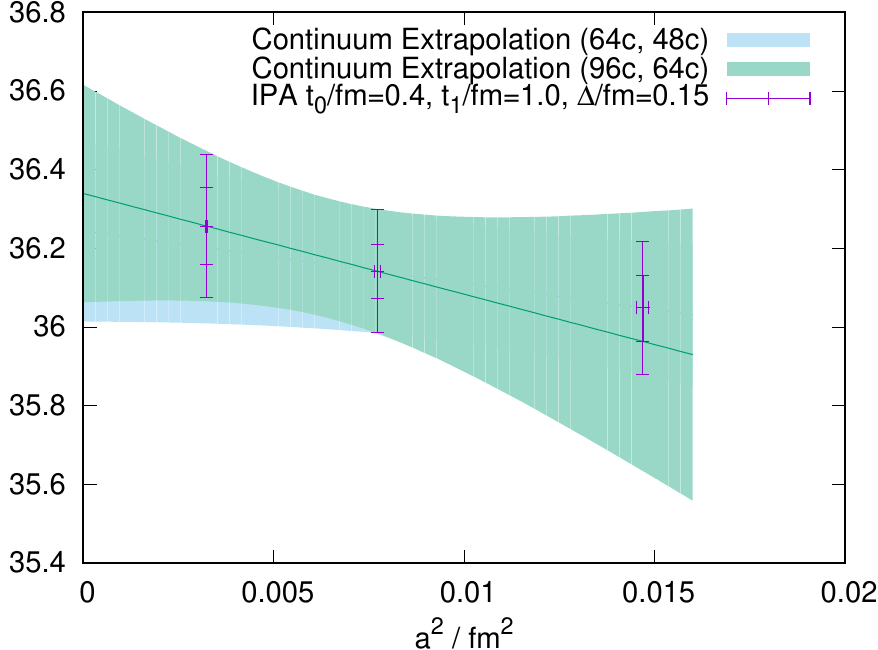}
    \caption{
    Same as Fig.~\ref{fig:continuumextrap_fullwindow1o1},
     but for $m_v/m_s=1/3$.
    }
    \label{fig:continuumextrap_fullwindow1o3}
\end{figure}

\begin{figure*}
 \centering
 \includegraphics[width=6.5cm,page=5]{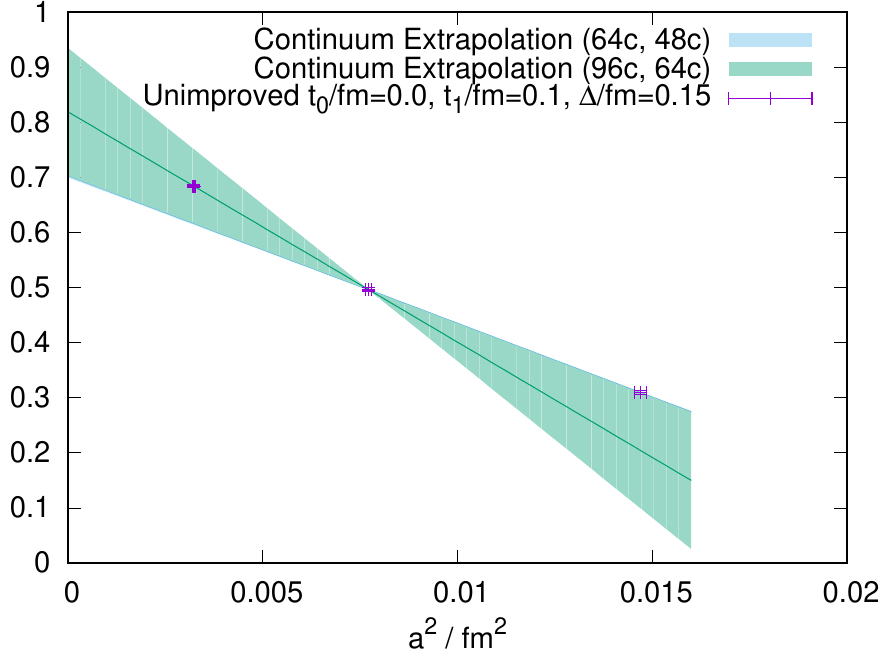}\hspace{1cm}
 \includegraphics[width=6.5cm,page=6]{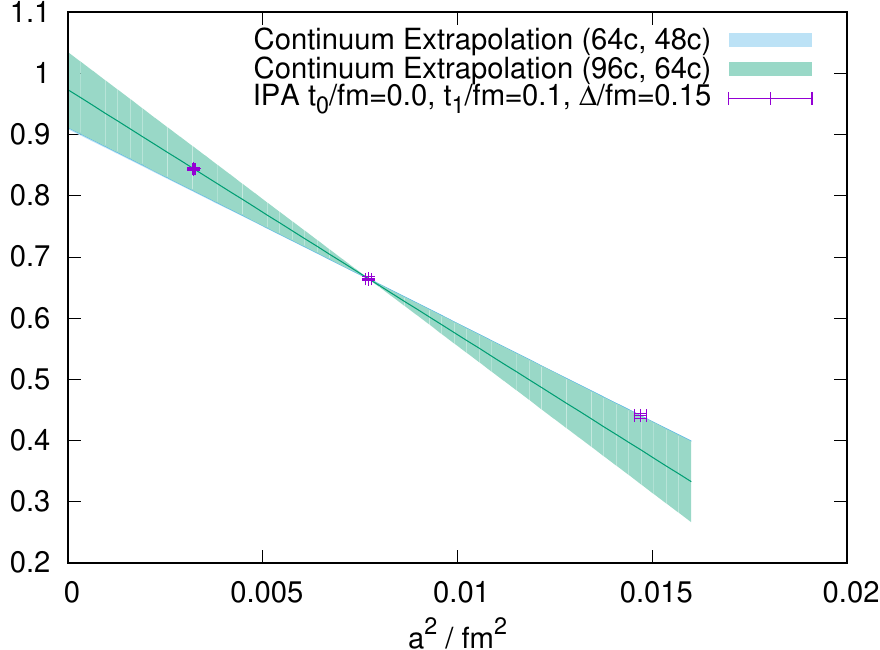}   \hspace{1cm} \includegraphics[width=6.5cm,page=11]{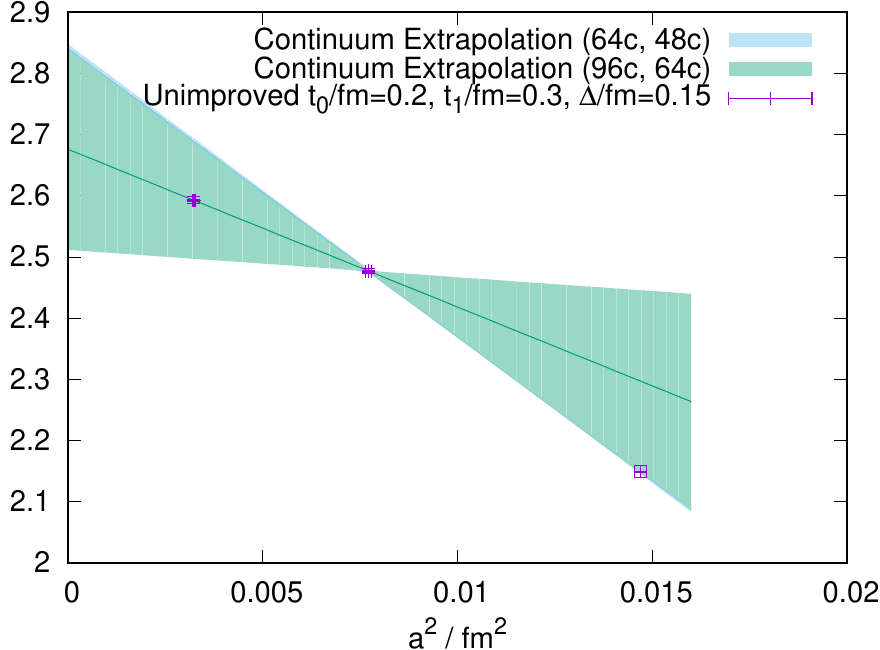}\hspace{1cm}
 \includegraphics[width=6.5cm,page=12]{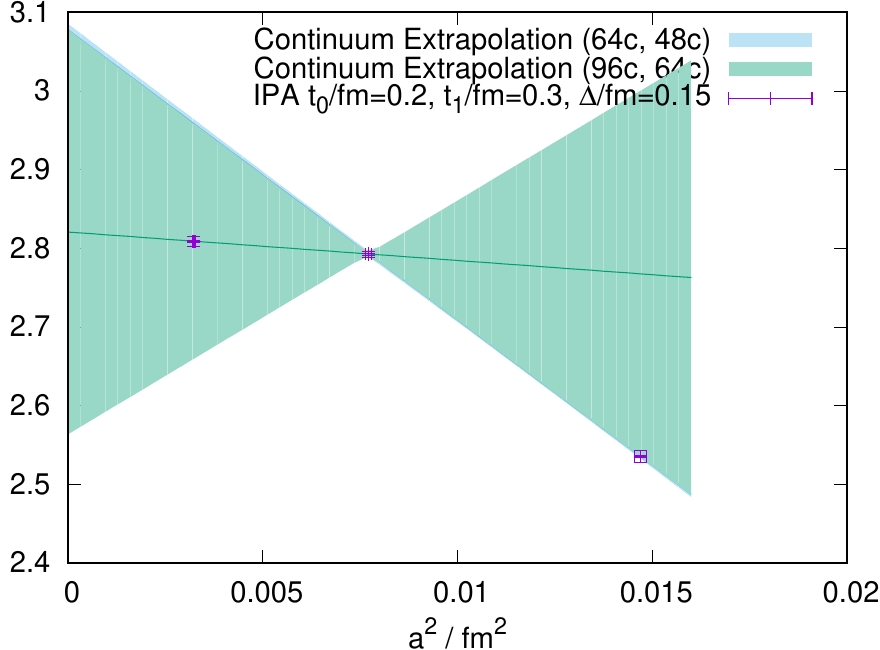}   \hspace{1cm} \includegraphics[width=6.5cm,page=17]{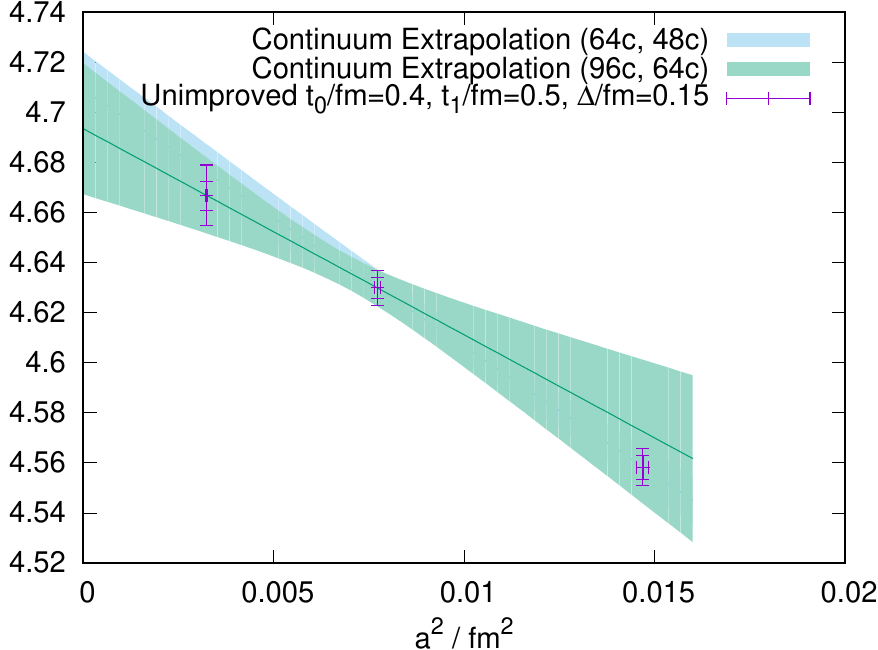}\hspace{1cm}
 \includegraphics[width=6.5cm,page=18]{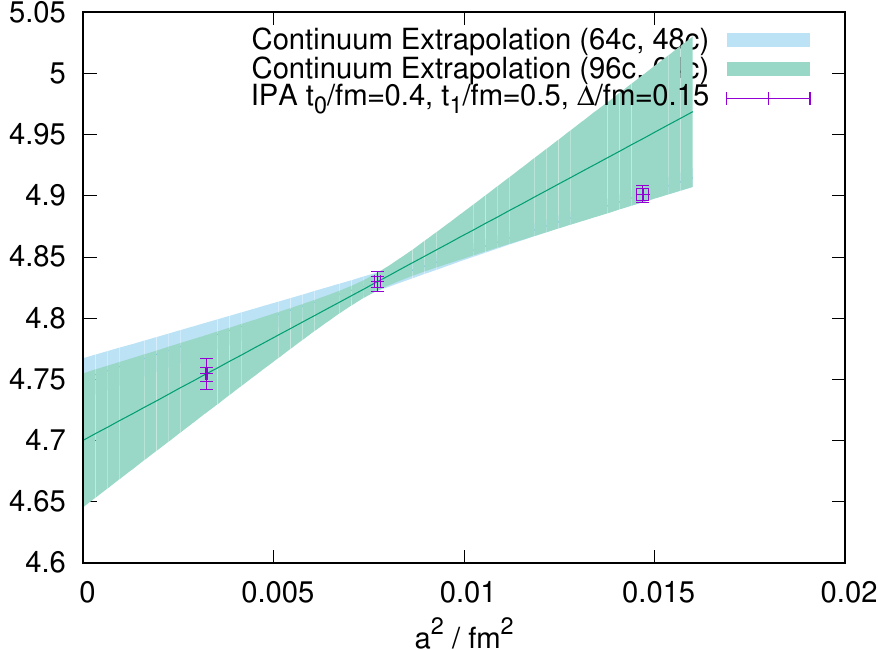}\hspace{1cm}
 \includegraphics[width=6.5cm,page=23]{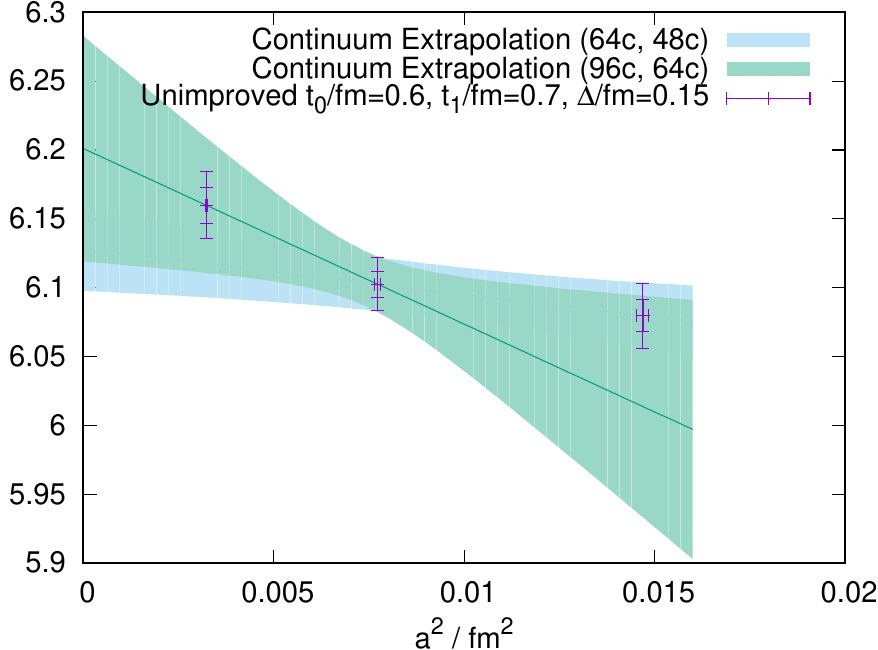}\hspace{1cm}
 \includegraphics[width=6.5cm,page=24]{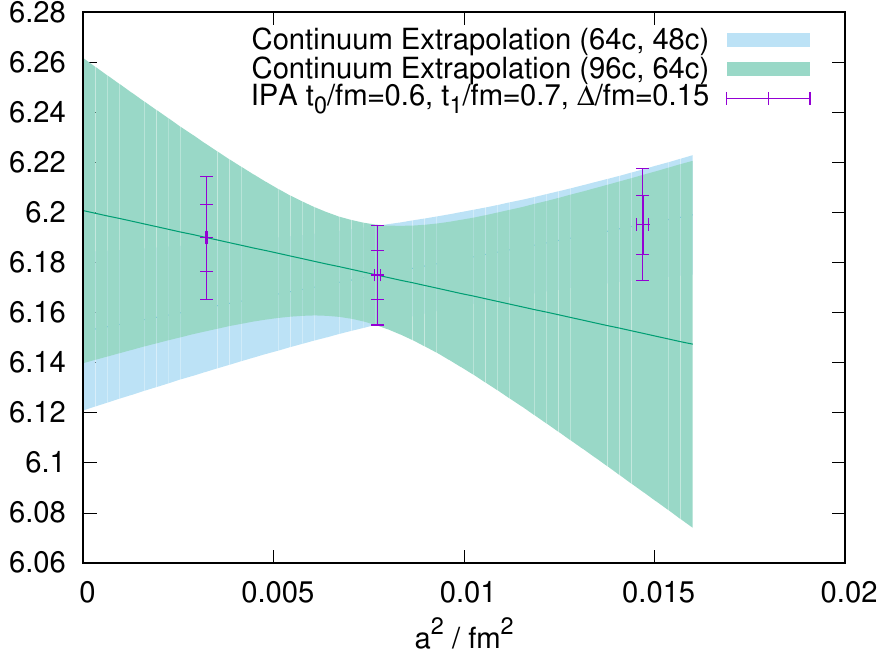}
 \caption{
 Windows with $t_1-t_0=0.1~{\rm fm}$ for the
 $m_v/m_s=1/3$ valence mass.
 Windows with $t_{0,1} \leq 0.8~{\rm fm}$ are included.
 We notice a significant uncertainty in the continuum extrapolation, which we may attribute to the somewhat small difference between $t_0$ and $t_1$.  We therefore also
 show results for $t_1-t_0=0.2$~fm in later tables.
 The left and right columns are for the unimproved
 data and for the data with the parity improvement,
 respectively.
 The parity improvement is only well-motivated in the
 $\rho$ resonance region, and so is not expected to work
 well for shorter distances.
 %continuum errors for narrow window are large;
 %therefore create broader windows as well 0-0.2,0.2-0.4,0.4-0.6,0.6-0.8.;
 %        additional note: IPA not well motivated for very short distances
 \label{fig:continuumextrap_smallwindow}
 }
\end{figure*}

\begin{figure*}
 \centering
 \includegraphics[width=6.5cm,page=29]{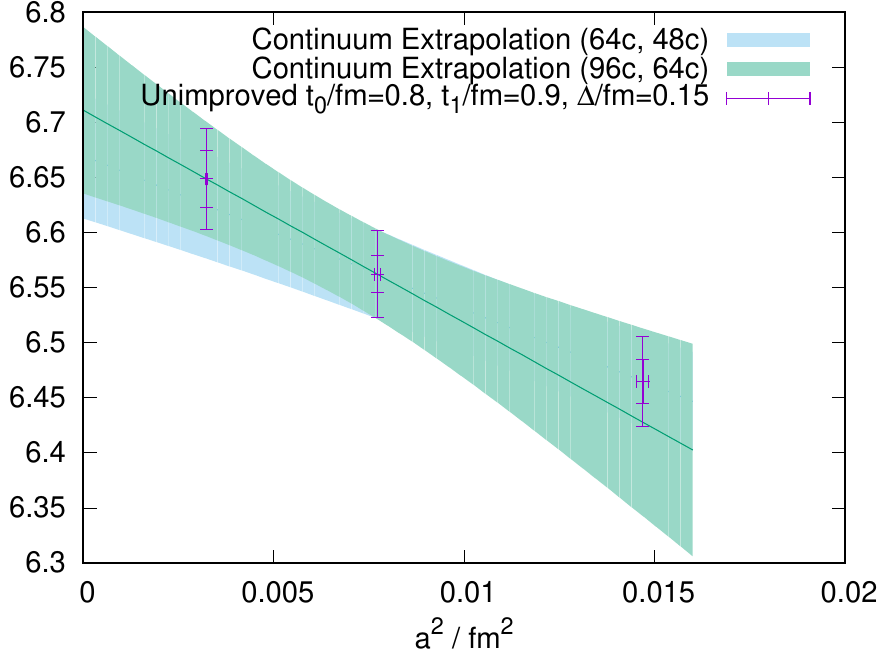}\hspace{1cm}
 \includegraphics[width=6.5cm,page=30]{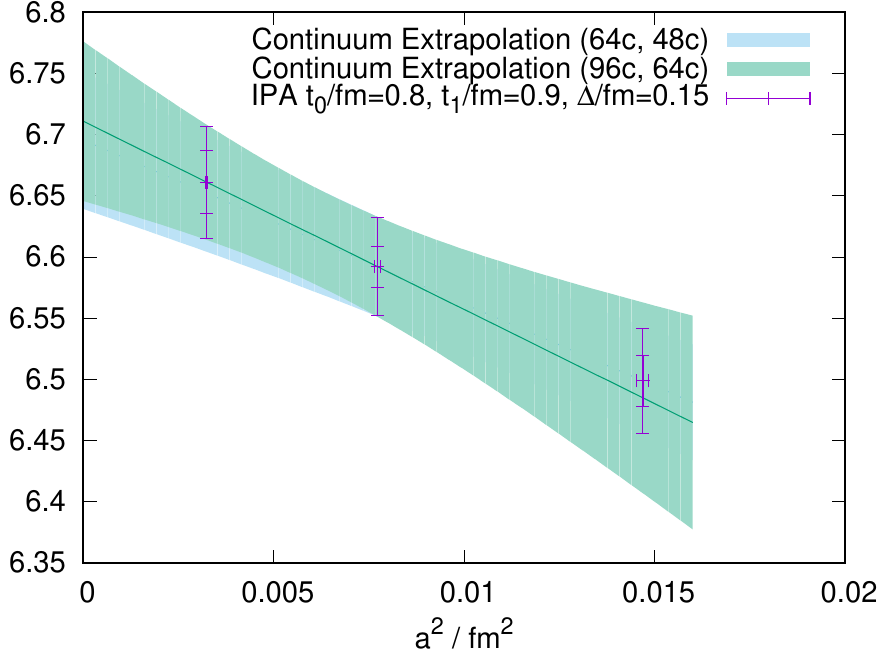}   \hspace{1cm} \includegraphics[width=6.5cm,page=35]{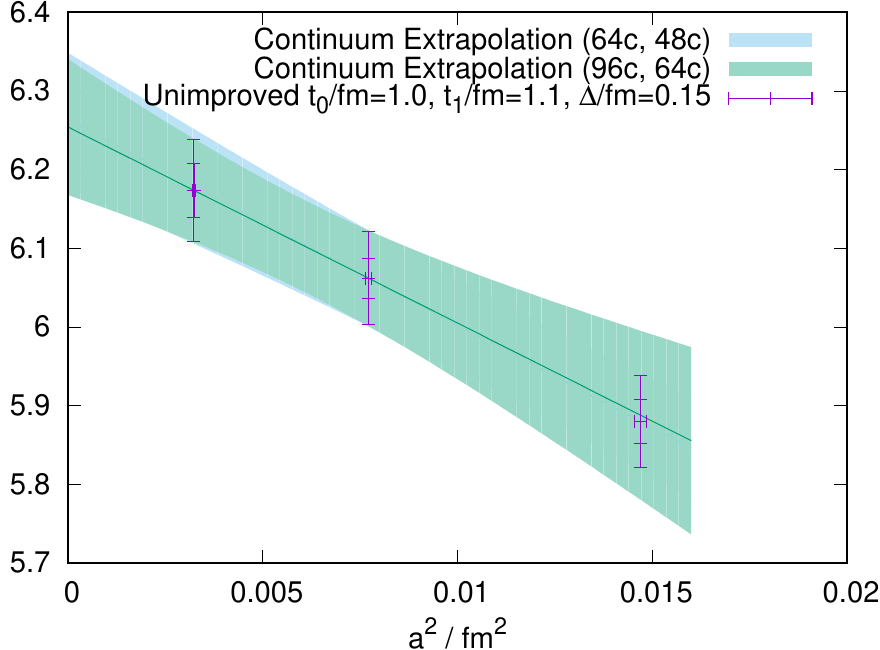}\hspace{1cm}
 \includegraphics[width=6.5cm,page=36]{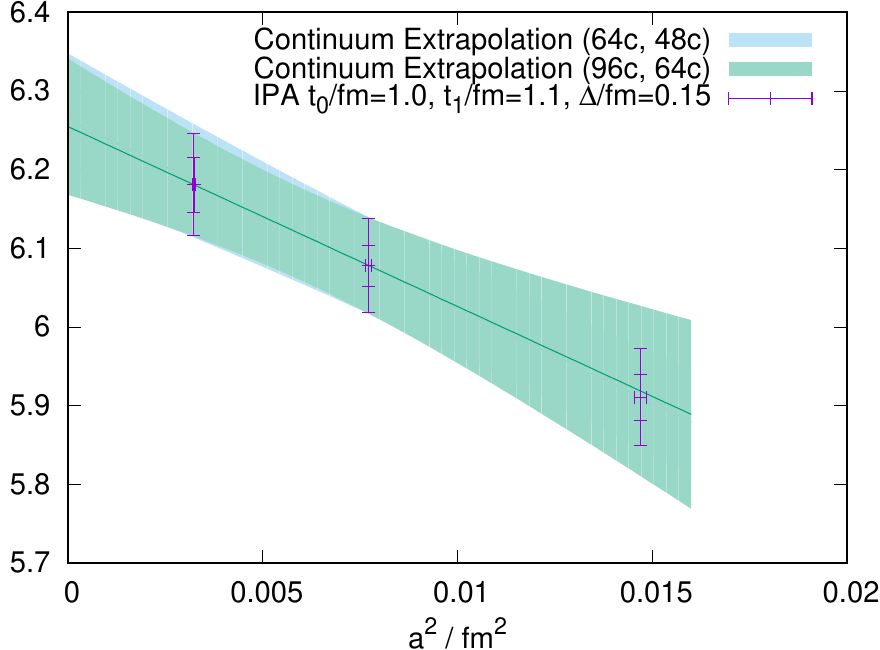}   \hspace{1cm} \includegraphics[width=6.5cm,page=41]{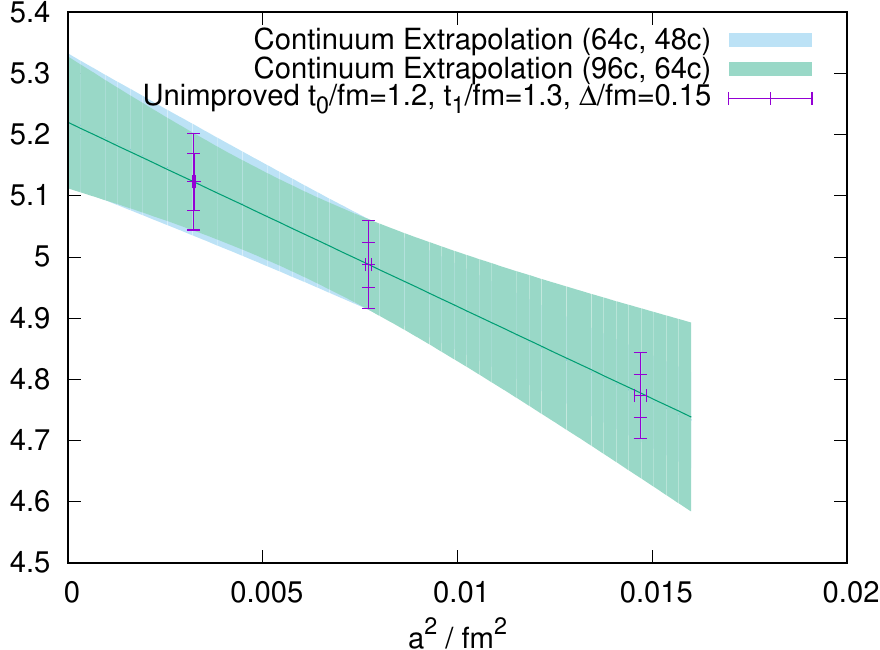}\hspace{1cm}
 \includegraphics[width=6.5cm,page=42]{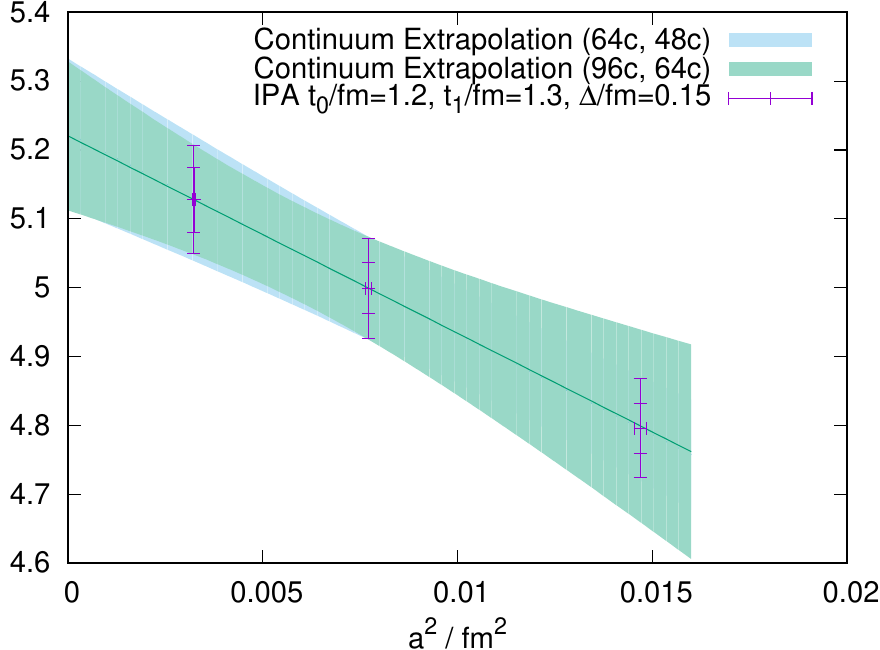}\hspace{1cm}
 \includegraphics[width=6.5cm,page=47]{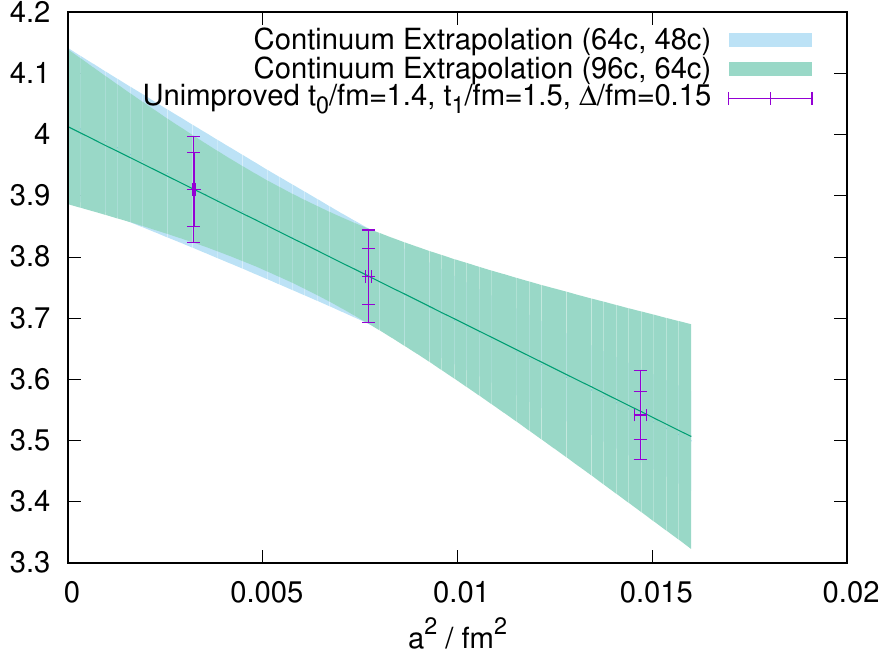}\hspace{1cm}
 \includegraphics[width=6.5cm,page=48]{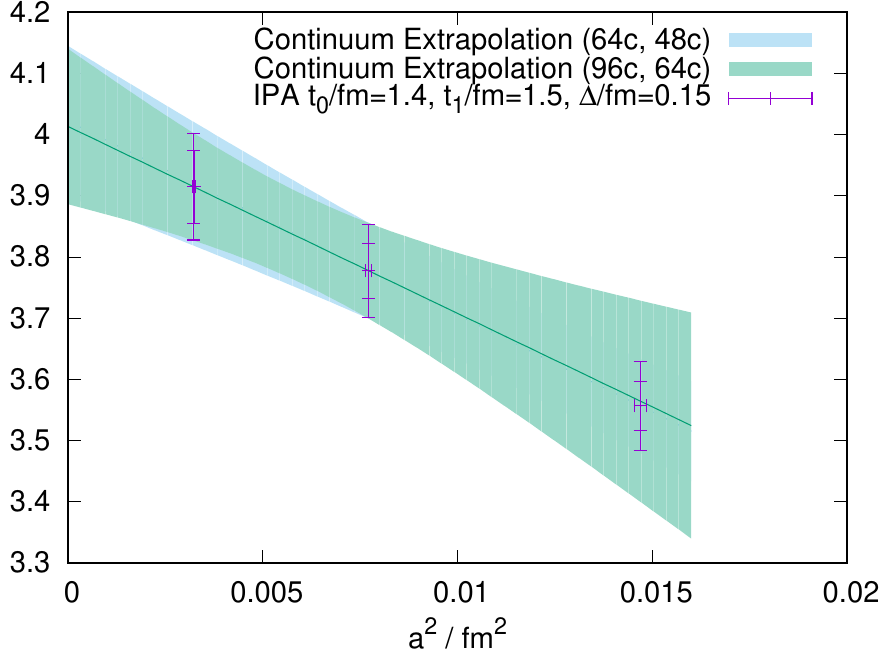}
 \caption{
 Same as Fig.~\ref{fig:continuumextrap_smallwindow},
 but for windows with $t_{0,1} \geq 0.8~{\rm fm}$.
 For long-distance windows, the continuum extrapolation
 is no longer an issue and fits are very consistent
 with data.
 %Cap; $m_v/m_s=1/3$; for windows from 0.8fm, even
 %narrow window continuum extrapolation errors appear small.
 \label{fig:continuumextrap_smallwindow2}
 }
\end{figure*}

Figs.~\ref{fig:continuumextrap_smallwindow}~and~%
 \ref{fig:continuumextrap_smallwindow2}
 show the continuum extrapolation for various choices
 of $t_1-t_0=0.1~{\rm fm}$ windows for both
 unimproved and improved data.
All of these windows are shown for the $m_v/m_s=1/3$
 mass point.
The continuum extrapolations for windows
 up to $t_0=0.8~{\rm fm}$ have visible
 contributions from discretization effects that are
 nonlinear in $a^2$.
Beyond $t_0=0.8~{\rm fm}$, the windows have no resolvable
 nonlinear discretization effects and are consistent
 well within statistical uncertainties for
 $t_0\geq1.0~{\rm fm}$.
The IPA is only well-motivated in the region of the
 $\rho$ resonance peak, and has larger discretization
 effects for short-distance windows.
In the long-distance windows, IPA becomes identical
 to the unimproved data.

%Fig.~\ref{fig:continuumextrap_midwindow}
% shows the effect of increasing the bounds of the
% windows from $t_1-t_0=0.1~{\rm fm}$ to $0.2~{\rm fm}$.
%The large nonlinear discretization effects seen
% in Fig.~\ref{fig:continuumextrap_smallwindow}
% are suppressed by increasing the window size,
% but overall are still large for short-distance windows.
%
%\begin{figure*}
%    \centering
%    \includegraphics[width=6.5cm,page=65]{figures/phat/ms1o3/extrap}\hspace{1cm}
%    \includegraphics[width=6.5cm,page=68]{figures/phat/ms1o3/extrap}\hspace{1cm}
%    \includegraphics[width=6.5cm,page=71]{figures/phat/ms1o3/extrap}\hspace{1cm}
%    \includegraphics[width=6.5cm,page=74]{figures/phat/ms1o3/extrap}
%    \caption{Cap; $m_v/m_s=1/3$; for broader windows the continuum extrapolation is generally better; still the IPA is not well motivated for too short distances;}
%    \label{fig:continuumextrap_midwindow}
%\end{figure*}

\begin{figure*}
 \centering
 \includegraphics[width=6.5cm,page=98]{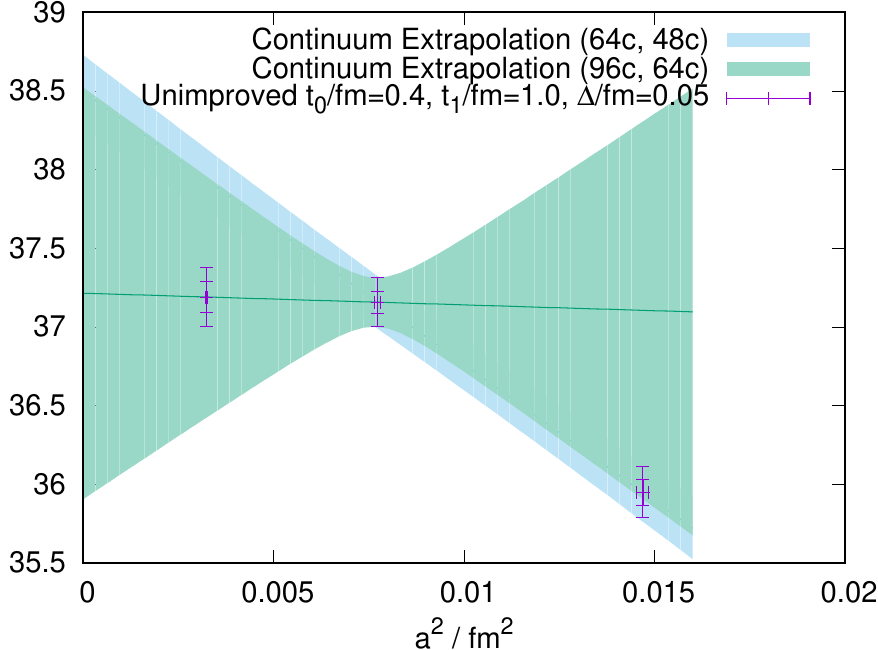}\hspace{1cm}
 \includegraphics[width=6.5cm,page=99]{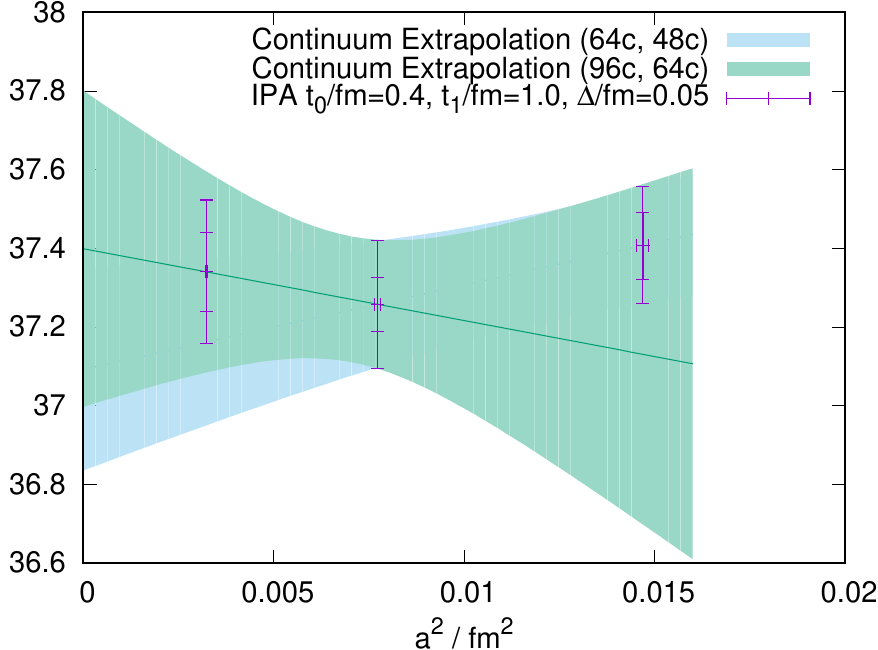}   \hspace{1cm} \includegraphics[width=6.5cm,page=101]{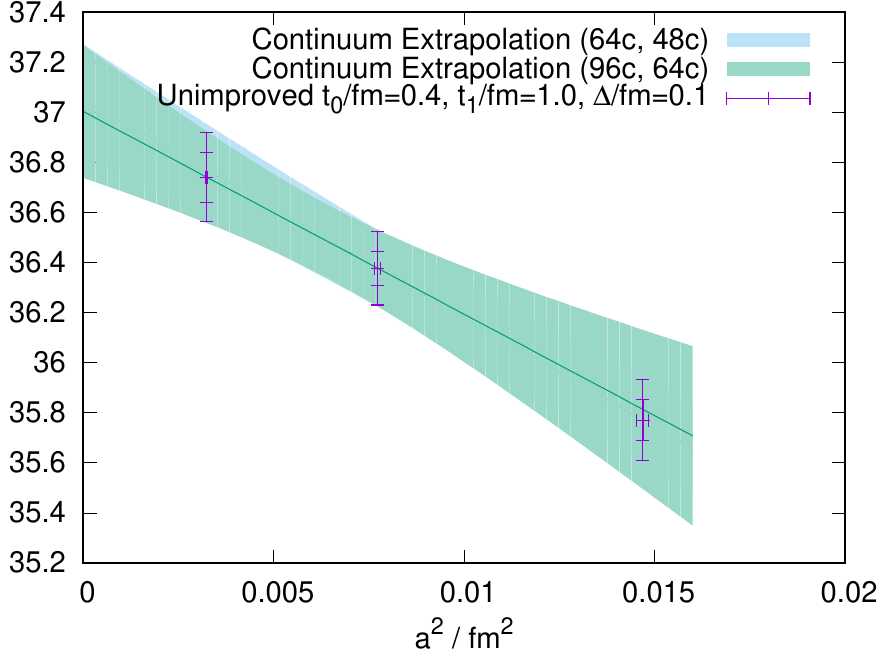}\hspace{1cm}
 \includegraphics[width=6.5cm,page=102]{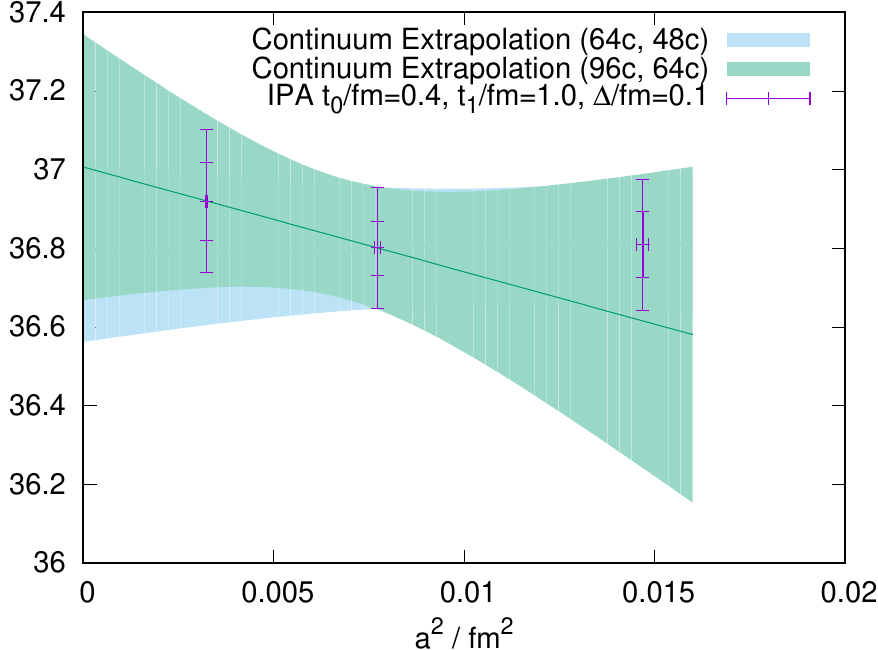}   \hspace{1cm}
 \includegraphics[width=6.5cm,page=89]{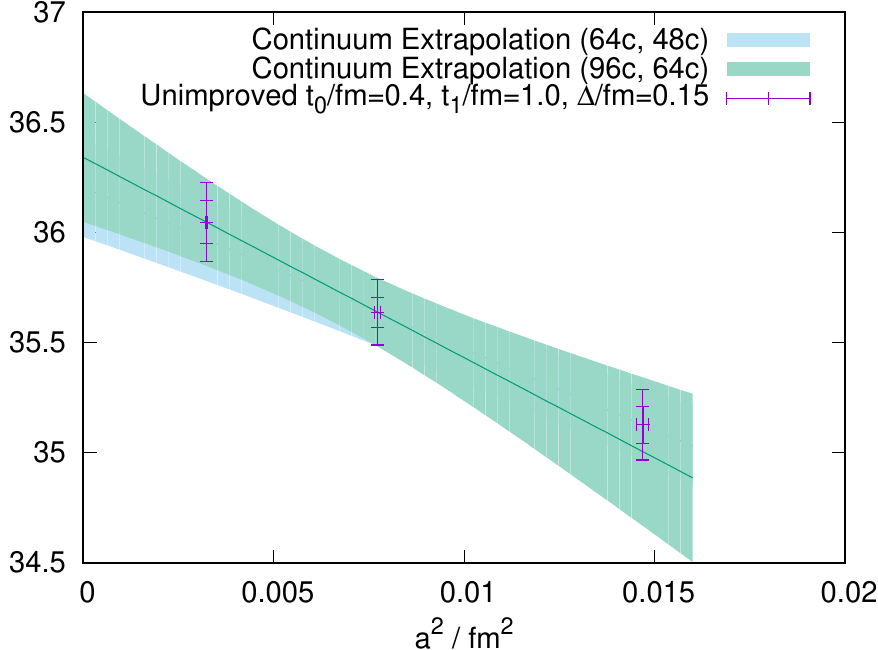}\hspace{1cm}
 \includegraphics[width=6.5cm,page=90]{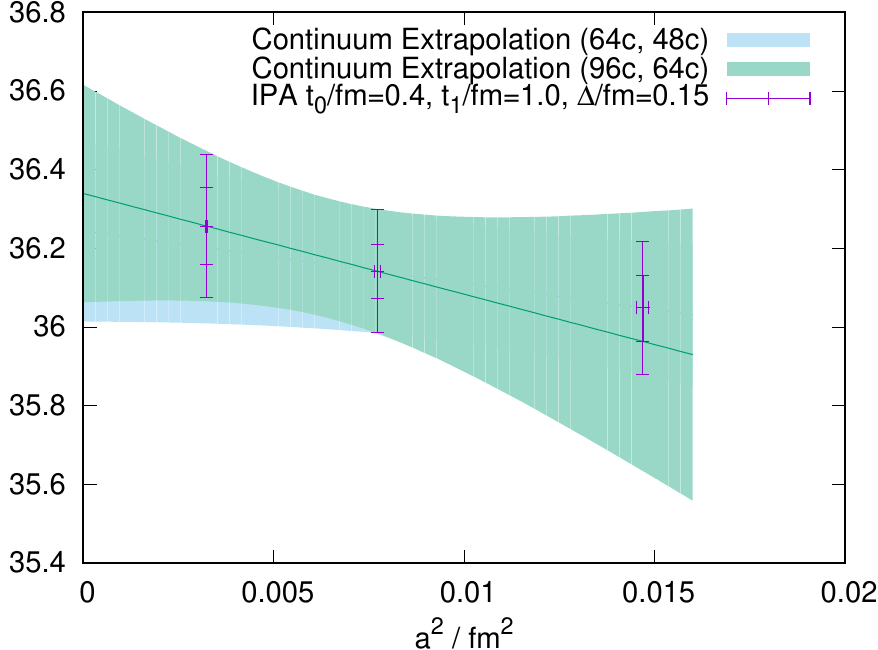}\hspace{1cm}
 \includegraphics[width=6.5cm,page=104]{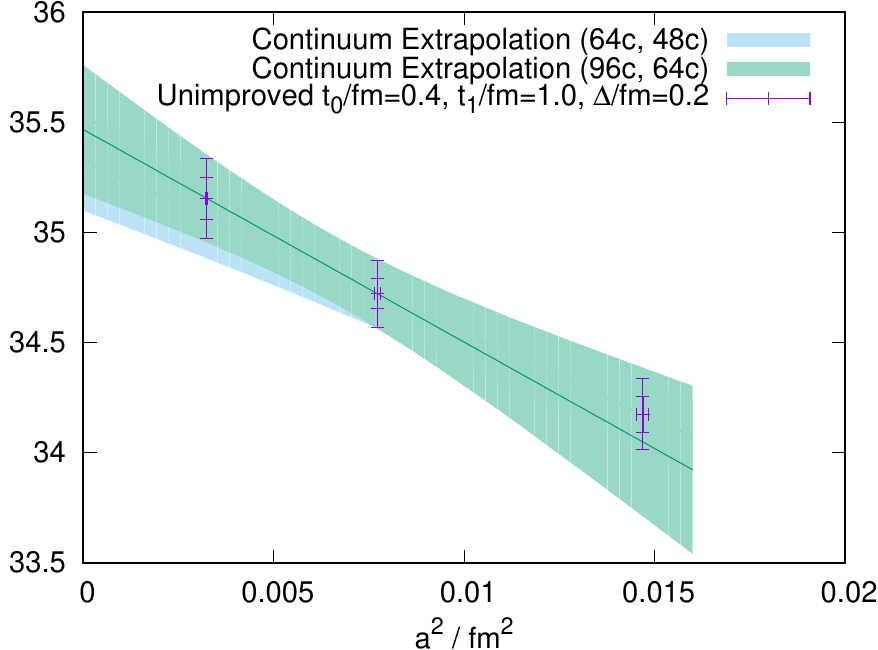}\hspace{1cm}
 \includegraphics[width=6.5cm,page=105]{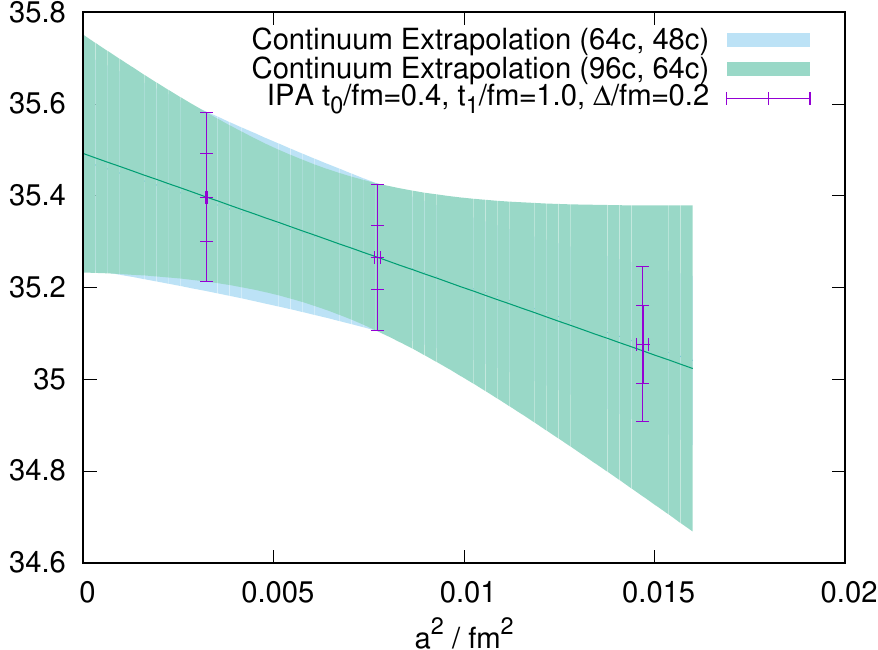}
 \caption{
 Continuum extrapolations with $m_v/m_s=1/3$ for various
 choices of window smearing $\Delta$.
 All windows have $(t_0,t_1)=(0.4,1.0)~{\rm fm}$.
 The left and right columns are for the unimproved
 data and for the data with the parity improvement,
 respectively.
 The continuum extrapolation errors are
 enhanced when $\Delta$ is too small,
 but the IPA softens this issue.
 \label{fig:continuumextrap_delta}
 }
\end{figure*}

Fig.~\ref{fig:continuumextrap_delta} demonstrates
 the effect of the window smearing parameter $\Delta$
 on the continuum extrapolation.
If $\Delta$ is smaller than the lattice spacing,
 the window turns on or off rapidly and can
 resolve contributions from individual timeslices.
These discretization effects are clearly visible
 for the coarsest ensembles when $\Delta$ is too small.
This is cleaned up by increasing the window smearing.
The IPA also smears neighboring timeslices,
 which reduces the effect of discretizations
 for even the smallest window smearings.

\begin{figure*}
 \centering
 \includegraphics[width=7cm,page=1]{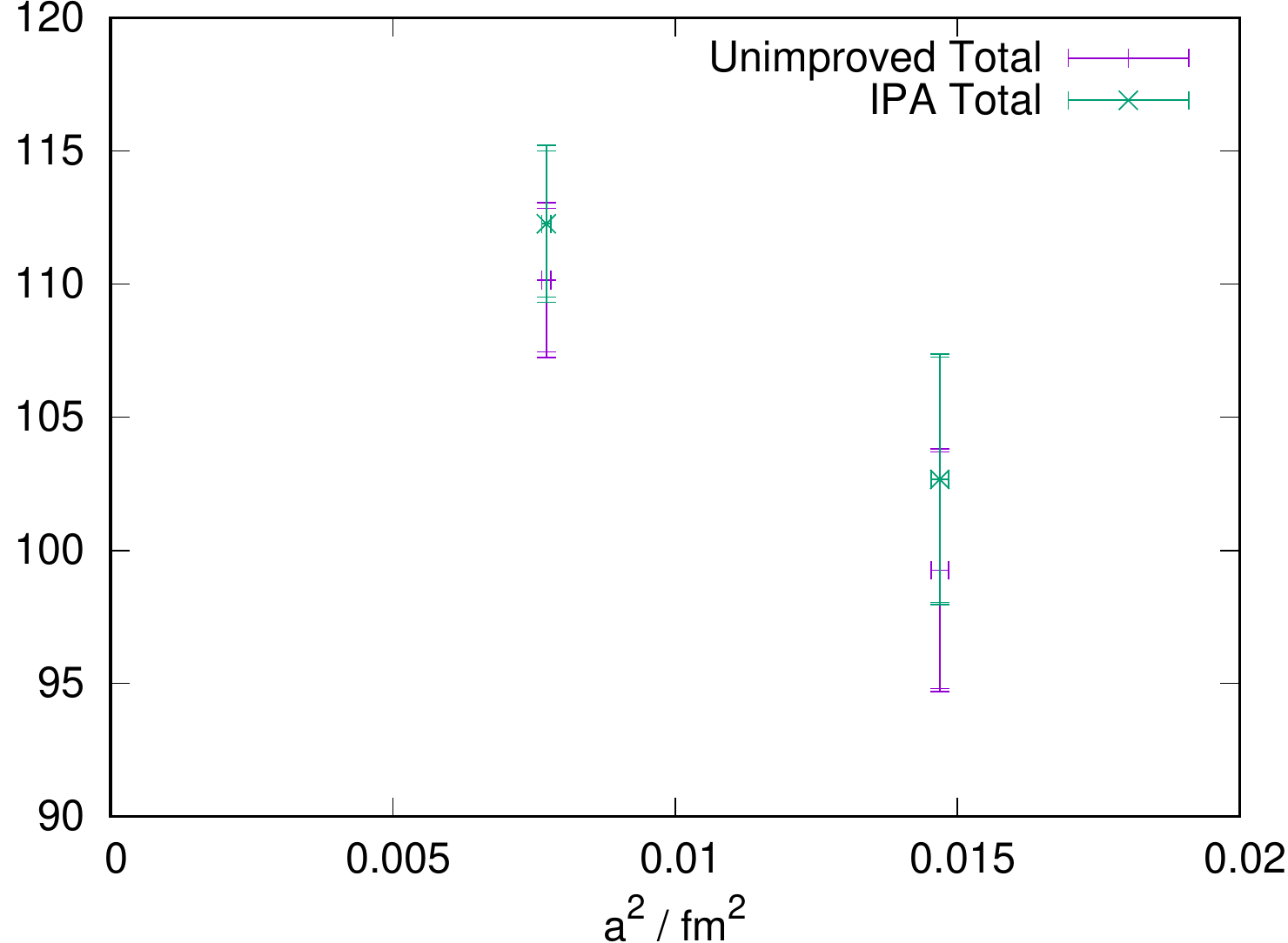}\hspace{1cm}
 \includegraphics[width=7cm,page=1]{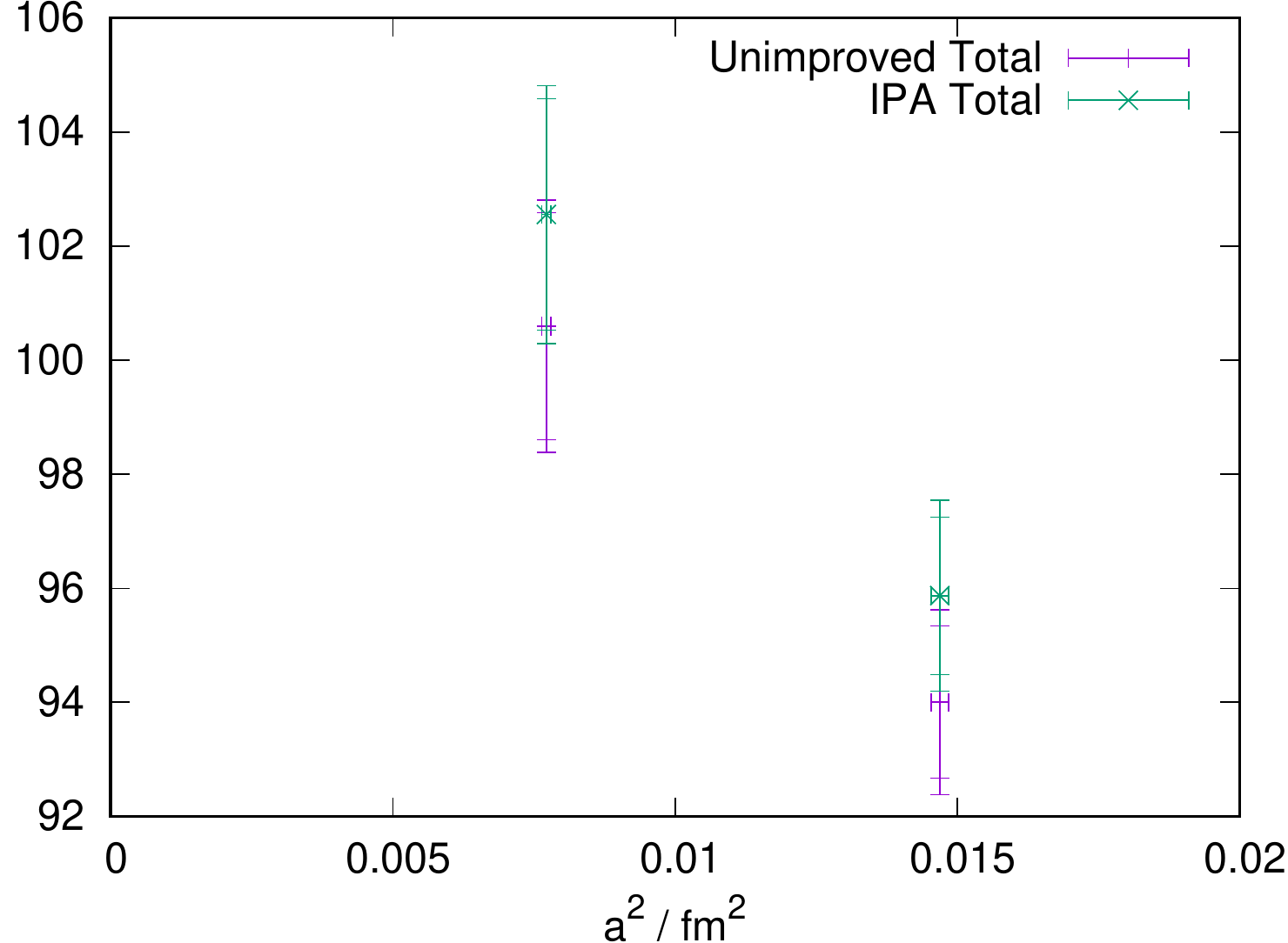}\hspace{1cm}
 \includegraphics[width=7cm,page=1]{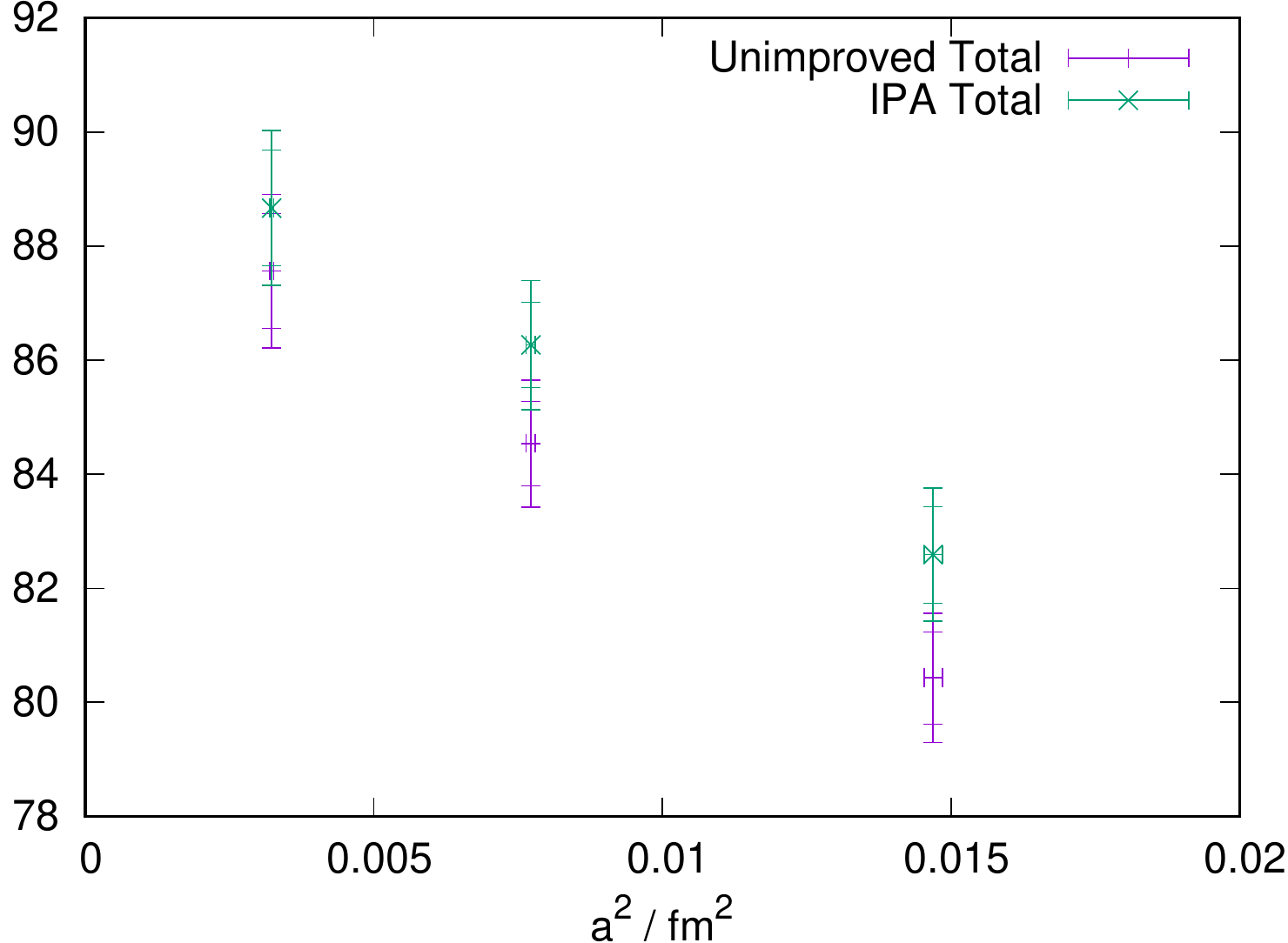}\hspace{1cm}
 \includegraphics[width=7cm,page=1]{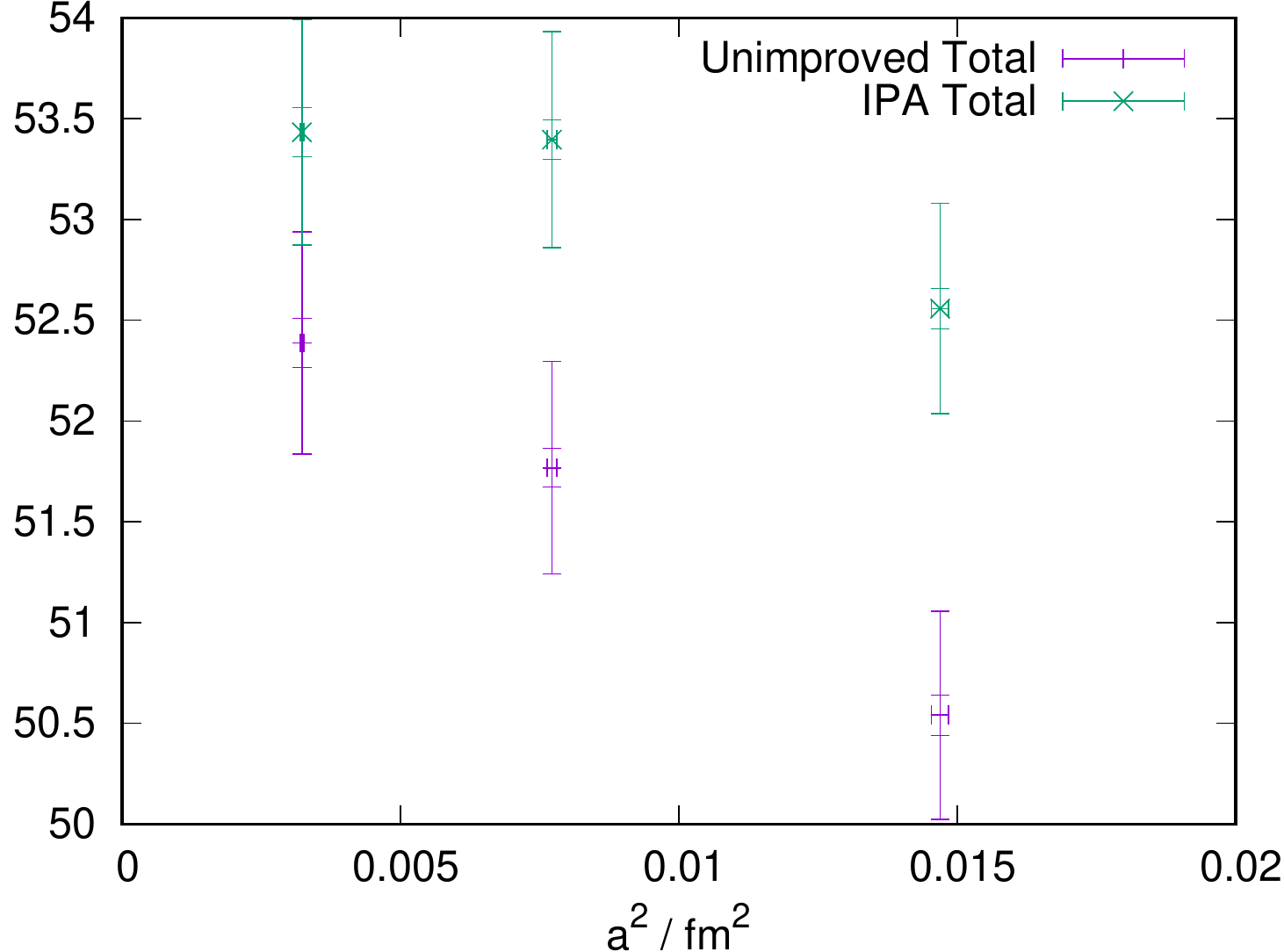}
 \caption{
 Unimproved versus parity improved data as a function
 of lattice spacing.
 The upper-left, upper-right, lower-left, and lower-right
 plots have $m_v/m_s=1/12,~1/6,~1/3,$~and~$1$, respectively.
 The parity improved data with $\rho$ parameters
 is not well motivated for the strange quark masses
 and is not used for the final analysis.
 }
 \label{fig:unimp_ipa}
\end{figure*}

Fig.~\ref{fig:unimp_ipa} shows the difference between
 the IPA procedure of Eq.~(\ref{eq:smoothing})
 and the unimproved total result for
 several choices of valence quark mass and lattice spacing.
The improvement with the $\rho$ resonance mass
 in Eq.~(\ref{eq:smoothing})
 is only well motivated for quark masses close to
 the isospin-symmetric valence quark mass limit.
Significant deviations from the unimproved data are
 seen in the $m_v/m_s=1$ data, while the $m_v/m_s=1/3$ 
 data are only exhibit tension at the $1-2\sigma$ level.
Good agreement between the two procedures is observed
 for the $m_v/m_s=1/12$ data.

\clearpage

\subsection{Valence Mass Extrapolation}
After the continuum extrapolation
 the valence quark masses are extrapolated to the
 isospin-symmetric light quark mass.
The value obtained from this extrapolation
 gives the connected light-quark contribution to the HVP
 from Eq.~\eqref{eqn:amuudconnisospin}.
In addition to the intercept of this extrapolation,
 the slope also provides information about the strong
 isospin-breaking given in Eq~\eqref{eqn:amusib}.
No extrapolation is needed to get to $m_v=m_s$ since
 this calculation is performed explicitly.

\begin{figure*}
 \centering
 \includegraphics[width=7cm]{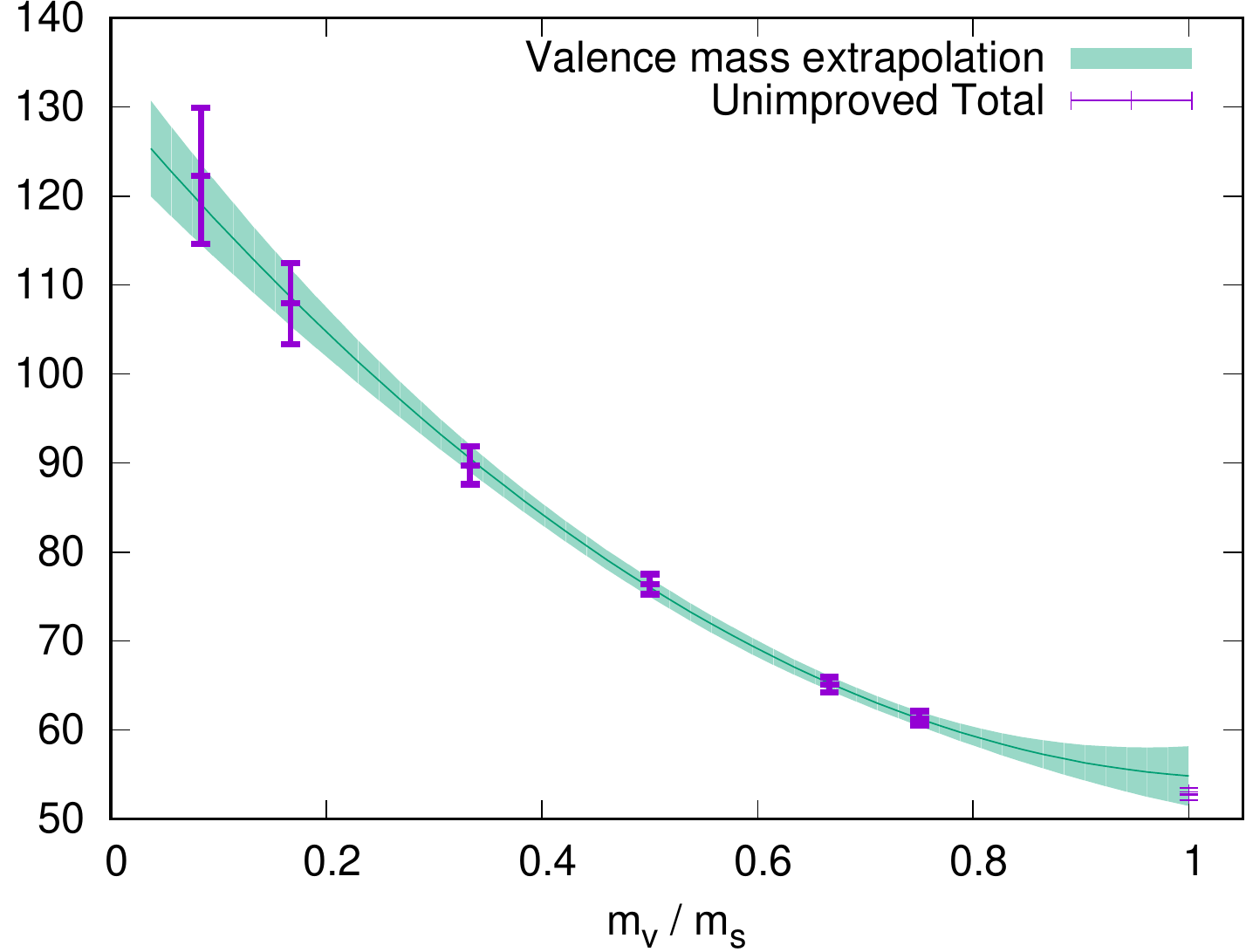}\hspace{1cm}
 \includegraphics[width=7cm]{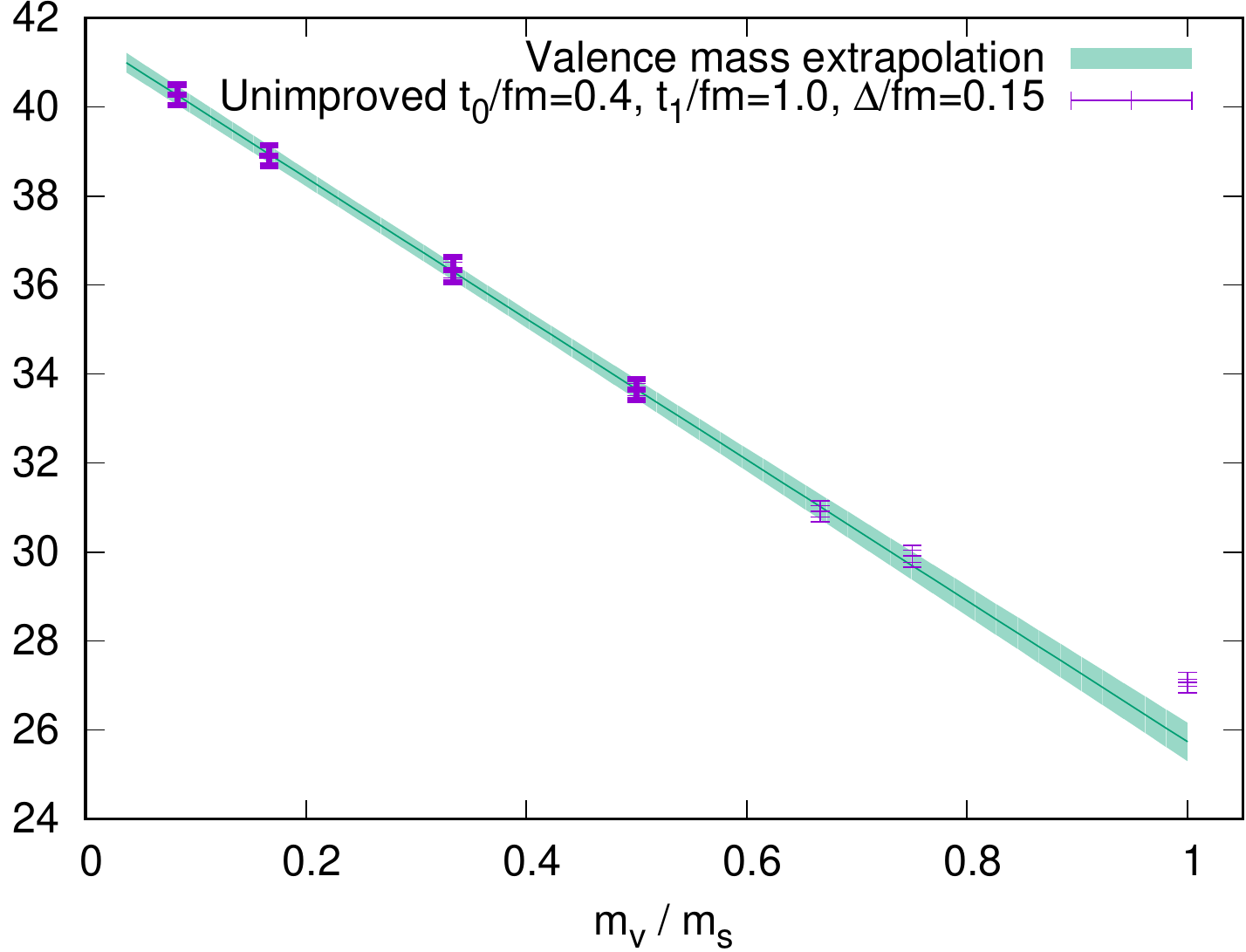}
 \caption{
 Valence mass extrapolation of $a_\mu^v$ for the
 total time extent (left) and for the window with
 $(t_0,t_1)=(0.4,1.0)~{\rm fm}$ (right).
 The points are the the continuum-extrapolated
 data and the shaded region is the mass extrapolation.
 For the rather short-distance window, a linear extrapolation
 is sufficient, while a quadratic fit is needed for
 the total result including long distances.
 The fit parameters for each window are given in
 Table~\ref{table:mvextrapparam} and the points included in the central
 fit are highlighted in the figure.
 The extrapolated band must be multiplied by a factor
 of 5 when comparing to the light quark results in
 Table~\ref{table:windowresults}, see Eq.~\eqref{eqn:amuudconnisospin}.
 }
 \label{fig:valenceextrap_fullwindow}
\end{figure*}

Fig.~\ref{fig:valenceextrap_fullwindow} shows the
 extrapolation in valence quark mass for
 both the total contribution and for the window
 with $(t_0,t_1)=(0.4,1.0)~{\rm fm}$.
Both statistical and systematic errors are included,
 and all data have been extrapolated to the continuum.
In this figure, the IPA procedure is not performed and the $\hat{p}$ prescription
 is used in all windows.
The mass dependence of the short-distance $\rho$ resonance
 states is very linear over most of the range
 between the light and strange quark masses,
 while the long-distance $\pi\pi$ states give a
 noticeable curvature.
The fit parameters for the extrapolation in valence
 quark masses for each of the windows is given in
 Table~\ref{table:mvextrapparam}.
 
\begin{table}
\begin{tabular}{ccc|ccc}
$t_0$/fm & $t_1$/fm & $\Delta$/fm & d & n & k \\\hline
Total &  & & 2 & 6 & 5 \\\hline
0.0 & 0.1 & 0.15 & 1 & 4 & 3 \\
0.1 & 0.2 & 0.15 & 1 & 4 & 3 \\
0.2 & 0.3 & 0.15 & 1 & 4 & 3 \\
0.3 & 0.4 & 0.15 & 1 & 4 & 3 \\
0.4 & 0.5 & 0.15 & 1 & 4 & 3 \\
0.5 & 0.6 & 0.15 & 1 & 4 & 3 \\
0.6 & 0.7 & 0.15 & 1 & 4 & 3 \\
0.7 & 0.8 & 0.15 & 1 & 4 & 3 \\
0.8 & 0.9 & 0.15 & 1 & 4 & 3 \\
0.9 & 1.0 & 0.15 & 1 & 4 & 3 \\
1.0 & 1.1 & 0.15 & 2 & 5 & 4 \\
1.1 & 1.2 & 0.15 & 2 & 5 & 4 \\
1.2 & 1.3 & 0.15 & 2 & 5 & 4 \\
1.3 & 1.4 & 0.15 & 2 & 5 & 4 \\
1.4 & 1.5 & 0.15 & 2 & 5 & 4 \\
1.5 & 1.6 & 0.15 & 2 & 5 & 4 \\
1.6 & 1.7 & 0.15 & 2 & 7 & 6 \\
1.7 & 1.8 & 0.15 & 2 & 7 & 6 \\
1.8 & 1.9 & 0.15 & 2 & 7 & 6 \\
1.9 & 2.0 & 0.15 & 2 & 7 & 6 \\
\hline
0.3 & 1.0 & 0.15 & 1 & 4 & 3 \\
0.3 & 1.3 & 0.15 & 2 & 5 & 4 \\
0.3 & 1.6 & 0.15 & 2 & 5 & 4 \\
\hline
0.4 & 1.0 & 0.15 & 1 & 4 & 3 \\
0.4 & 1.3 & 0.15 & 2 & 5 & 4 \\
0.4 & 1.6 & 0.15 & 2 & 5 & 4 \\
\hline
0.4 & 1.0 & 0.05 & 1 & 4 & 3 \\
0.4 & 1.0 & 0.1 & 1 & 4 & 3 \\
0.4 & 1.0 & 0.2 & 1 & 4 & 3 \\
\end{tabular}
\caption{
 List of fit parameters for the valence mass extrapolation
 in each window.
 The fits are parameterized as degree $d$ polynomials
 of $m_v/m_s$, including the $n$ lightest masses.
 A fit is repeated with the $k$ lightest masses to
 estimate the systematic error due to the valence
 extrapolation and is added as a systematic uncertainty.
 This uncertainty is included in the systematic error
 band in Figs.~\ref{fig:valenceextrap_fullwindow}~and~%
 \ref{fig:valenceextrap_smallwindow}.
 %$m_v/m_s$ fit parameters;
 %polynomial in $m_v/m_s$ of degree $d$,
 %including the $n$ lightest masses for the central fit
 %(marked bold in figures);
 %repeat fit with $k$ lightest masses included
 %and add shift as a systematic to the extrapolation
 %(included in error band of plots)
 }
\label{table:mvextrapparam}
\end{table}

\begin{figure*}
 \centering
 \includegraphics[width=7cm]{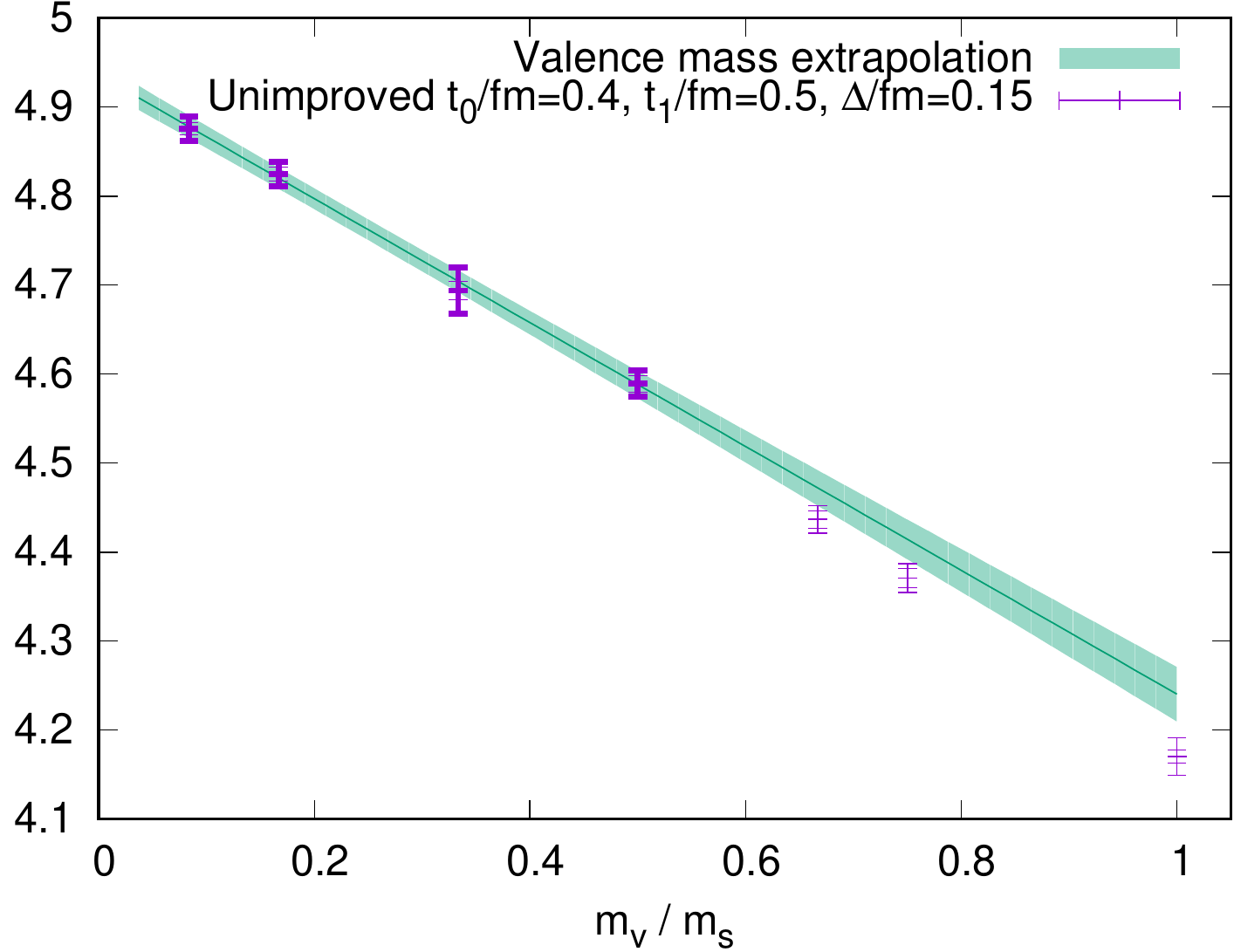}\hspace{1cm}
 \includegraphics[width=7cm]{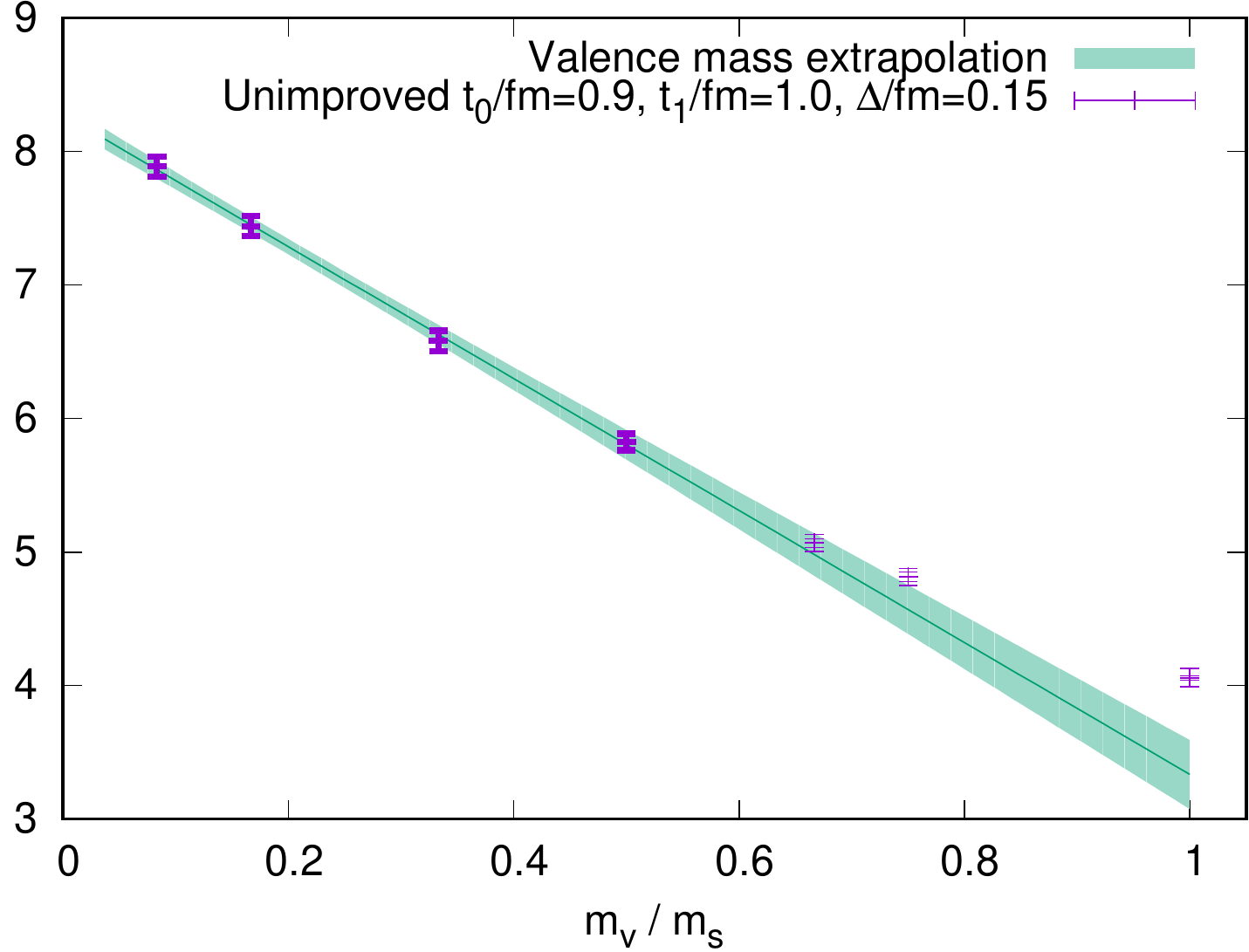}\hspace{1cm}
 \includegraphics[width=7cm]{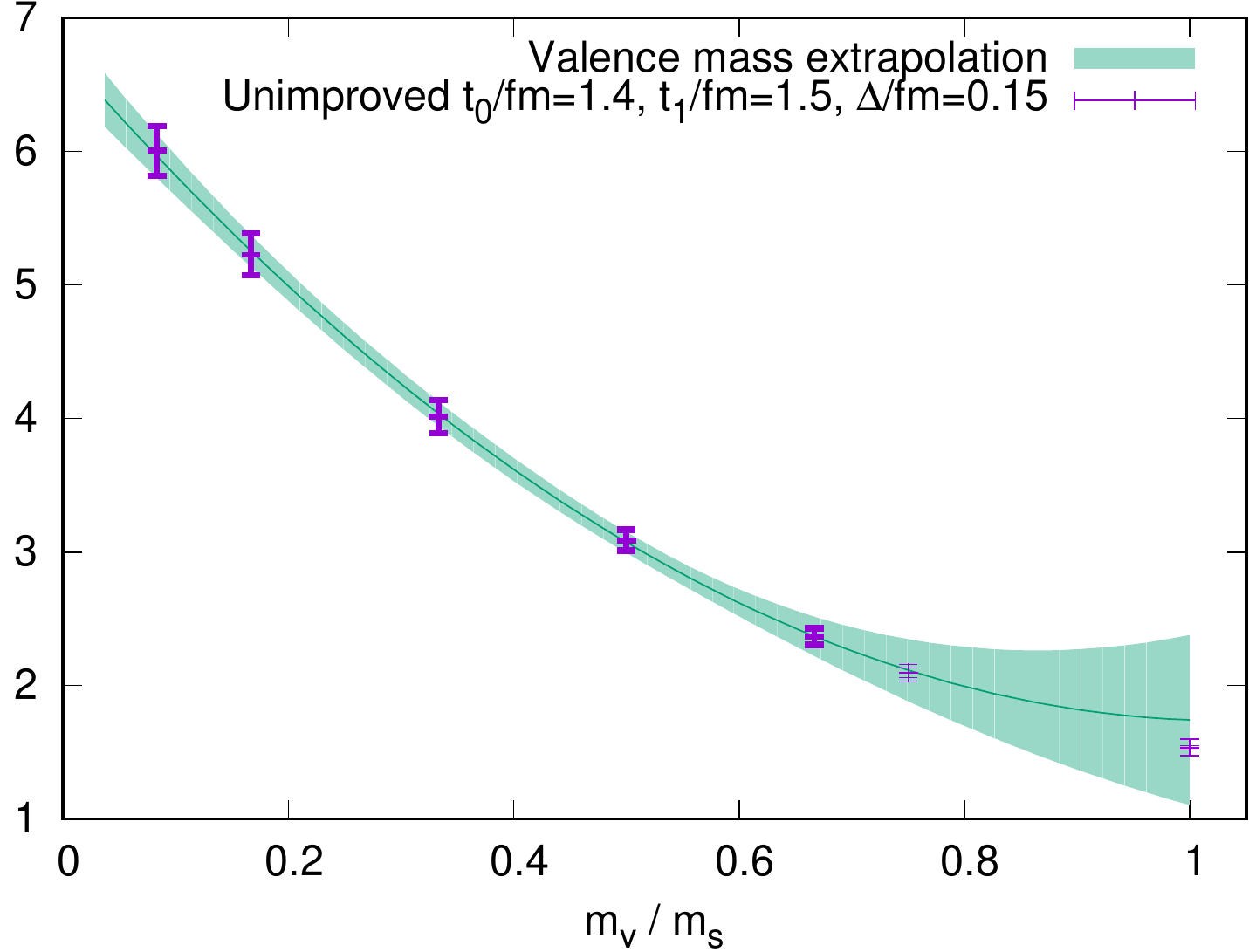}\hspace{1cm}
 \includegraphics[width=7cm]{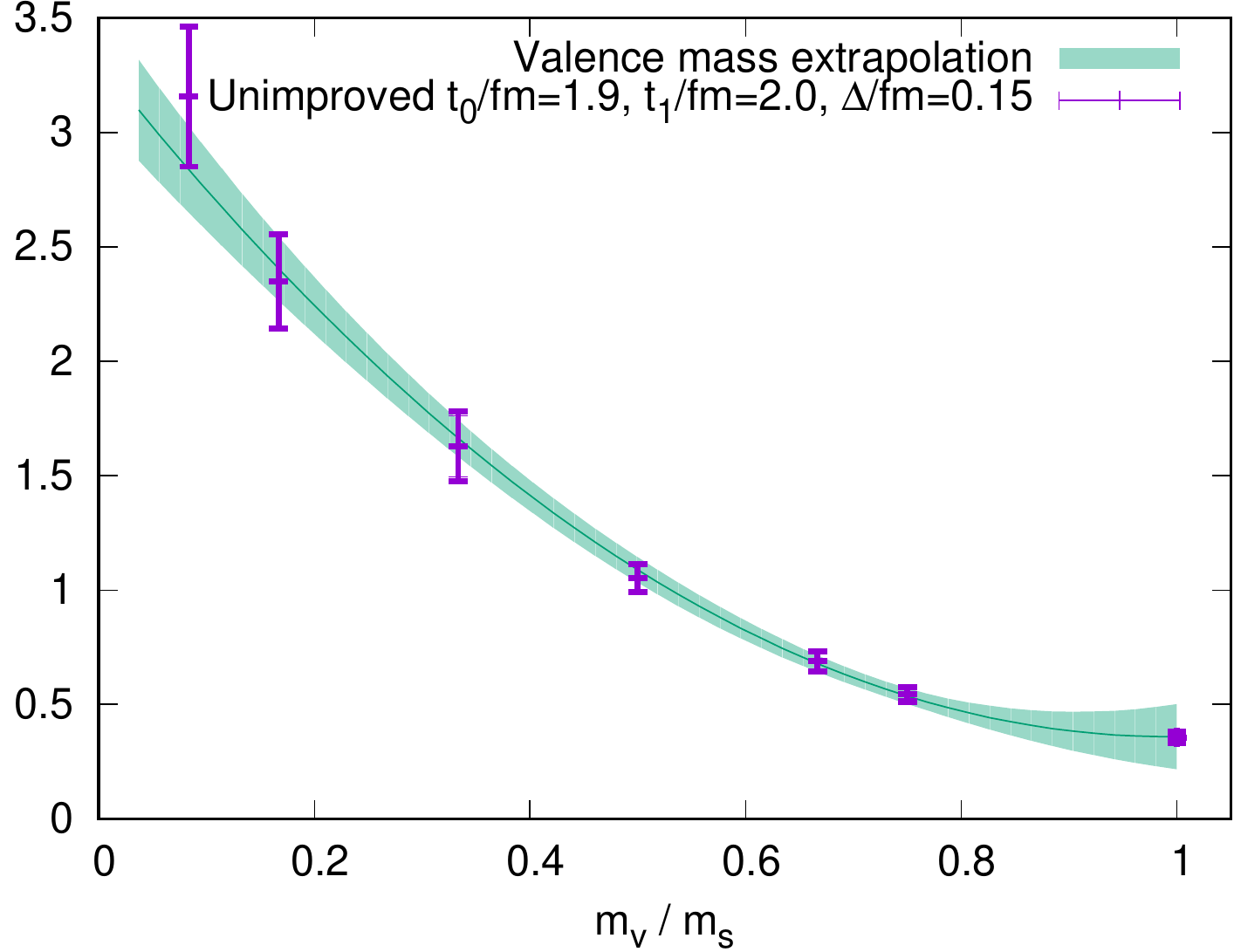}
 \caption{
 Valence mass extrapolation for four different windows.
 The points are the the continuum-extrapolated
 data and the shaded region is the mass extrapolation.
 For short-distance windows, a linear extrapolation
 is sufficient, while a quadratic fit is needed for
 long-distance windows.
 The fit parameters for each window are given in
 Table~\ref{table:mvextrapparam} and the points included in the central
 fit are bold.
  \label{fig:valenceextrap_smallwindow}
 }
\end{figure*}

Fig.~\ref{fig:valenceextrap_smallwindow} shows
 the valence quark data extrapolated to the isospin-symmetric
 limit for a few choices of $t_1-t_0 = 0.1~{\rm fm}$
 windows.
The first two windows with $t_0=0.4~{\rm fm}$ and $t_0=0.9~{\rm fm}$ are
 linear over a large mass range.
The third window shows clear evidence of curvature in the valence quark mass,
 and therefore motivates the switch from a linear
 to a quadratic fit ansatz.
The curvature continues to increase with increasing $t_0$.

Table~\ref{table:windowresults}
 shows the results for a series of time ranges and widths
 after extrapolation to the light quark mass.
These results are repeated in
 Tables~\ref{table:windowphatrxunimp}%
 -\ref{table:windowprxipa}
 in Appendix~\ref{sec:appresults}
 with all systematic uncertainties shown in detail.
For all windows, the $p$ and $\hat{p}$ prescriptions
 give results that are $1\sigma$ consistent.
The parity improvement also gives consistent results
 with unimproved data for windows with
 $t_0 \geq 0.4~{\rm fm}$.
Short windows with $t_0$ and $t_1$ close to $0~{\rm fm}$
 are also subject to much larger discretization errors
 from the continuum extrapolation than wider window
 regions or those farther from $0~{\rm fm}$.
For instance, the window with $(t_0,t_1)=(0.0,0.2)~{\rm fm}$
 has a smaller relative systematic uncertainty than
 the window with $(t_0,t_1)=(0.0,0.1)~{\rm fm}$.
The parity improvement is not well motivated for
 very short distance windows and we give the results only
 for completeness.

\begin{table*}
\begin{tabular}{ccc|rrrr|rr}
$t_0$/fm & $t_1$/fm & $\Delta$/fm & $l$,$\hat{p}$U & $l$,$\hat{p}$I & $l$,$p$U & $l$,$p$I & $s$,$\hat{p}$U & $s$,$p$U \\\hline
% root.phat-0 root.phat-1p root.p-0 root.p-1p root.phat-0-strange root.p-0-strange 
Total &  &  & 627(26)(08)& 632(27)(07)& 628(26)(07)& 634(27)(07)& 52.83(22)(65)& 53.08(22)(82)\\
\hline
0.0 & 0.1 & 0.15 & 3.59(00)(59)& 4.60(00)(31)& 4.32(00)(20)& 5.69(00)(24)& 0.81(00)(12)& 0.887(00)(40)\\
0.1 & 0.2 & 0.15 & 8.633(03)(73)& 9.93(00)(52)& 9.29(00)(46)& 11.0(0.0)(1.3)& 1.666(01)(12)& 1.728(01)(86)\\
0.2 & 0.3 & 0.15 & 14.24(01)(82)& 15.4(0.0)(1.3)& 14.7(0.0)(1.2)& 16.0(0.0)(1.8)& 2.57(00)(16)& 2.59(00)(24)\\
0.3 & 0.4 & 0.15 & 18.62(02)(35)& 20.2(0.0)(1.1)& 18.71(02)(27)& 20.3(0.0)(1.3)& 3.448(05)(65)& 3.451(05)(49)\\
0.4 & 0.5 & 0.15 & 24.552(35)(60)& 24.71(04)(24)& 24.518(36)(59)& 24.65(04)(21)& 4.170(07)(20)& 4.169(08)(24)\\
0.5 & 0.6 & 0.15 & 29.38(06)(29)& 29.42(06)(26)& 29.36(06)(31)& 29.36(06)(31)& 4.666(10)(59)& 4.665(11)(64)\\
0.6 & 0.7 & 0.15 & 33.72(10)(36)& 33.87(10)(26)& 33.70(10)(39)& 33.82(10)(29)& 4.866(13)(74)& 4.866(13)(79)\\
0.7 & 0.8 & 0.15 & 37.54(14)(14)& 37.30(15)(19)& 37.54(15)(14)& 37.28(15)(20)& 4.799(16)(39)& 4.799(16)(39)\\
0.8 & 0.9 & 0.15 & 39.32(20)(20)& 39.52(21)(18)& 39.33(21)(20)& 39.52(21)(18)& 4.505(17)(44)& 4.504(18)(44)\\
0.9 & 1.0 & 0.15 & 40.47(27)(29)& 40.43(28)(28)& 40.47(27)(30)& 40.44(28)(28)& 4.058(19)(65)& 4.058(19)(65)\\
1.0 & 1.1 & 0.15 & 40.47(44)(39)& 40.56(46)(37)& 40.49(45)(39)& 40.57(46)(37)& 3.527(19)(76)& 3.527(19)(76)\\
1.1 & 1.2 & 0.15 & 39.34(54)(39)& 39.47(56)(43)& 39.35(55)(39)& 39.48(57)(44)& 2.973(19)(75)& 2.973(19)(75)\\
1.2 & 1.3 & 0.15 & 37.53(65)(49)& 37.54(67)(48)& 37.55(66)(49)& 37.56(68)(48)& 2.441(18)(77)& 2.440(18)(77)\\
1.3 & 1.4 & 0.15 & 34.88(77)(49)& 34.98(79)(51)& 34.89(77)(49)& 35.00(80)(52)& 1.955(17)(67)& 1.955(17)(67)\\
1.4 & 1.5 & 0.15 & 31.94(88)(52)& 31.98(91)(53)& 31.96(89)(52)& 31.99(92)(53)& 1.534(15)(60)& 1.534(15)(60)\\
1.5 & 1.6 & 0.15 & 28.66(100)(52)& 28.7(1.0)(0.5)& 28.7(1.0)(0.5)& 28.7(1.0)(0.5)& 1.181(13)(52)& 1.181(13)(52)\\
1.6 & 1.7 & 0.15 & 24.58(81)(61)& 24.64(82)(62)& 24.59(81)(61)& 24.65(83)(62)& 0.894(12)(44)& 0.894(12)(44)\\
1.7 & 1.8 & 0.15 & 21.20(85)(60)& 21.28(86)(62)& 21.21(85)(60)& 21.20(86)(55)& 0.667(10)(37)& 0.667(10)(37)\\
1.8 & 1.9 & 0.15 & 18.13(86)(59)& 18.22(88)(63)& 18.13(87)(60)& 18.23(88)(63)& 0.491(08)(30)& 0.491(08)(30)\\
1.9 & 2.0 & 0.15 & 15.49(89)(66)& 15.42(91)(50)& 15.37(89)(46)& 15.57(91)(70)& 0.357(07)(24)& 0.357(07)(25)\\
\hline
0.0 & 0.2 & 0.15 & 12.22(00)(52)& 14.53(00)(21)& 13.61(00)(26)& 16.7(0.0)(1.5)& 2.48(00)(11)& 2.615(02)(47)\\
0.2 & 0.4 & 0.15 & 32.87(03)(48)& 35.6(0.0)(2.4)& 33.41(03)(94)& 36.3(0.0)(3.0)& 6.02(01)(10)& 6.05(01)(19)\\
0.4 & 0.6 & 0.15 & 53.93(10)(28)& 54.12(10)(13)& 53.88(10)(33)& 54.00(10)(17)& 8.837(18)(74)& 8.834(18)(85)\\
0.6 & 0.8 & 0.15 & 71.26(24)(37)& 71.16(25)(44)& 71.23(24)(40)& 71.11(25)(49)& 9.666(29)(91)& 9.665(29)(97)\\
0.8 & 1.0 & 0.15 & 79.80(47)(42)& 79.96(49)(44)& 79.81(48)(42)& 79.96(49)(44)& 8.56(04)(10)& 8.56(04)(10)\\
\hline
0.3 & 1.0 & 0.15 & 223.6(0.8)(1.1)& 225.5(0.8)(1.2)& 223.6(0.8)(1.1)& 225.4(0.8)(1.2)& 30.51(08)(25)& 30.51(09)(26)\\
0.3 & 1.3 & 0.15 & 340.7(2.6)(1.9)& 343.4(2.7)(2.7)& 340.7(2.6)(1.9)& 343.3(2.7)(2.7)& 39.45(13)(35)& 39.45(13)(35)\\
0.3 & 1.6 & 0.15 & 436.2(5.1)(3.1)& 439.0(5.3)(4.2)& 436.3(5.1)(3.2)& 439.5(5.3)(4.0)& 44.12(17)(49)& 44.12(17)(49)\\
\hline
0.4 & 1.0 & 0.15 & 204.99(79)(85)& 205.25(82)(77)& 204.93(80)(90)& 205.08(83)(83)& 27.06(08)(21)& 27.06(08)(22)\\
0.4 & 1.3 & 0.15 & 322.2(2.6)(1.8)& 322.8(2.7)(1.9)& 322.2(2.6)(1.8)& 321.6(2.7)(2.1)& 36.01(13)(36)& 36.00(13)(35)\\
0.4 & 1.6 & 0.15 & 417.7(5.1)(3.2)& 418.5(5.3)(3.4)& 417.9(5.1)(3.2)& 418.3(5.3)(3.3)& 40.68(17)(51)& 40.67(17)(50)\\
\hline
0.4 & 1.0 & 0.05 & 215.5(0.8)(6.2)& 208.5(0.8)(1.5)& 215.8(0.8)(6.4)& 208.4(0.8)(1.6)& 27.9(0.1)(1.1)& 27.9(0.1)(1.1)\\
0.4 & 1.0 & 0.1 & 208.85(77)(74)& 207.6(0.8)(1.1)& 208.76(78)(70)& 207.4(0.8)(1.3)& 27.70(08)(21)& 27.69(08)(20)\\
0.4 & 1.0 & 0.2 & 201.08(82)(85)& 201.86(85)(77)& 201.10(84)(86)& 201.83(86)(76)& 26.24(08)(21)& 26.24(08)(21)\\

\end{tabular}
\caption{
Results for $a_\mu^{{\rm ud,}\conniso}$, labelled by "l", and $a_\mu^{{\rm s,}\conniso}$,
labelled by "s", for the total contribution as well as different windows.
We compare results obtained from the $\hat{p}$ and $p$ prescriptions
as well as unimproved (U) and IPA (I) results.  The IPA
procedure is defined in Section~\ref{sec:ipa}.
 The value for the full time range is given in the first row,
 labeled ``Total''.
 All windows with $t_0 \geq 0.4~{\rm fm}$ are consistent
 for all 4 choices of prescription and improvement.
 For windows with lower $t_0$, discretization effects
 account for significant differences in the window sums.
 The IPA prescription, however, is not well motivated for
 small distances.  We use the $\hat{p}$ and unimproved prescriptions for
 our central values further discussed in the following.
 We notice that broader windows with $t_1-t_0=0.2$~fm have smaller
 relative systematic uncertainties for small $t_0$ compared to the narrower windows.
 \label{table:windowresults}
 }
\end{table*}

\clearpage

\subsection{Corrections for Finite Volume}

The finite-volume correction (FVC)
 is a correction to the long-distance physics
 in the correlation function due to the finite
 spatial extent of the lattice.
The lattice states in the long-distance region
 are mostly composed of two-pion scattering states
 with zero center-of-mass momentum, up to mixing
 with other states that share the same quantum numbers.
The finite spatial extent imposes a lower limit
 on the size of a unit of momentum, $p=2\pi/L$,
 which discretizes the spectrum of states that
 satisfy the periodic boundary conditions.
The lowest-energy state in the spectrum of the
 connected isospin-1 channel is a two-pion state
 with both pions having one unit of momentum back-to-back.
In infinite-volume, where there is no minimum momentum,
 two-pion states would contribute all the way down to
 the two-pion threshold energy $s=4M_\pi^2$, significantly
 changing the long-distance exponential tail of
 the correlation function.
The FVC attempts to fix this mismatch of the finite-
 and infinite-volume.

The estimate of the FVC
 is obtained from the
 Lellouch-L\"uscher-Gounaris-Sakurai procedure~%
 \cite{Gounaris:1968mw,Luscher:1990ux,Lellouch:2000pv,Meyer:2011um}.
This is carried out by combining the pion form factor
 with estimates of the finite- and infinite-volume
 spectrum and matrix elements.
In the infinite volume, the pion form factor
 is related to $R(s)$ via
\begin{equation}
 R(s) = \frac{1}{4}
 \left( 1-\frac{4M_\pi^2}{s} \right)^{3/2}
 |F_\pi(s)|^2
\end{equation}
 and $R(s)$ is inserted into Eq.~(\ref{eq:rratio})
 to obtain the $\pi\pi$ contribution to the correlation
 function $C(t)$.
The pion form factor $F_\pi(s)$ used in $R(s)$
 is obtained from the Gounaris-Sakurai (GS)
 parameterization~\cite{Gounaris:1968mw}.
The pion scattering phase shift for obtaining
 the finite-volume spectrum and matrix elements is
 computed with the GS parameterization as well,
 which has the simple relation
\begin{align}
 F_\pi(s) &= f(0)/f(s)
\end{align}
with
\begin{align}
    f(s) = (-i + \cot \delta_1(s)) \frac{k_\pi(s)^3}{\sqrt{s}} \,,
\end{align}
 where
\begin{equation}
 k_\pi(s) = \sqrt{\frac{1}{4}s -M_\pi^2}
\end{equation}
 is the pion momentum for the center of mass energy $\sqrt{s}$.
The phase shift $\delta_1(s)$ is related to the finite-volume
 spectrum according to the relation~\cite{Luscher:1990ux}
\begin{equation}
 \delta_1(s) + \phi(q) = n\pi ,\, n\in\mathds{Z}
 \label{eq:pipiquant}
\end{equation}
 where $q(s)= (L/2\pi) k_\pi(s)$
 and $\phi(q)$ is determined from
\begin{equation}
 \tan\phi(q) = -\frac{\pi^{3/2}q}{{\cal Z}_{00}(1;q)}
\end{equation}
 with the analytic continuation of the zeta function
\begin{equation}
 {\cal Z}_{00}(s;q)
 = \frac{1}{\sqrt{4\pi}}
 \sum_{\vec{n}\in\mathds{Z}^3}
 \frac{1}{(|\vec{n}|^2-q^2)^s}.
\end{equation}
For a set of finite-volume states obtained by solving
 Eq.~(\ref{eq:pipiquant}), the corresponding 
 vector current amplitudes are obtained
 from~\cite{Lellouch:2000pv,Meyer:2011um}
\begin{align}
 &\big|\langle 0 | V_i |
 \pi\pi(\sqrt{s}=E_{\pi\pi}) \rangle
 \big|^2
 \nonumber\\
 &\quad = | F_\pi(E_{\pi\pi}^2) |^2
 \frac{2k_\pi^4}{3\pi E_{\pi\pi}^2}
 \left[\frac{\partial}{\partial k_\pi}
 ( \delta_1 + \phi) \right]^{-1}.
\end{align}

The $C(t)$ obtained from both the infinite-volume (IV)
 parameterization of $F_\pi$ and from the explicit
 reconstruction of a finite number of states $N$,
\begin{equation}
 C^{\rm FV}(t) = \sum_n^N 
 \big|\langle 0 | V_i | \pi\pi_n \rangle \big|^2
 e^{-E_{\pi\pi_n}t},
\end{equation}
 are both summed with the Bernecker-Meyer kernel
$w_t$ as in Eq.~(\ref{eq:amuhvp}).
The finite volume correction for the connected diagram is then
\begin{equation}
 \Delta a_\mu^{\rm FVC}
 = \frac{10}{9}\sum_{t=0}^\infty w_t [C^{\rm IV}(t) - C^{\rm FV}(t)] \,,
\end{equation}
where $C^{\rm IV}$ denotes the infinite-volume correlaion function.
The factor $10/9$ arises for the connected compared to the total
contribution \cite{DellaMorte:2010aq,Aubin:2019usy}.

These numbers are listed in Table~\ref{table:fvc}
 for the full result and windows.
To estimate the finite volume correction,
 $C^{\rm FV}(t)$ is reconstructed with 12 states.
The difference between the 11- and 12-state reconstructions
 is added as a systematic error.
An additional $30\%$ uncertainty on the correction
 is applied to cover other uncontrolled systematic effects
 associated with the finite volume corrections.
 
\begin{figure}
    \centering
    \includegraphics[width=7cm]{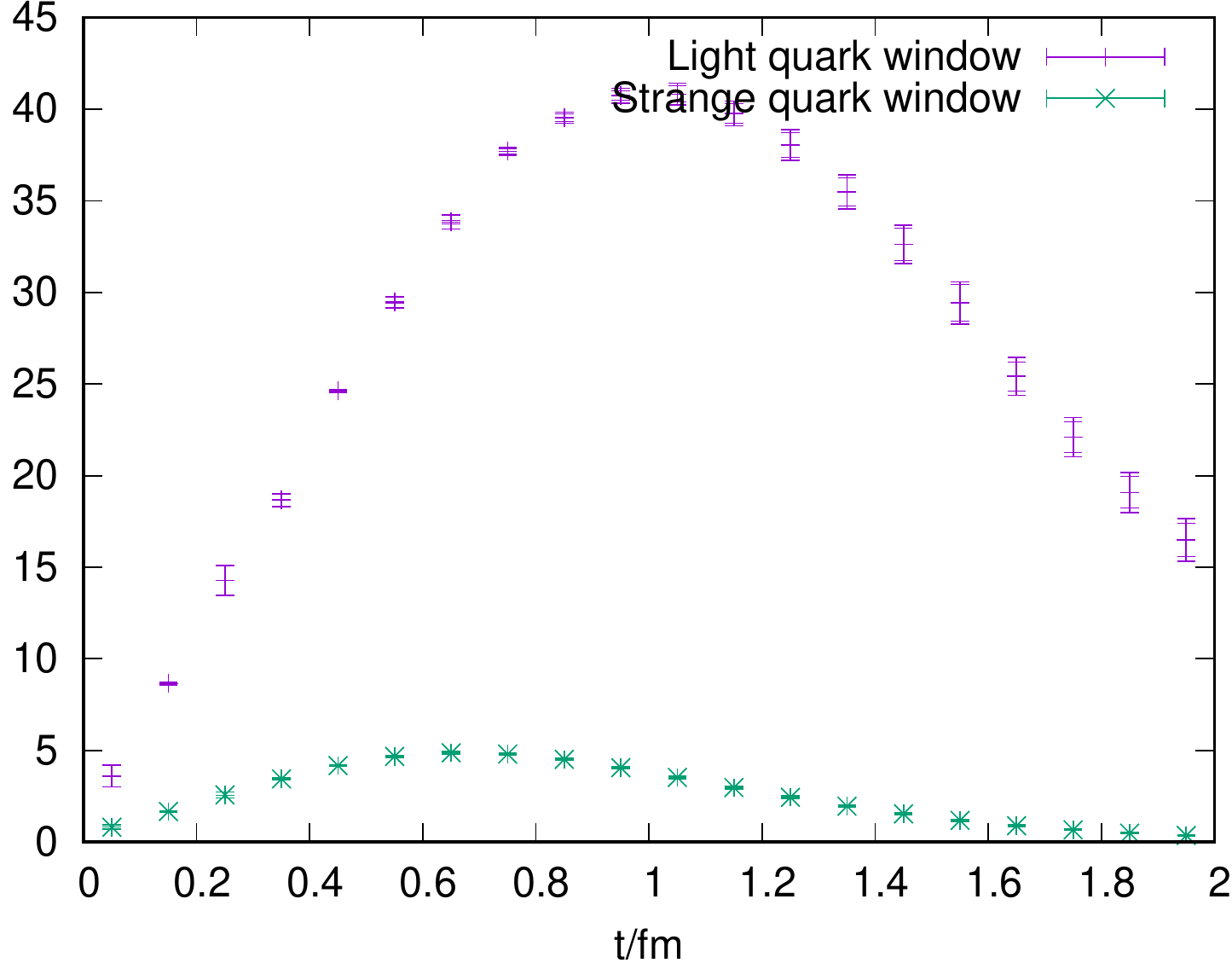}
    \includegraphics[width=7cm]{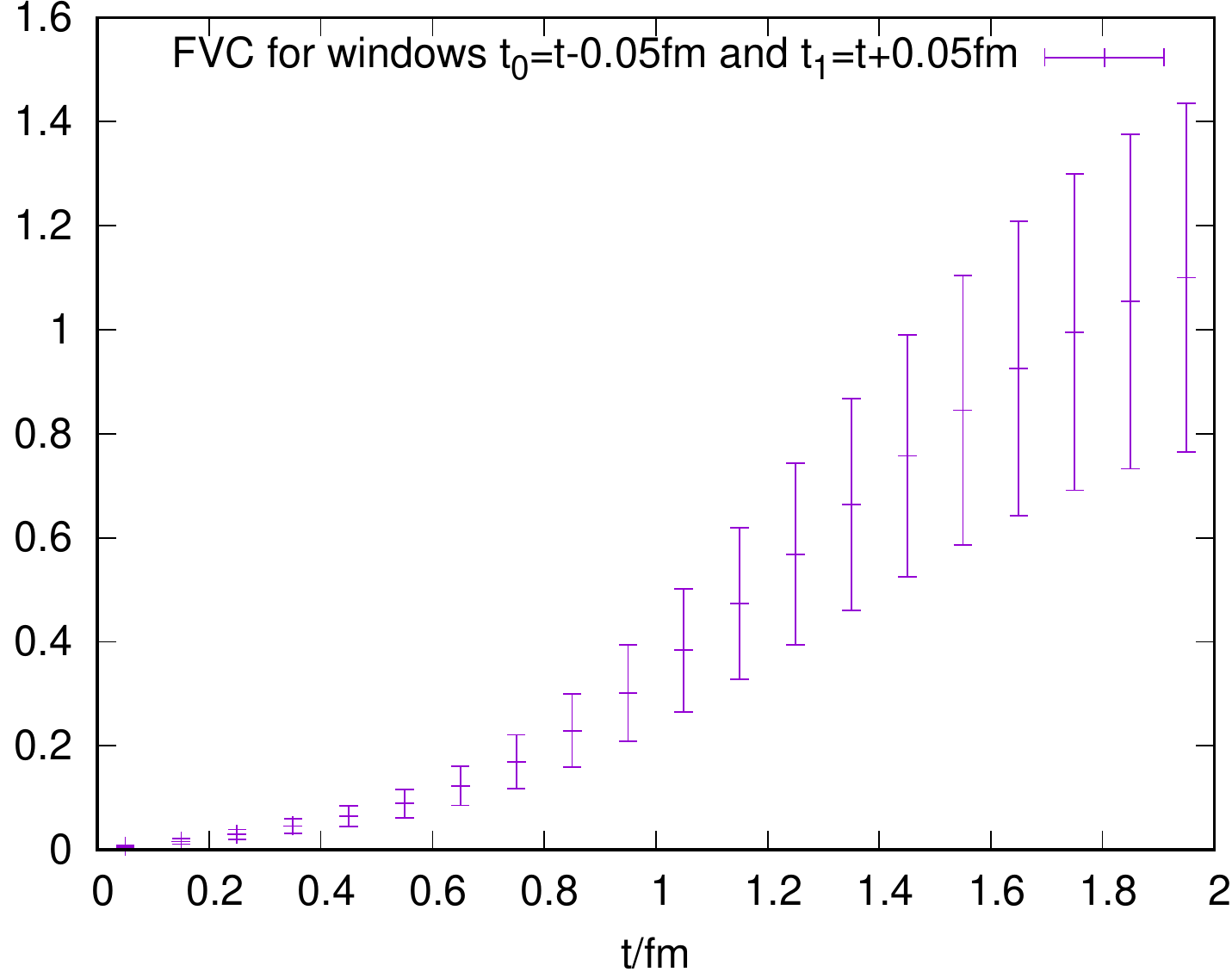}
    \caption{
    The top panel shows the continuum, infinite-volume limit
    of the connected isospin-symmetric light and strange windows with
    $t_0=t-0.05$~fm and $t_1=t+0.05$~fm and $\Delta=0.15$~fm.
    The bottom panel shows the FVC for light valence
    quark mass in the isospin-symmetric limit.
    The continuum limit of the very short-distance windows is
    difficult to control as described in Sec.~\ref{sec:methodology}.
    }
    \label{fig:fvc}
\end{figure}

The size of the contribution from each light and strange
 window and the size of the finite-volume correction
 for the light quark mass for each of
 the $t_1-t_0=0.1~{\rm fm}$ windows
 are shown in Fig.~\ref{fig:fvc}.
The upper panel shows the window result from $t_0=t-0.05$~fm to $t_1=t+0.05$~fm
with $\Delta=0.15$~fm for both the isospin-symmetric connected light-quark as
well as strange quark contribution.
The lower panel shows the size of the
 FVC to the light quark mass contribution
 for each of the windows.
These numbers are provided in Table~\ref{table:fvc}.  For the strange quark contribution,
we assume FVC to be negligible.
We combine these FVC with the finite-volume isospin-symmetric light-quark connected windows to our final results for $a_\mu^{\rm ud,conn.,isospin}$ and $a_\mu^{\rm s,conn.,isospin}$ in Table~\ref{table:finalresult}.

\begin{table}
\begin{tabular}{ccc|r|r}
$t_0$/fm & $t_1$/fm & $\Delta$/fm & $\Delta a_\mu^{\rm ud,conn.,isospin} 10^{10}$ & $\Delta a_\mu^{\rm SIB,conn.} 10^{10}$\\\hline
Total &  &  & 29.9(9.0) & 3.8(1.1)\\
\hline
0.0 & 0.1 & 0.15 & 0.0068(21) & 0.000103(32)\\
0.1 & 0.2 & 0.15 & 0.0162(50) & 0.000320(100)\\
0.2 & 0.3 & 0.15 & 0.0299(92) & 0.00085(26)\\
0.3 & 0.4 & 0.15 & 0.046(14) & 0.00195(60)\\
0.4 & 0.5 & 0.15 & 0.065(20) & 0.0039(12)\\
0.5 & 0.6 & 0.15 & 0.089(27) & 0.0069(21)\\
0.6 & 0.7 & 0.15 & 0.123(38) & 0.0112(34)\\
0.7 & 0.8 & 0.15 & 0.169(52) & 0.0168(52)\\
0.8 & 0.9 & 0.15 & 0.229(70) & 0.0238(73)\\
0.9 & 1.0 & 0.15 & 0.301(93) & 0.0320(98)\\
1.0 & 1.1 & 0.15 & 0.38(12) & 0.041(13)\\
1.1 & 1.2 & 0.15 & 0.47(15) & 0.051(16)\\
1.2 & 1.3 & 0.15 & 0.57(17) & 0.062(19)\\
1.3 & 1.4 & 0.15 & 0.66(20) & 0.072(22)\\
1.4 & 1.5 & 0.15 & 0.76(23) & 0.083(25)\\
1.5 & 1.6 & 0.15 & 0.85(26) & 0.093(28)\\
1.6 & 1.7 & 0.15 & 0.93(28) & 0.102(31)\\
1.7 & 1.8 & 0.15 & 1.00(30) & 0.110(34)\\
1.8 & 1.9 & 0.15 & 1.05(32) & 0.117(36)\\
1.9 & 2.0 & 0.15 & 1.10(33) & 0.123(38)\\
\hline
0.0 & 0.2 & 0.15 & 0.0230(71) & 0.00042(13)\\
0.2 & 0.4 & 0.15 & 0.076(23) & 0.00280(87)\\
0.4 & 0.6 & 0.15 & 0.154(47) & 0.0108(33)\\
0.6 & 0.8 & 0.15 & 0.293(90) & 0.0279(86)\\
0.8 & 1.0 & 0.15 & 0.53(16) & 0.056(17)\\
\hline
0.3 & 1.0 & 0.15 & 1.02(31) & 0.096(30)\\
0.3 & 1.3 & 0.15 & 2.45(75) & 0.251(77)\\
0.3 & 1.6 & 0.15 & 4.7(1.4) & 0.50(15)\\
\hline
0.4 & 1.0 & 0.15 & 0.98(30) & 0.094(29)\\
0.4 & 1.3 & 0.15 & 2.40(74) & 0.249(76)\\
0.4 & 1.6 & 0.15 & 4.7(1.4) & 0.50(15)\\
\hline
0.4 & 1.0 & 0.05 & 0.92(28) & 0.088(27)\\
0.4 & 1.0 & 0.1 & 0.94(29) & 0.091(28)\\
0.4 & 1.0 & 0.2 & 1.02(31) & 0.099(31)\\

\end{tabular}
\caption{
 Finite volume corrections for each window.
 The numbers for the light-quark connected isospin-symmetric contribution are plotted in the bottom panel of
 Fig.~\ref{fig:fvc}.
}
\label{table:fvc}
\end{table}

\begin{table}
\begin{tabular}{ccc|r|r}
$t_0$/fm & $t_1$/fm & $\Delta$/fm & $a_\mu^{\rm ud,conn.,isospin} 10^{10}$ & $a_\mu^{\rm s,conn.,isospin} 10^{10}$\\\hline
% root.phat-0 root.phat-0-strange 
Total &  &  & 657(26)(12)& 52.83(22)(65)\\
\hline
0.0 & 0.1 & 0.15 & 3.60(00)(59)& 0.81(00)(12)\\
0.1 & 0.2 & 0.15 & 8.649(03)(73)& 1.666(01)(12)\\
0.2 & 0.3 & 0.15 & 14.27(01)(82)& 2.57(00)(16)\\
0.3 & 0.4 & 0.15 & 18.67(02)(35)& 3.448(05)(65)\\
0.4 & 0.5 & 0.15 & 24.617(35)(63)& 4.170(07)(20)\\
0.5 & 0.6 & 0.15 & 29.47(06)(29)& 4.666(10)(59)\\
0.6 & 0.7 & 0.15 & 33.85(10)(37)& 4.866(13)(74)\\
0.7 & 0.8 & 0.15 & 37.71(14)(15)& 4.799(16)(39)\\
0.8 & 0.9 & 0.15 & 39.55(20)(21)& 4.505(17)(44)\\
0.9 & 1.0 & 0.15 & 40.77(27)(31)& 4.058(19)(65)\\
1.0 & 1.1 & 0.15 & 40.86(44)(41)& 3.527(19)(76)\\
1.1 & 1.2 & 0.15 & 39.81(54)(42)& 2.973(19)(75)\\
1.2 & 1.3 & 0.15 & 38.10(65)(51)& 2.441(18)(77)\\
1.3 & 1.4 & 0.15 & 35.54(77)(53)& 1.955(17)(67)\\
1.4 & 1.5 & 0.15 & 32.70(88)(56)& 1.534(15)(60)\\
1.5 & 1.6 & 0.15 & 29.50(100)(58)& 1.181(13)(52)\\
1.6 & 1.7 & 0.15 & 25.51(81)(66)& 0.894(12)(44)\\
1.7 & 1.8 & 0.15 & 22.20(85)(66)& 0.667(10)(37)\\
1.8 & 1.9 & 0.15 & 19.18(86)(67)& 0.491(08)(30)\\
1.9 & 2.0 & 0.15 & 16.59(89)(75)& 0.357(07)(24)\\
\hline
0.0 & 0.2 & 0.15 & 12.25(00)(52)& 2.48(00)(11)\\
0.2 & 0.4 & 0.15 & 32.95(03)(48)& 6.02(01)(10)\\
0.4 & 0.6 & 0.15 & 54.08(10)(29)& 8.837(18)(74)\\
0.6 & 0.8 & 0.15 & 71.55(24)(38)& 9.666(29)(91)\\
0.8 & 1.0 & 0.15 & 80.33(47)(44)& 8.56(04)(10)\\
\hline
0.3 & 1.0 & 0.15 & 224.6(0.8)(1.1)& 30.51(08)(25)\\
0.3 & 1.3 & 0.15 & 343.1(2.6)(2.0)& 39.45(13)(35)\\
0.3 & 1.6 & 0.15 & 441.0(5.1)(3.4)& 44.12(17)(49)\\
\hline
0.4 & 1.0 & 0.15 & 205.97(79)(90)& 27.06(08)(21)\\
0.4 & 1.3 & 0.15 & 324.6(2.6)(1.9)& 36.01(13)(36)\\
0.4 & 1.6 & 0.15 & 422.4(5.1)(3.5)& 40.68(17)(51)\\
\hline
0.4 & 1.0 & 0.05 & 216.5(0.8)(6.2)& 27.9(0.1)(1.1)\\
0.4 & 1.0 & 0.1 & 209.80(77)(79)& 27.70(08)(21)\\
0.4 & 1.0 & 0.2 & 202.10(82)(91)& 26.24(08)(21)\\

\end{tabular}
\caption{Final results, including finite-volume corrections, for connected isospin-symmetric light and strange quark contributions.}
\label{table:finalresult}
\end{table}

\begin{table}
\begin{tabular}{l|r|l}
Contribution& Result $\times 10^{10}$  & From\\\hline
Total & 714(27)(13) & \\\hline
ud, conn., isospin & 657(26)(12) & Table~\ref{table:finalresult} \\
s, conn., isospin & 52.83(22)(65) & Table~\ref{table:finalresult} \\
c, conn., isospin & 14.3(0.0)(0.7) & Ref.~\cite{Blum:2018mom} \\
uds, disc., isospin & -11.2(3.3)(2.3) & Ref.~\cite{Blum:2018mom} \\
SIB, conn. & 9.0(0.8)(1.2) & Table~\ref{table:sib} \\
SIB, disc. & -6.9(0.0)(3.5) & Eq.~\eqref{eqn:sibdisc} \\
QED, conn. & 5.9(5.7)(1.7) & Ref.~\cite{Blum:2018mom} \\
QED, disc. & -6.9(2.1)(2.0) & Ref.~\cite{Blum:2018mom} \\
\end{tabular}
\caption{We combine new results obtained in this paper with results for the missing contributions from RBC/UKQCD \cite{Blum:2018mom} to our total result for $a_\mu^{\rm HVP~LO}$.}
\label{table:totalhvp}
\end{table}

Since at leading order in $\Delta m$ the pion-mass splitting is a pure QED effect, it is expected that two-pion
contributions to the total SIB contribution largely cancel.  This is, however, not
true for the connected and disconnected pieces (diagrams M and O of Fig.~\ref{fig:mdiag})
separately.  In particular, when connected and disconnected SIB corrections are compared between different lattice collaborations, performed at different volumes, this is important to take into account.

In order to address this issue, we use NLO PQChPT \cite{DellaMorte:2010aq,Aubin:2019usy},
which yields a correlator for the connected and disconnected diagrams of Fig.~\ref{fig:diagramcd},
\begin{align}\label{eqn:nlochpt}
C^{\rm NLO,PQ{\chi}PT,conn.}(t) &= \frac{10}{9}\frac13 \frac1{L^3}\sum_{\vec{p}}\frac{\vec{p}^2}{(E^{vl}_p)^2} e^{-2 E^{vl}_p t} \,, \\
C^{\rm NLO,PQ{\chi}PT,disc.}(t) &= -\frac{1}{9}\frac13 \frac1{L^3}\sum_{\vec{p}}\frac{\vec{p}^2}{(E^{vv}_p)^2} e^{-2 E^{vv}_p t} \,,
\end{align}
with
\begin{align}
  E^{vl}_p &= \sqrt{ (m_\pi^{vl})^2 + \vec{p}^2 } \,, &\notag
  E^{vv}_p &= \sqrt{ (m_\pi^{vv})^2 + \vec{p}^2 } \,, \\
  (m^{vl}_\pi)^2 &= B(m_l + m_v) \,, &
  (m^{vv}_\pi)^2 &= 2Bm_v \,.
\end{align}
In these expressions $L^3$ is the spatial volume.  We then use Eqs.~\eqref{eqn:Mderivc} and \eqref{eqn:Oderivd}, to relate this correlator to diagram M and O, for which
we can the compute finite-volume corrections.  We find
\begin{align}
 \frac{\partial}{\partial m_v}C^{\rm NLO,PQ{\chi}PT,conn.} &= \frac{5}{9}  \frac{\partial}{\partial m_v}c = -\frac{10}{9}M \,, \\
 \frac{\partial}{\partial m_v}C^{\rm NLO,PQ{\chi}PT,disc.} &= -\frac{1}{9} \frac{\partial}{\partial m_v}d = \frac{2}{9}O \,.
\end{align}
From Eq.~\eqref{eqn:nlochpt} it then follows that
\begin{align}
&\frac{\partial}{\partial m_v}C^{\rm NLO,PQ{\chi}PT,conn.}\Biggr\vert_{m_v=m_l} \notag\\
&= -5\frac{\partial}{\partial m_v}C^{\rm NLO,PQ{\chi}PT,disc.}\Biggr\vert_{m_v=m_l}
\end{align}
and therefore that within NLO PQChPT
\begin{align}
 M = O \,.
\end{align}
Since the connected plus disconnected SIB enters as $M-O$, indeed the total two-pion contributions
cancel.
In this work, we use the separate expressions for the
 connected and disconnected SIB FVC and quote the
 appropriate infinite-volume result for
 $a_\mu^{\rm SIB,conn.}$ in addition to the
 finite-volume result $a_\mu^{\rm SIB,conn.,fv}$
 for different windows in Tab.~\ref{table:sib}.

It is instructive to consider the infinite-volume NLO PQChPT results
\begin{align}\label{eqn:sibdisc}
 a_\mu^{\rm SIB,conn.,NLO PQ{\chi}PT} &= -a_\mu^{\rm SIB,disc.,NLO PQ{\chi}PT} \\
 &= 6.9(3.5) 10^{-10} \,,
\end{align}
where we add a $50\%$ systematic error.  In these expressions, we use $m_\pi=135$ MeV
since in this context we are interested in the evaluation of mass-derivatives at the isospin symmetric limit ($m_v=m_l$).

% 
% 
% M = O -> M-O = 0

\begin{table}
\begin{tabular}{ccc|r|r}
$t_0$/fm & $t_1$/fm & $\Delta$/fm & $a_\mu^{\rm SIB,conn.,fv} 10^{10}$ & $a_\mu^{\rm SIB,conn.} 10^{10}$\\\hline
% root.phat-0 
Total &  &  &  5.25(76)(29) &  9.0(0.8)(1.2)\\
\hline
0.0 & 0.1 & 0.15 &  -0.002(00)(17) &  -0.002(00)(17)\\
0.1 & 0.2 & 0.15 &  0.0015(01)(23) &  0.0019(01)(23)\\
0.2 & 0.3 & 0.15 &  0.007(00)(23) &  0.008(00)(23)\\
0.3 & 0.4 & 0.15 &  0.009(01)(11) &  0.011(01)(11)\\
0.4 & 0.5 & 0.15 &  0.0266(10)(16) &  0.0305(10)(22)\\
0.5 & 0.6 & 0.15 &  0.0462(16)(91) &  0.0531(16)(93)\\
0.6 & 0.7 & 0.15 &  0.077(02)(11) &  0.088(02)(12)\\
0.7 & 0.8 & 0.15 &  0.1159(35)(66) &  0.1327(35)(90)\\
0.8 & 0.9 & 0.15 &  0.1502(46)(76) &  0.174(05)(11)\\
0.9 & 1.0 & 0.15 &  0.189(06)(14) &  0.221(06)(18)\\
1.0 & 1.1 & 0.15 &  0.255(20)(19) &  0.296(20)(24)\\
1.1 & 1.2 & 0.15 &  0.296(24)(22) &  0.348(24)(28)\\
1.2 & 1.3 & 0.15 &  0.331(27)(29) &  0.393(27)(36)\\
1.3 & 1.4 & 0.15 &  0.348(31)(29) &  0.420(31)(38)\\
1.4 & 1.5 & 0.15 &  0.356(34)(30) &  0.439(34)(41)\\
1.5 & 1.6 & 0.15 &  0.351(37)(27) &  0.443(37)(41)\\
1.6 & 1.7 & 0.15 &  0.297(18)(26) &  0.399(18)(42)\\
1.7 & 1.8 & 0.15 &  0.270(18)(25) &  0.381(18)(43)\\
1.8 & 1.9 & 0.15 &  0.243(18)(24) &  0.361(18)(44)\\
1.9 & 2.0 & 0.15 &  0.219(18)(26) &  0.342(18)(47)\\
\hline
0.0 & 0.2 & 0.15 &  -0.001(00)(14) &  -0.000(00)(14)\\
0.2 & 0.4 & 0.15 &  0.016(01)(13) &  0.019(01)(13)\\
0.4 & 0.6 & 0.15 &  0.0729(26)(83) &  0.0836(26)(91)\\
0.6 & 0.8 & 0.15 &  0.193(06)(12) &  0.221(06)(16)\\
0.8 & 1.0 & 0.15 &  0.339(10)(19) &  0.395(10)(27)\\
\hline
0.3 & 1.0 & 0.15 &  0.615(19)(35) &  0.711(19)(49)\\
0.3 & 1.3 & 0.15 &  1.47(12)(10) &  1.72(12)(13)\\
0.3 & 1.6 & 0.15 &  2.53(21)(17) &  3.03(21)(24)\\
\hline
0.4 & 1.0 & 0.15 &  0.606(18)(31) &  0.700(18)(46)\\
0.4 & 1.3 & 0.15 &  1.47(12)(10) &  1.72(12)(13)\\
0.4 & 1.6 & 0.15 &  2.53(21)(18) &  3.03(21)(25)\\
\hline
0.4 & 1.0 & 0.05 &  0.63(02)(19) &  0.72(02)(20)\\
0.4 & 1.0 & 0.1 &  0.603(18)(35) &  0.693(18)(48)\\
0.4 & 1.0 & 0.2 &  0.615(19)(31) &  0.715(19)(47)\\

\end{tabular}
\caption{
We provide results for the connected SIB contribution both at finite volume ($a_\mu^{\rm SIB,conn.,fv}$) and at infinite volume $a_\mu^{\rm SIB,conn.}$.  While the sum of the connected and disconnected SIB contribution have likely small finite-volume corrections, $a_\mu^{\rm SIB,conn.}$ itself receives a significant correction.  This is important for comparisons of this contribution between different lattice results.
}
\label{table:sib}
\end{table}

\begin{figure}[t]
 \centering
 \includegraphics[page=2,width=8.5cm]{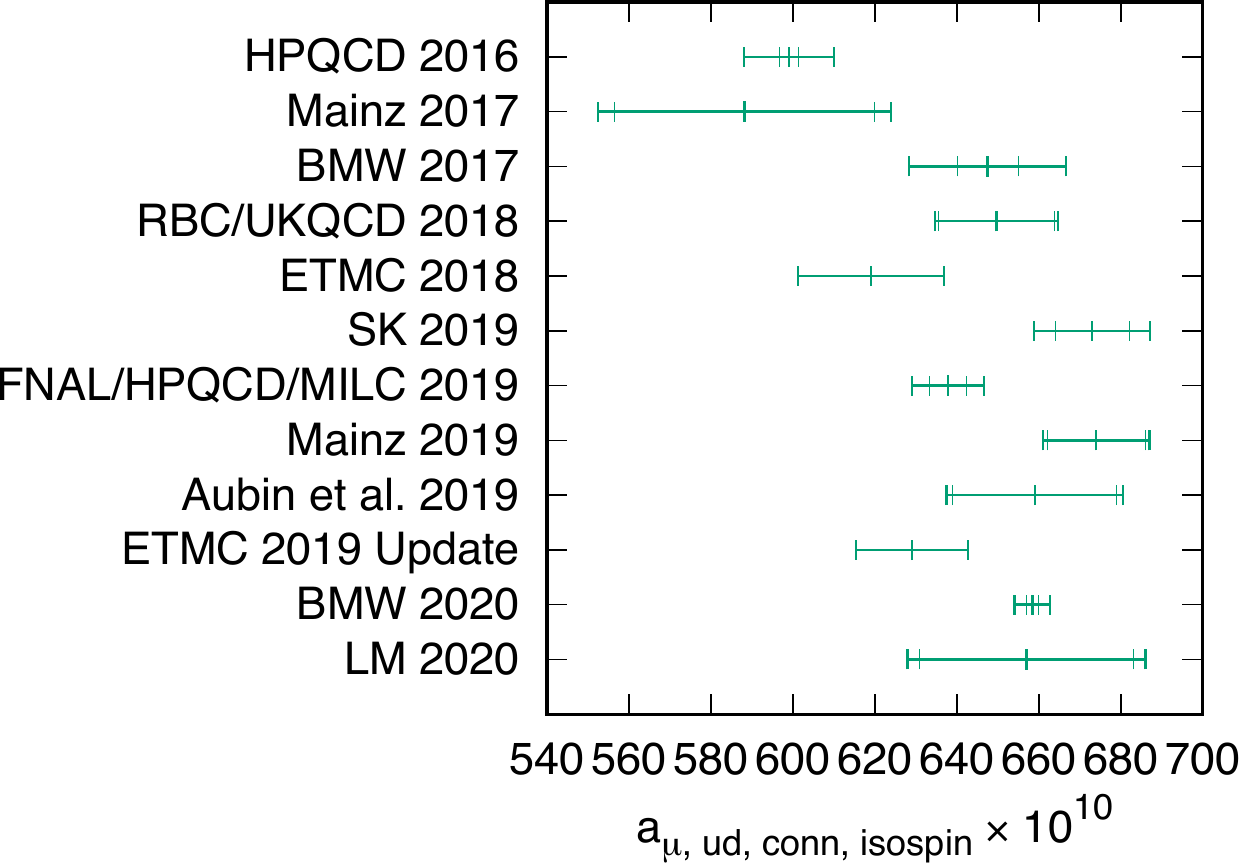}\\
 \vspace{0.2cm}
   \includegraphics[page=6,width=8.5cm]{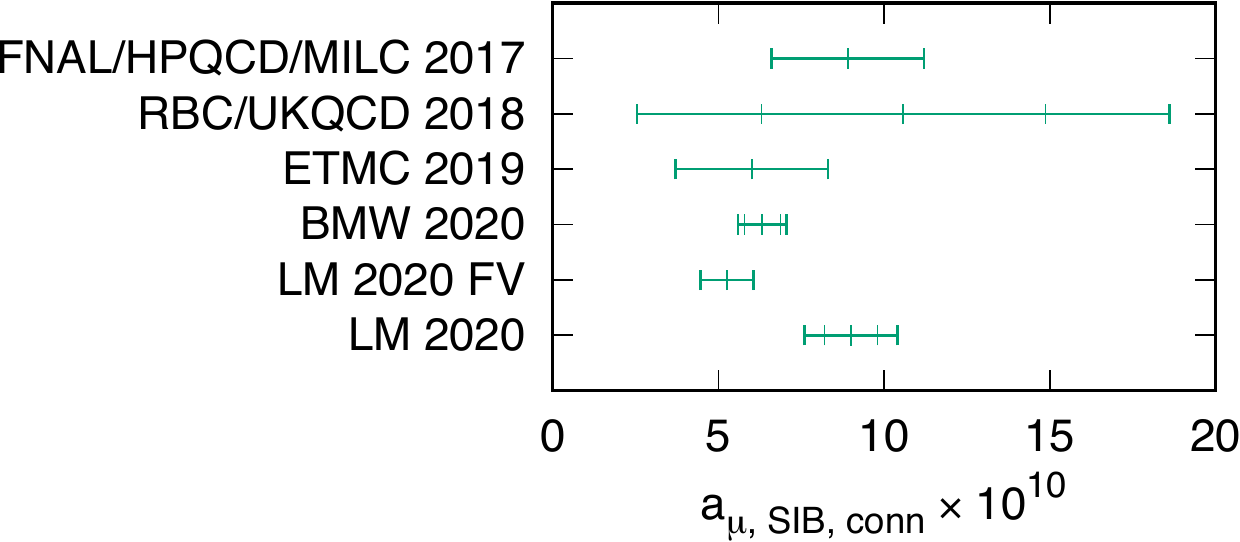}\\
   \vspace{0.2cm}
   \includegraphics[page=8,width=8.5cm]{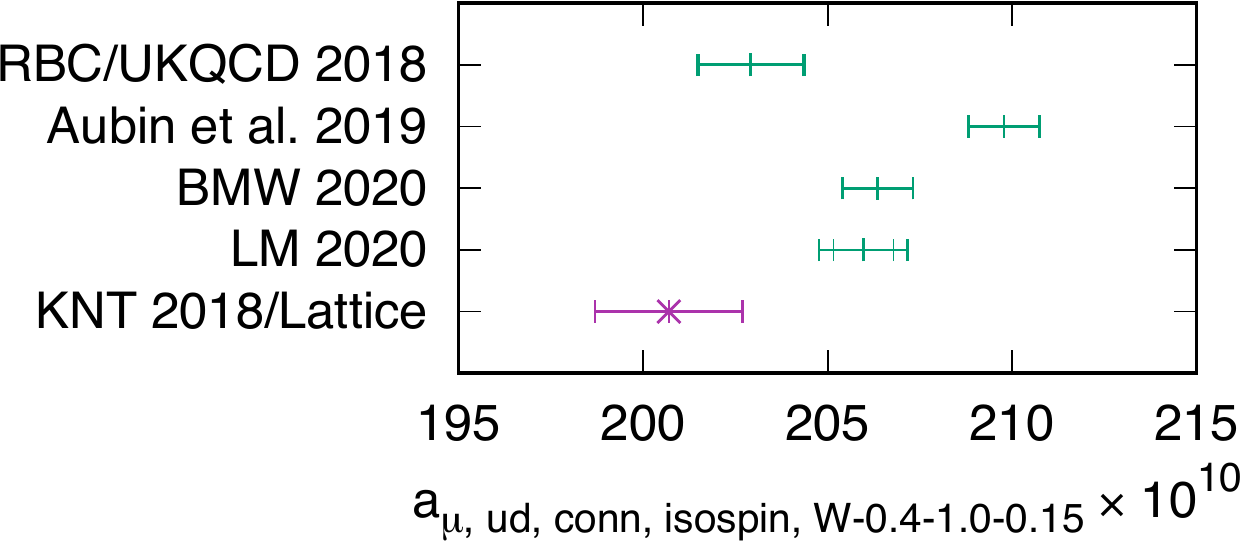}
 \caption{
  Overview of results for $a_\mu^{\rm ud,conn.,isospin}$, $a_\mu^{\rm SIB,conn.}$,
  and the window $a_\mu^{\rm ud,conn.,isospin,W}$ with $t_0=0.4$ fm, $t_1=1.0$ fm, and $\Delta=0.15$ fm.  The referenced contributions are listed in the caption of Fig.~\ref{fig:overview} apart from ETMC 2019 \cite{Giusti:2019xct} and FNAL/HPQCD/MILC 2017 \cite{Chakraborty:2017tqp}.
  The result of this work is labelled ``LM 2020''.
  For $a_\mu^{\rm ud,conn.,isospin}$ the precise BMW 2020 is higher in particular compared to values by ETMC 2019 Update as well as FNAL/HPQCD/MILC 2019.  For $a_\mu^{\rm SIB,conn.}$, we provide also a finite-volume result ``LM 2020 FV'' to compare to the other results obtained at similar volume.  ``LM 2020'' is the value corrected to infinite volume.
  For the window $a_\mu^{\rm ud,conn.,isospin,W}$ there is a clear tension between Aubin {\it et al.}~2019 and the R-ratio as well as RBC/UKQCD 2018.  In this work, we perform a calculation on the same gauge configurations as Aubin {\it et al.}~2019 but with a different discretization of the vector current and find a substantially lower value.  This difference may therefore originate from difficulties associated with the continuum limit.
  In Fig.~\ref{fig:overviewall}, we provide an overview of a broader
  set of individual contributions to the HVP.
  }
  \label{fig:overviewmain}
 \end{figure}

\section{Discussion and Conclusion}
\label{sec:conclusion}

  \begin{figure}[t]
 \centering
 \includegraphics[page=1,width=8.5cm]{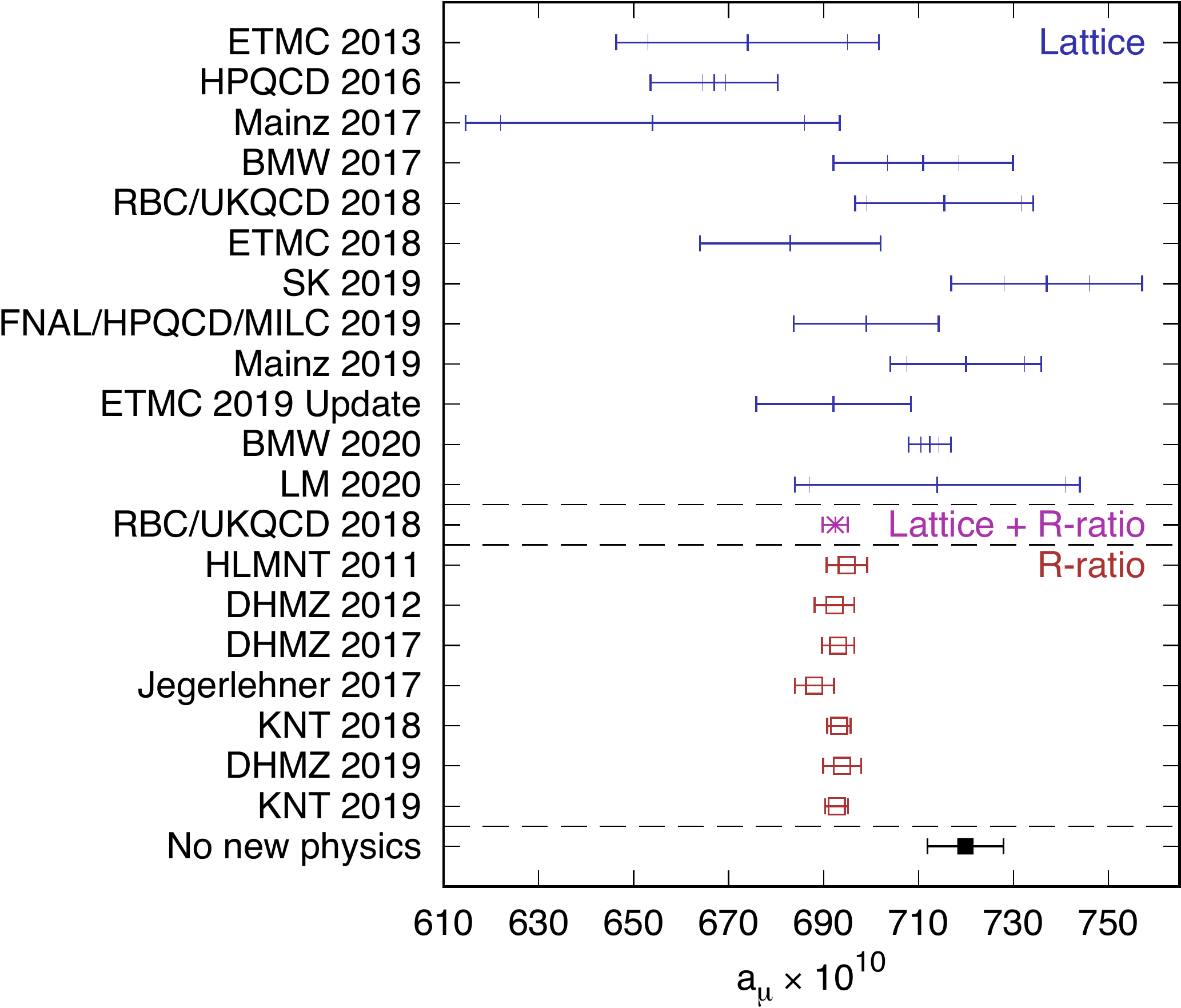}
 \caption{Overview of total results for $a_\mu^{\rm HVP}$.
 The referenced contributions are:
 ETMC 2013~\cite{Burger:2013jya},
 HPQCD 2016~\cite{Chakraborty:2016mwy},
 Mainz 2017~\cite{DellaMorte:2017dyu},
 BMW 2017~\cite{Borsanyi:2017zdw},
 RBC/UKQCD 2018~\cite{Blum:2018mom},
 ETMC 2018~\cite{Giusti:2018mdh},
 SK 2019~\cite{Shintani:2019wai},
 FNAL/HPQCD/MILC 2019~\cite{Davies:2019efs},
 Mainz 2019~\cite{Gerardin:2019rua},
 ETMC 2019 Update~\cite{Giusti:2019hkz},
 BMW 2020~\cite{Borsanyi:2020mff},
 HLMNT 2011~\cite{Hagiwara:2011af},
 DHMZ 2012~\cite{Davier:2010nc},
 DHMZ 2017~\cite{Davier:2017zfy},
 Jegerlehner 2017~\cite{Jegerlehner:2017lbd},
 KNT 2018~\cite{Keshavarzi:2018mgv},
 DHMZ 2019~\cite{Davier:2019can}, and
 KNT 2019~\cite{Keshavarzi:2019abf}.
 The result of this work is labelled ``LM 2020''.
 }
  \label{fig:overview}
 \end{figure}

We summarize our results for the total contributions to $a_\mu^{\rm HVP~LO}$ in Tab.~\ref{table:totalhvp} and compare this result in Fig.~\ref{fig:overview} to results
by other collaborations.  In Fig.~\ref{fig:overviewmain}, we compare our results
for $a_\mu^{\rm ud,conn.,isospin}$, $a_\mu^{\rm SIB,conn.}$,
and the window $a_\mu^{\rm ud,conn.,isospin,W}$ with $t_0=0.4$ fm, $t_1=1.0$ fm, and $\Delta=0.15$ fm to other collaborations.
We would like to stress in particular the difference between the window results from Aubin {\it et al.} and this work, which is especially noteworthy since they were performed on the same gauge configurations.  Apart from the small valence mass-extrapolation, which we suggest to be mild for the window, the main difference between this work and Aubin {\it et al.} is the choice of a site-local compared to a conserved current.  This suggests that properly estimating the uncertainties associated with the continuum limit may be challenging.  We note that in this work, Aubin {\it et al.}, as well as the recent BMW result, results at similar inverse lattice spacings from $a^{-1}\approx 1.6$ GeV to $a^{-1}\approx 3.5$ GeV were used, all with staggered sea quark ensembles at physical pion mass. 

We hope that the larger set of window results provided in this work can be useful to further scrutinize the emerging tensions within the lattice QCD community and within lattice QCD and the R-ratio.  We expect that in the near future other lattice collaborations will also provide results for the total $a_\mu^{\rm HVP~LO}$ with uncertainties close to $5\times 10^{-10}$, which may shed further light on the emerging
tensions.  It will be particularly important that such results include different lattice discretizations. 

\begin{acknowledgments}
We thank our colleagues in the RBC \& UKQCD collaborations
 for interesting discussions.  
We thank the MILC collaboration for the ensembles
 used in this analysis.
Inversions and contractions were performed with the
 MILC code version 7.
This work was supported by resources provided
 by the Scientific Data and Computing Center (SDCC)
 at Brookhaven National Laboratory (BNL),
 a DOE Office of Science User Facility
 supported by the Office of Science of the
 US Department of Energy.
The SDCC is a major component of the
 Computational Science Initiative at BNL.
We gratefully acknowledge computing resources
 provided through USQCD clusters at BNL
 and Jefferson Lab.
CL and ASM are supported in part by
 US DOE Contract DESC0012704(BNL)
 and by a DOE Office of Science Early Career Award.

\end{acknowledgments}

\clearpage

\appendix

%\nocite{*}

\section{Results}
\label{sec:appresults}

\begin{figure*}[h]
 \centering
 \includegraphics[page=2,width=8.5cm]{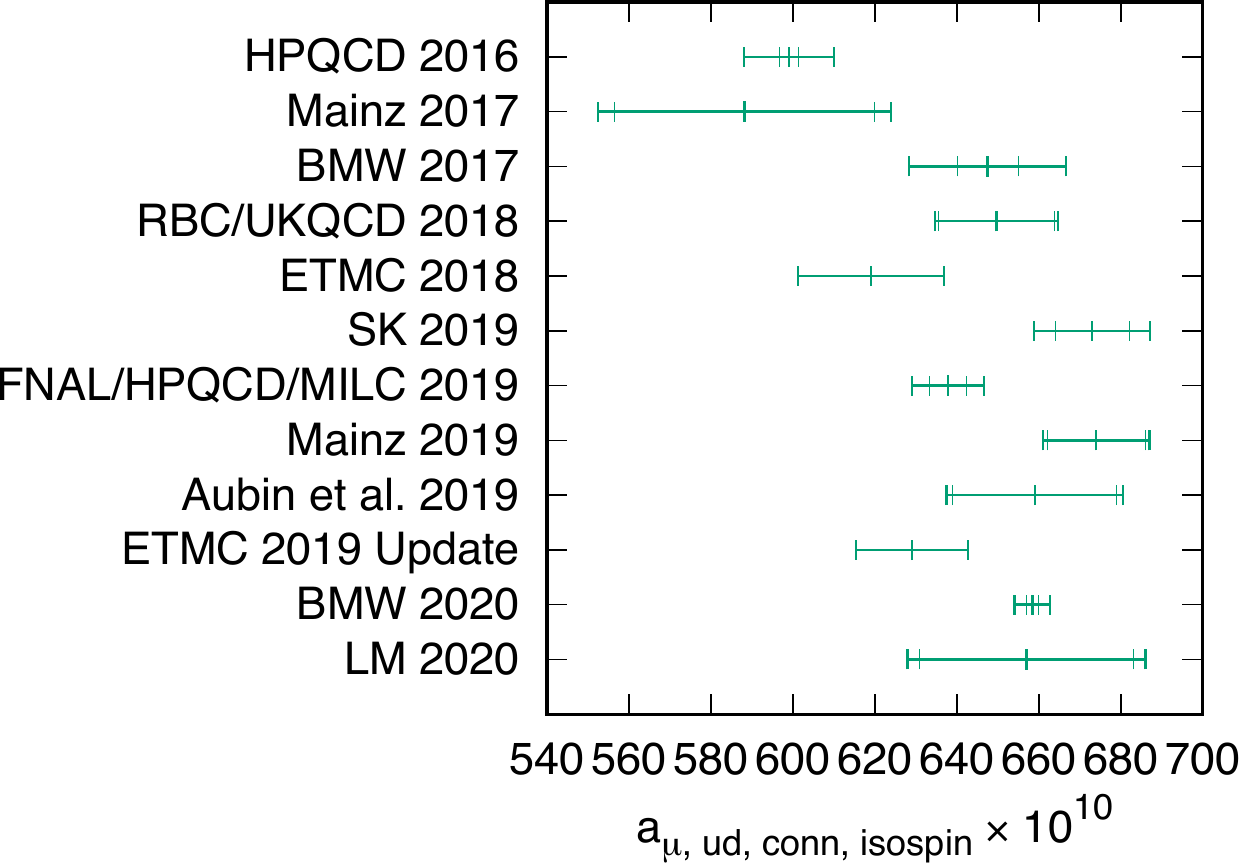}
 \includegraphics[page=3,width=8.5cm]{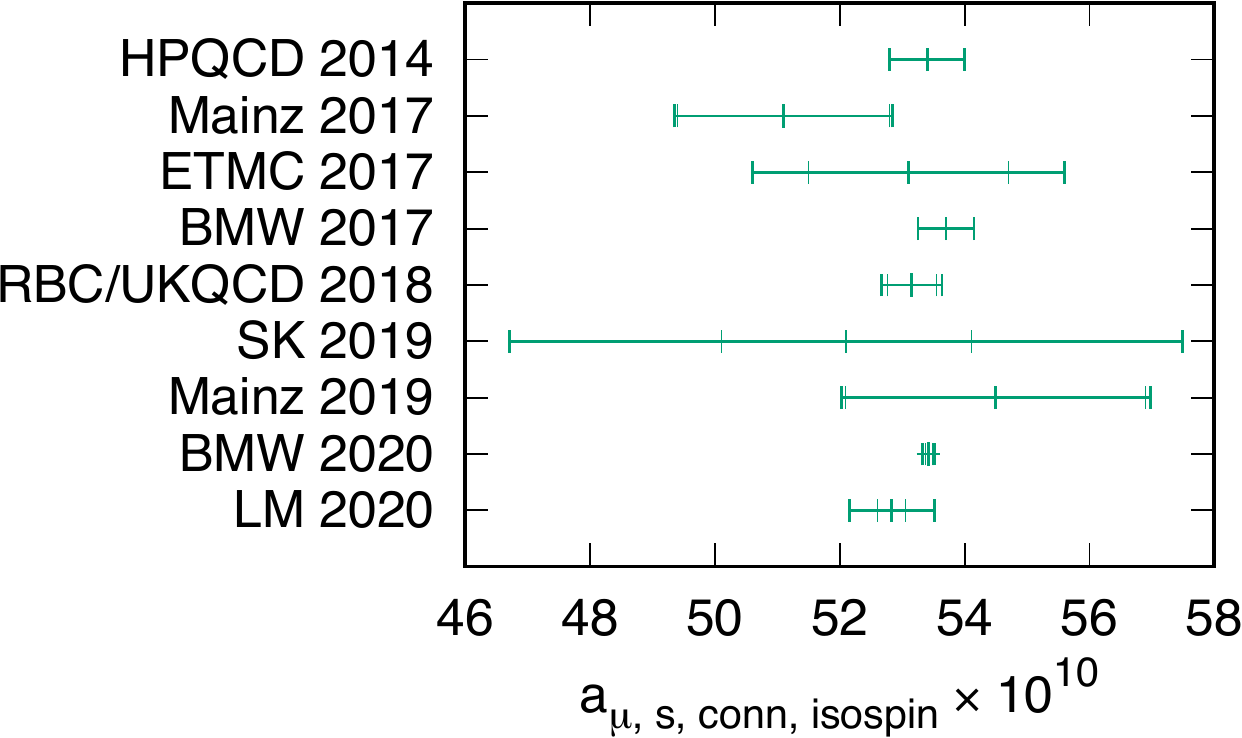}\\\vspace{0.2cm}
 \includegraphics[page=4,width=8.5cm]{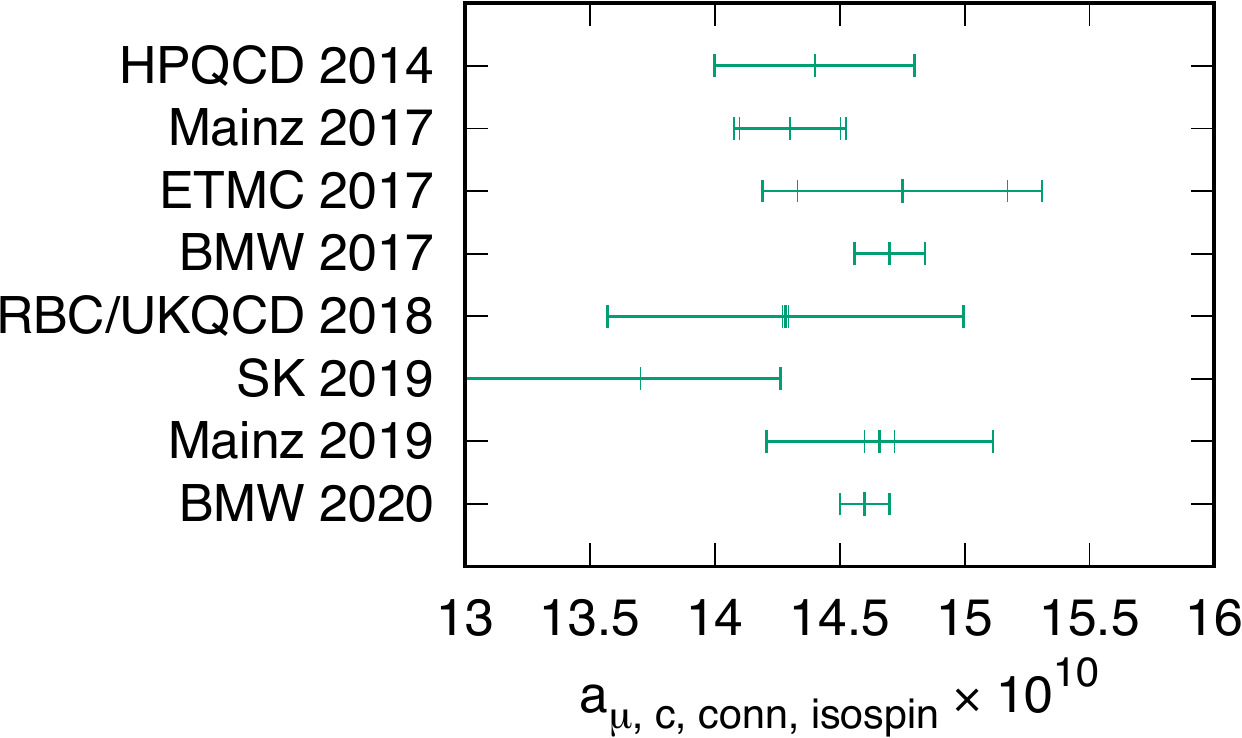}
  \includegraphics[page=5,width=8.5cm]{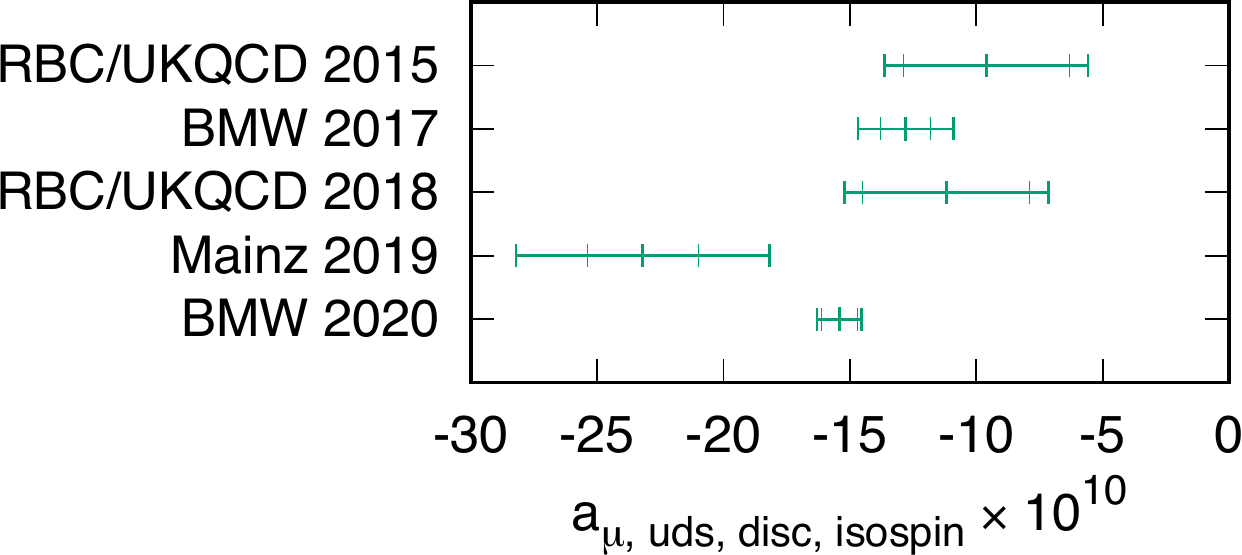}\\\vspace{0.2cm}
   \includegraphics[page=6,width=8.5cm]{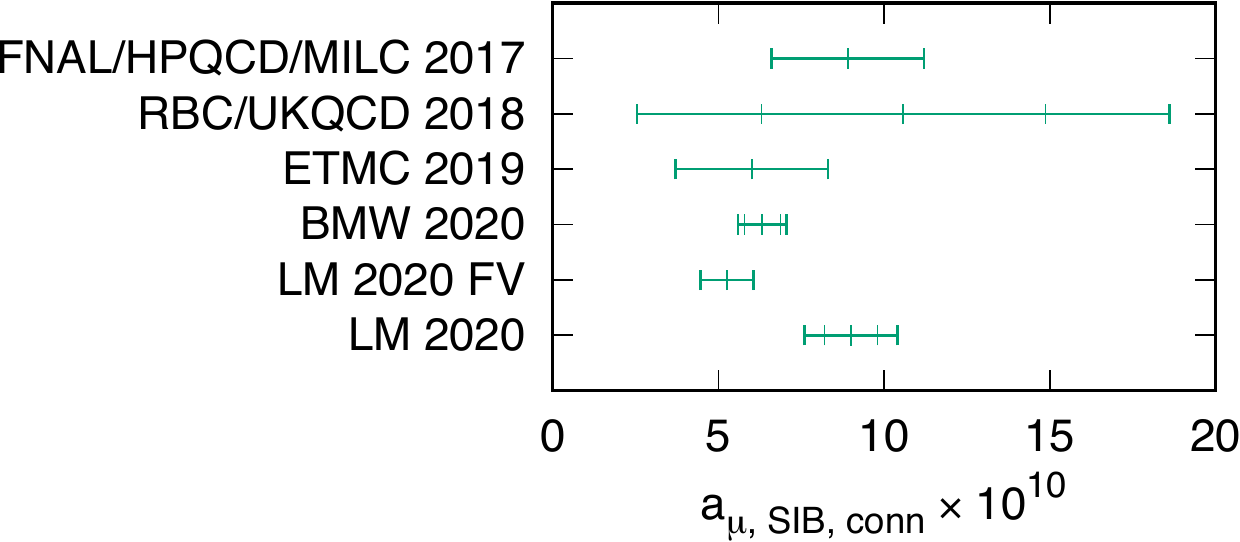}\hspace{0.5cm}
  \includegraphics[page=7,width=7.0cm]{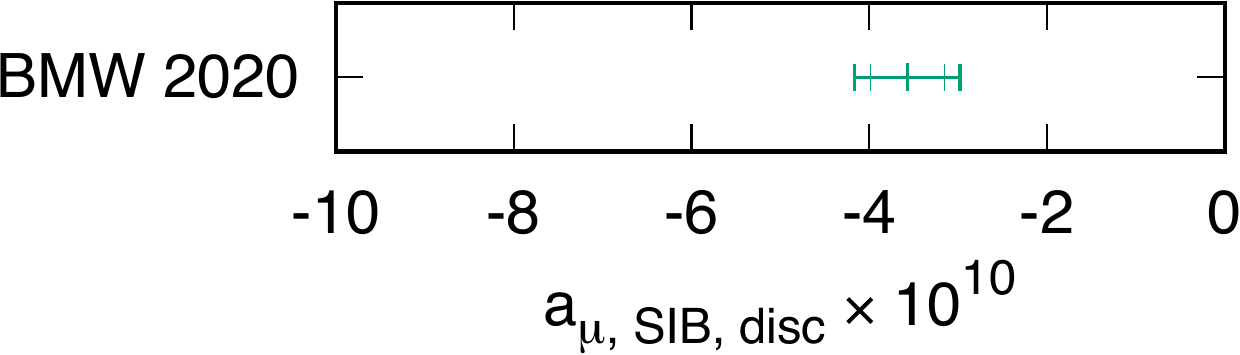}\\\vspace{0.2cm}
   \includegraphics[page=8,width=8.5cm]{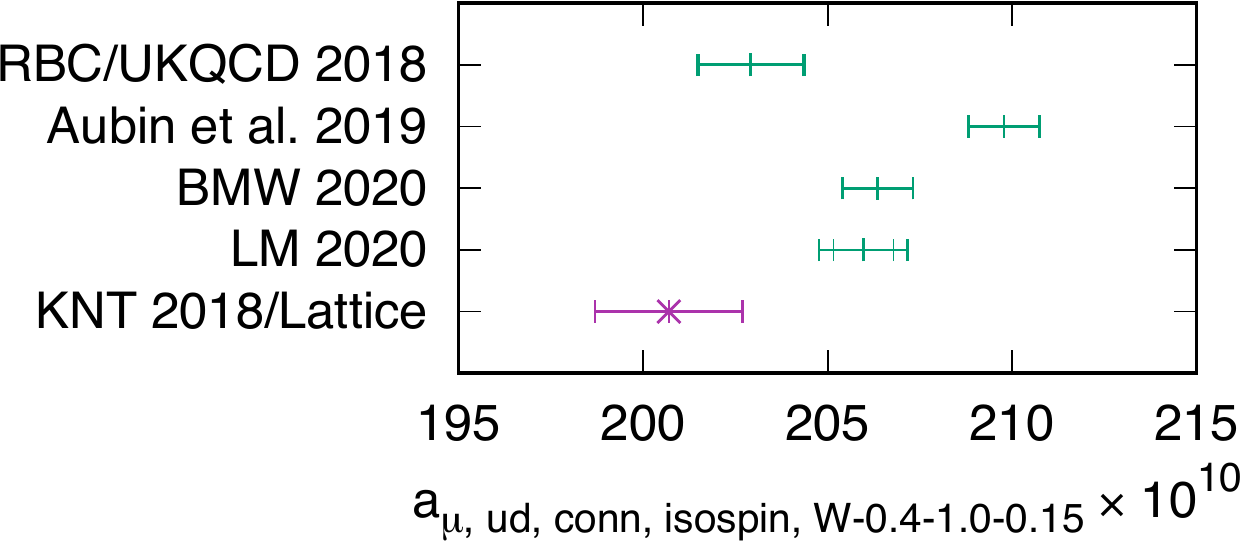}\hspace{0.5cm}
  \includegraphics[page=9,width=8.5cm]{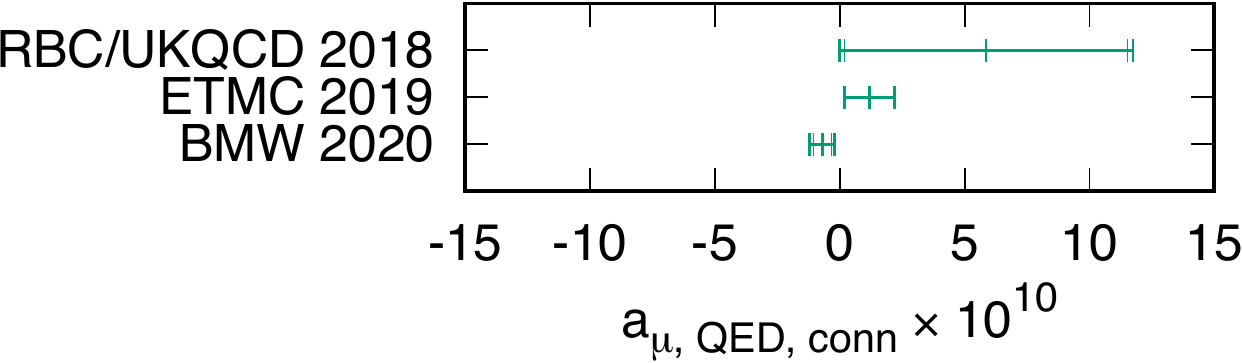}\\\vspace{0.2cm}
   \includegraphics[page=10,width=8.5cm]{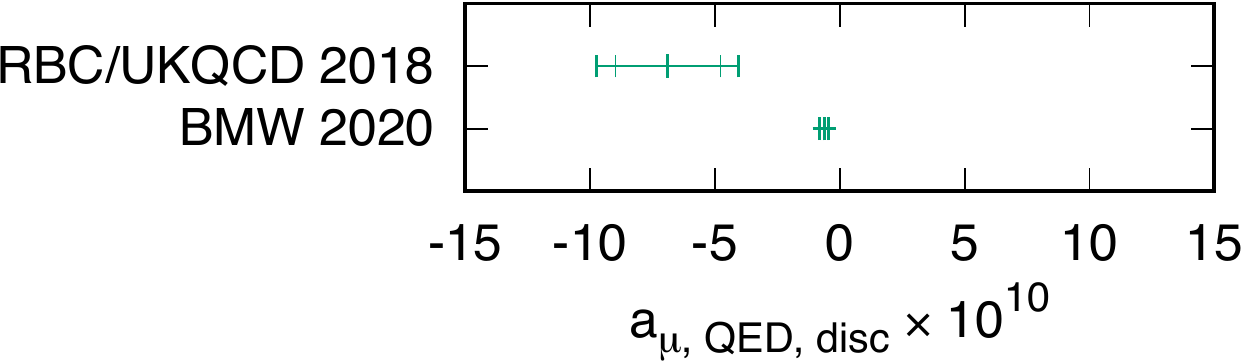}
 \caption{Overview of individual contributions to $a_\mu^{\rm HVP}$.  The result of this work is labelled ``LM 2020''.  The references are defined in the captions of Figs.~\ref{fig:overviewmain} and \ref{fig:overview}.}
  \label{fig:overviewall}
 \end{figure*}

This section contains tables with a detailed breakdown
 of systematic errors from the
 window data that appear in Table~\ref{table:windowresults}.
%Each table contains either the $p$ or $\hat{p}$
% prescription and unimproved or parity improved data.
Tables~\ref{table:windowphatrxunimp}~%
 and~\ref{table:windowphatrxipa} give the systematic
 error breakdown for the $\hat{p}$ prescription,
 while Tables~\ref{table:windowprxunimp}~%
 and~\ref{table:windowprxipa} give the $p$ prescription.
Tables~\ref{table:windowphatrxunimp}~%
 and~\ref{table:windowprxunimp}~%
 both contain unimproved data while
 Tables~\ref{table:windowphatrxipa}~%
 and~\ref{table:windowprxipa} both use the IPA
 procedure of Section~\ref{sec:ipa}.
All of Tables~\ref{table:windowphatrxunimp}-%
 \ref{table:windowprxipa} use the light quark mass data.
Tables~\ref{table:windowphatrxunimpstrange}~%
 and~\ref{table:windowprxunimpstrange}
 give the unimproved data with the $\hat{p}$ and $p$
 prescriptions for the strange quark mass, respectively.

\begin{table*}
\begin{tabular}{ccc|r}
$t_0$/fm & $t_1$/fm & $\Delta$/fm & $a^{\rm ud,conn.,isospin,W}_\mu  \times 10^{10}$ \\\hline
Total &  &  & $ 627(26)_{\rm S}(00)_{Z_V^{48}}(00)_{Z_V^{64}}(02)_{Z_V^{96}}(03)_{\rm C_3}(00)_{\rm C}(03)_{m_l}(00)_{\frac{m_l}{m_s}}(02)_{\frac{m_v}{m_s}}(00)_{\frac{w_0}{a^{48}}}(00)_{\frac{w_0}{a^{64}}}(01)_{\frac{w_0}{a^{96}}}(06)_{w_0} $ \\
\hline
0.0 & 0.1 & 0.15 & $ 3.59(00)_{\rm S}(00)_{Z_V^{48}}(00)_{Z_V^{64}}(00)_{Z_V^{96}}(06)_{\rm C_3}(52)_{\rm C}(00)_{m_l}(00)_{\frac{m_l}{m_s}}(26)_{\frac{m_v}{m_s}}(00)_{\frac{w_0}{a^{48}}}(00)_{\frac{w_0}{a^{64}}}(00)_{\frac{w_0}{a^{96}}}(01)_{w_0} $ \\
0.1 & 0.2 & 0.15 & $ 8.633(03)_{\rm S}(04)_{Z_V^{48}}(13)_{Z_V^{64}}(02)_{Z_V^{96}}(06)_{\rm C_3}(63)_{\rm C}(01)_{m_l}(00)_{\frac{m_l}{m_s}}(32)_{\frac{m_v}{m_s}}(00)_{\frac{w_0}{a^{48}}}(01)_{\frac{w_0}{a^{64}}}(00)_{\frac{w_0}{a^{96}}}(10)_{w_0} $ \\
0.2 & 0.3 & 0.15 & $ 14.24(01)_{\rm S}(01)_{Z_V^{48}}(02)_{Z_V^{64}}(01)_{Z_V^{96}}(09)_{\rm C_3}(73)_{\rm C}(00)_{m_l}(00)_{\frac{m_l}{m_s}}(36)_{\frac{m_v}{m_s}}(00)_{\frac{w_0}{a^{48}}}(00)_{\frac{w_0}{a^{64}}}(00)_{\frac{w_0}{a^{96}}}(02)_{w_0} $ \\
0.3 & 0.4 & 0.15 & $ 18.62(02)_{\rm S}(01)_{Z_V^{48}}(03)_{Z_V^{64}}(01)_{Z_V^{96}}(03)_{\rm C_3}(30)_{\rm C}(00)_{m_l}(00)_{\frac{m_l}{m_s}}(16)_{\frac{m_v}{m_s}}(00)_{\frac{w_0}{a^{48}}}(00)_{\frac{w_0}{a^{64}}}(00)_{\frac{w_0}{a^{96}}}(01)_{w_0} $ \\
0.4 & 0.5 & 0.15 & $ 24.552(35)_{\rm S}(20)_{Z_V^{48}}(49)_{Z_V^{64}}(03)_{Z_V^{96}}(03)_{\rm C_3}(19)_{\rm C}(14)_{m_l}(00)_{\frac{m_l}{m_s}}(10)_{\frac{m_v}{m_s}}(01)_{\frac{w_0}{a^{48}}}(01)_{\frac{w_0}{a^{64}}}(00)_{\frac{w_0}{a^{96}}}(07)_{w_0} $ \\
0.5 & 0.6 & 0.15 & $ 29.38(06)_{\rm S}(02)_{Z_V^{48}}(05)_{Z_V^{64}}(01)_{Z_V^{96}}(03)_{\rm C_3}(25)_{\rm C}(02)_{m_l}(00)_{\frac{m_l}{m_s}}(13)_{\frac{m_v}{m_s}}(00)_{\frac{w_0}{a^{48}}}(00)_{\frac{w_0}{a^{64}}}(00)_{\frac{w_0}{a^{96}}}(03)_{w_0} $ \\
0.6 & 0.7 & 0.15 & $ 33.72(10)_{\rm S}(03)_{Z_V^{48}}(06)_{Z_V^{64}}(01)_{Z_V^{96}}(03)_{\rm C_3}(32)_{\rm C}(04)_{m_l}(00)_{\frac{m_l}{m_s}}(15)_{\frac{m_v}{m_s}}(01)_{\frac{w_0}{a^{48}}}(01)_{\frac{w_0}{a^{64}}}(00)_{\frac{w_0}{a^{96}}}(02)_{w_0} $ \\
0.7 & 0.8 & 0.15 & $ 37.54(14)_{\rm S}(03)_{Z_V^{48}}(07)_{Z_V^{64}}(01)_{Z_V^{96}}(00)_{\rm C_3}(03)_{\rm C}(06)_{m_l}(00)_{\frac{m_l}{m_s}}(04)_{\frac{m_v}{m_s}}(01)_{\frac{w_0}{a^{48}}}(01)_{\frac{w_0}{a^{64}}}(00)_{\frac{w_0}{a^{96}}}(08)_{w_0} $ \\
0.8 & 0.9 & 0.15 & $ 39.32(20)_{\rm S}(03)_{Z_V^{48}}(07)_{Z_V^{64}}(01)_{Z_V^{96}}(02)_{\rm C_3}(09)_{\rm C}(08)_{m_l}(00)_{\frac{m_l}{m_s}}(02)_{\frac{m_v}{m_s}}(01)_{\frac{w_0}{a^{48}}}(02)_{\frac{w_0}{a^{64}}}(00)_{\frac{w_0}{a^{96}}}(14)_{w_0} $ \\
0.9 & 1.0 & 0.15 & $ 40.47(27)_{\rm S}(03)_{Z_V^{48}}(07)_{Z_V^{64}}(02)_{Z_V^{96}}(01)_{\rm C_3}(12)_{\rm C}(10)_{m_l}(00)_{\frac{m_l}{m_s}}(16)_{\frac{m_v}{m_s}}(02)_{\frac{w_0}{a^{48}}}(02)_{\frac{w_0}{a^{64}}}(01)_{\frac{w_0}{a^{96}}}(17)_{w_0} $ \\
1.0 & 1.1 & 0.15 & $ 40.47(44)_{\rm S}(04)_{Z_V^{48}}(10)_{Z_V^{64}}(03)_{Z_V^{96}}(01)_{\rm C_3}(20)_{\rm C}(14)_{m_l}(00)_{\frac{m_l}{m_s}}(11)_{\frac{m_v}{m_s}}(03)_{\frac{w_0}{a^{48}}}(05)_{\frac{w_0}{a^{64}}}(02)_{\frac{w_0}{a^{96}}}(25)_{w_0} $ \\
1.1 & 1.2 & 0.15 & $ 39.34(54)_{\rm S}(04)_{Z_V^{48}}(09)_{Z_V^{64}}(02)_{Z_V^{96}}(00)_{\rm C_3}(17)_{\rm C}(16)_{m_l}(01)_{\frac{m_l}{m_s}}(14)_{\frac{m_v}{m_s}}(04)_{\frac{w_0}{a^{48}}}(06)_{\frac{w_0}{a^{64}}}(01)_{\frac{w_0}{a^{96}}}(26)_{w_0} $ \\
1.2 & 1.3 & 0.15 & $ 37.53(65)_{\rm S}(03)_{Z_V^{48}}(09)_{Z_V^{64}}(01)_{Z_V^{96}}(00)_{\rm C_3}(19)_{\rm C}(18)_{m_l}(01)_{\frac{m_l}{m_s}}(23)_{\frac{m_v}{m_s}}(04)_{\frac{w_0}{a^{48}}}(06)_{\frac{w_0}{a^{64}}}(01)_{\frac{w_0}{a^{96}}}(32)_{w_0} $ \\
1.3 & 1.4 & 0.15 & $ 34.88(77)_{\rm S}(03)_{Z_V^{48}}(07)_{Z_V^{64}}(01)_{Z_V^{96}}(00)_{\rm C_3}(12)_{\rm C}(19)_{m_l}(01)_{\frac{m_l}{m_s}}(23)_{\frac{m_v}{m_s}}(04)_{\frac{w_0}{a^{48}}}(06)_{\frac{w_0}{a^{64}}}(01)_{\frac{w_0}{a^{96}}}(35)_{w_0} $ \\
1.4 & 1.5 & 0.15 & $ 31.94(88)_{\rm S}(02)_{Z_V^{48}}(06)_{Z_V^{64}}(00)_{Z_V^{96}}(00)_{\rm C_3}(14)_{\rm C}(19)_{m_l}(01)_{\frac{m_l}{m_s}}(26)_{\frac{m_v}{m_s}}(04)_{\frac{w_0}{a^{48}}}(06)_{\frac{w_0}{a^{64}}}(00)_{\frac{w_0}{a^{96}}}(37)_{w_0} $ \\
1.5 & 1.6 & 0.15 & $ 28.66(100)_{\rm S}(02)_{Z_V^{48}}(05)_{Z_V^{64}}(00)_{Z_V^{96}}(00)_{\rm C_3}(13)_{\rm C}(19)_{m_l}(01)_{\frac{m_l}{m_s}}(23)_{\frac{m_v}{m_s}}(04)_{\frac{w_0}{a^{48}}}(06)_{\frac{w_0}{a^{64}}}(00)_{\frac{w_0}{a^{96}}}(40)_{w_0} $ \\
1.6 & 1.7 & 0.15 & $ 24.58(81)_{\rm S}(01)_{Z_V^{48}}(02)_{Z_V^{64}}(03)_{Z_V^{96}}(12)_{\rm C_3}(00)_{\rm C}(16)_{m_l}(01)_{\frac{m_l}{m_s}}(42)_{\frac{m_v}{m_s}}(02)_{\frac{w_0}{a^{48}}}(02)_{\frac{w_0}{a^{64}}}(05)_{\frac{w_0}{a^{96}}}(39)_{w_0} $ \\
1.7 & 1.8 & 0.15 & $ 21.20(85)_{\rm S}(01)_{Z_V^{48}}(02)_{Z_V^{64}}(03)_{Z_V^{96}}(10)_{\rm C_3}(01)_{\rm C}(15)_{m_l}(00)_{\frac{m_l}{m_s}}(43)_{\frac{m_v}{m_s}}(02)_{\frac{w_0}{a^{48}}}(02)_{\frac{w_0}{a^{64}}}(05)_{\frac{w_0}{a^{96}}}(37)_{w_0} $ \\
1.8 & 1.9 & 0.15 & $ 18.13(86)_{\rm S}(01)_{Z_V^{48}}(02)_{Z_V^{64}}(02)_{Z_V^{96}}(04)_{\rm C_3}(00)_{\rm C}(13)_{m_l}(00)_{\frac{m_l}{m_s}}(47)_{\frac{m_v}{m_s}}(02)_{\frac{w_0}{a^{48}}}(02)_{\frac{w_0}{a^{64}}}(04)_{\frac{w_0}{a^{96}}}(33)_{w_0} $ \\
1.9 & 2.0 & 0.15 & $ 15.49(89)_{\rm S}(01)_{Z_V^{48}}(01)_{Z_V^{64}}(02)_{Z_V^{96}}(06)_{\rm C_3}(03)_{\rm C}(12)_{m_l}(00)_{\frac{m_l}{m_s}}(58)_{\frac{m_v}{m_s}}(02)_{\frac{w_0}{a^{48}}}(02)_{\frac{w_0}{a^{64}}}(04)_{\frac{w_0}{a^{96}}}(30)_{w_0} $ \\
\hline
0.0 & 0.2 & 0.15 & $ 12.22(00)_{\rm S}(00)_{Z_V^{48}}(02)_{Z_V^{64}}(00)_{Z_V^{96}}(06)_{\rm C_3}(46)_{\rm C}(00)_{m_l}(00)_{\frac{m_l}{m_s}}(23)_{\frac{m_v}{m_s}}(00)_{\frac{w_0}{a^{48}}}(00)_{\frac{w_0}{a^{64}}}(00)_{\frac{w_0}{a^{96}}}(02)_{w_0} $ \\
0.2 & 0.4 & 0.15 & $ 32.87(03)_{\rm S}(02)_{Z_V^{48}}(06)_{Z_V^{64}}(01)_{Z_V^{96}}(05)_{\rm C_3}(43)_{\rm C}(01)_{m_l}(00)_{\frac{m_l}{m_s}}(20)_{\frac{m_v}{m_s}}(00)_{\frac{w_0}{a^{48}}}(00)_{\frac{w_0}{a^{64}}}(00)_{\frac{w_0}{a^{96}}}(01)_{w_0} $ \\
0.4 & 0.6 & 0.15 & $ 53.93(10)_{\rm S}(04)_{Z_V^{48}}(10)_{Z_V^{64}}(02)_{Z_V^{96}}(02)_{\rm C_3}(23)_{\rm C}(04)_{m_l}(00)_{\frac{m_l}{m_s}}(11)_{\frac{m_v}{m_s}}(00)_{\frac{w_0}{a^{48}}}(00)_{\frac{w_0}{a^{64}}}(00)_{\frac{w_0}{a^{96}}}(02)_{w_0} $ \\
0.6 & 0.8 & 0.15 & $ 71.26(24)_{\rm S}(06)_{Z_V^{48}}(14)_{Z_V^{64}}(02)_{Z_V^{96}}(03)_{\rm C_3}(29)_{\rm C}(10)_{m_l}(00)_{\frac{m_l}{m_s}}(10)_{\frac{m_v}{m_s}}(01)_{\frac{w_0}{a^{48}}}(02)_{\frac{w_0}{a^{64}}}(00)_{\frac{w_0}{a^{96}}}(10)_{w_0} $ \\
0.8 & 1.0 & 0.15 & $ 79.80(47)_{\rm S}(06)_{Z_V^{48}}(15)_{Z_V^{64}}(03)_{Z_V^{96}}(02)_{\rm C_3}(04)_{\rm C}(18)_{m_l}(01)_{\frac{m_l}{m_s}}(14)_{\frac{m_v}{m_s}}(03)_{\frac{w_0}{a^{48}}}(04)_{\frac{w_0}{a^{64}}}(01)_{\frac{w_0}{a^{96}}}(30)_{w_0} $ \\
\hline
0.3 & 1.0 & 0.15 & $ 223.61(81)_{\rm S}(18)_{Z_V^{48}}(42)_{Z_V^{64}}(06)_{Z_V^{96}}(08)_{\rm C_3}(79)_{\rm C}(33)_{m_l}(01)_{\frac{m_l}{m_s}}(23)_{\frac{m_v}{m_s}}(05)_{\frac{w_0}{a^{48}}}(06)_{\frac{w_0}{a^{64}}}(01)_{\frac{w_0}{a^{96}}}(41)_{w_0} $ \\
0.3 & 1.3 & 0.15 & $ 340.7(2.6)_{\rm S}(0.3)_{Z_V^{48}}(0.9)_{Z_V^{64}}(0.2)_{Z_V^{96}}(0.2)_{\rm C_3}(0.5)_{\rm C}(0.8)_{m_l}(0.0)_{\frac{m_l}{m_s}}(0.4)_{\frac{m_v}{m_s}}(0.2)_{\frac{w_0}{a^{48}}}(0.3)_{\frac{w_0}{a^{64}}}(0.1)_{\frac{w_0}{a^{96}}}(1.2)_{w_0} $ \\
0.3 & 1.6 & 0.15 & $ 436.2(5.1)_{\rm S}(0.4)_{Z_V^{48}}(1.0)_{Z_V^{64}}(0.2)_{Z_V^{96}}(0.1)_{\rm C_3}(0.0)_{\rm C}(1.4)_{m_l}(0.0)_{\frac{m_l}{m_s}}(1.0)_{\frac{m_v}{m_s}}(0.3)_{\frac{w_0}{a^{48}}}(0.4)_{\frac{w_0}{a^{64}}}(0.1)_{\frac{w_0}{a^{96}}}(2.3)_{w_0} $ \\
\hline
0.4 & 1.0 & 0.15 & $ 204.99(79)_{\rm S}(16)_{Z_V^{48}}(39)_{Z_V^{64}}(05)_{Z_V^{96}}(05)_{\rm C_3}(48)_{\rm C}(33)_{m_l}(01)_{\frac{m_l}{m_s}}(07)_{\frac{m_v}{m_s}}(05)_{\frac{w_0}{a^{48}}}(06)_{\frac{w_0}{a^{64}}}(01)_{\frac{w_0}{a^{96}}}(42)_{w_0} $ \\
0.4 & 1.3 & 0.15 & $ 322.2(2.6)_{\rm S}(0.3)_{Z_V^{48}}(0.8)_{Z_V^{64}}(0.2)_{Z_V^{96}}(0.1)_{\rm C_3}(0.1)_{\rm C}(0.8)_{m_l}(0.0)_{\frac{m_l}{m_s}}(0.5)_{\frac{m_v}{m_s}}(0.2)_{\frac{w_0}{a^{48}}}(0.3)_{\frac{w_0}{a^{64}}}(0.1)_{\frac{w_0}{a^{96}}}(1.2)_{w_0} $ \\
0.4 & 1.6 & 0.15 & $ 417.7(5.1)_{\rm S}(0.4)_{Z_V^{48}}(1.0)_{Z_V^{64}}(0.2)_{Z_V^{96}}(0.1)_{\rm C_3}(0.4)_{\rm C}(1.4)_{m_l}(0.0)_{\frac{m_l}{m_s}}(1.2)_{\frac{m_v}{m_s}}(0.3)_{\frac{w_0}{a^{48}}}(0.4)_{\frac{w_0}{a^{64}}}(0.1)_{\frac{w_0}{a^{96}}}(2.3)_{w_0} $ \\
\hline
0.4 & 1.0 & 0.05 & $ 215.5(0.8)_{\rm S}(0.2)_{Z_V^{48}}(0.4)_{Z_V^{64}}(0.1)_{Z_V^{96}}(0.6)_{\rm C_3}(5.4)_{\rm C}(0.3)_{m_l}(0.0)_{\frac{m_l}{m_s}}(2.9)_{\frac{m_v}{m_s}}(0.1)_{\frac{w_0}{a^{48}}}(0.1)_{\frac{w_0}{a^{64}}}(0.0)_{\frac{w_0}{a^{96}}}(0.5)_{w_0} $ \\
0.4 & 1.0 & 0.1 & $ 208.85(77)_{\rm S}(17)_{Z_V^{48}}(40)_{Z_V^{64}}(05)_{Z_V^{96}}(02)_{\rm C_3}(23)_{\rm C}(33)_{m_l}(01)_{\frac{m_l}{m_s}}(25)_{\frac{m_v}{m_s}}(05)_{\frac{w_0}{a^{48}}}(06)_{\frac{w_0}{a^{64}}}(01)_{\frac{w_0}{a^{96}}}(36)_{w_0} $ \\
0.4 & 1.0 & 0.2 & $ 201.08(82)_{\rm S}(16)_{Z_V^{48}}(38)_{Z_V^{64}}(06)_{Z_V^{96}}(06)_{\rm C_3}(46)_{\rm C}(33)_{m_l}(01)_{\frac{m_l}{m_s}}(03)_{\frac{m_v}{m_s}}(05)_{\frac{w_0}{a^{48}}}(07)_{\frac{w_0}{a^{64}}}(01)_{\frac{w_0}{a^{96}}}(46)_{w_0} $ \\

\end{tabular}
\caption{
 %Window results, $\hat{p}$ prescription, unimproved
 A detailed breakdown of the systematic uncertainties
 for the data in the column labeled $l,\hat{p}{\rm U}$
 in Table~\ref{table:windowresults}.
 These are the unimproved data with the $\hat{p}$
 prescription applied.
 The subscripts on each uncertainty denote the different
 sources of uncertainty.
 In cases where each ensemble has a different uncertainty,
 a superscript of 48, 64, or 96 is included to indicate
 the 48c, 64c, or 96c ensemble, respectively.
 The subscript $S$ denotes the statistical error.
 The subscript $Z_V$ indicates the uncertainty
 on the vector current renormalization factors,
 which are given in Table~\ref{table:ensembledata}.
 The subscript $C_3$ indicates the continuum limit
 uncertainty on the $m_v/m_s=1/3$ ensemble,
 which is used to inform the shift on the other
 valence quark masses.
 The uncertainty obtained from the continuum limit of
 the other valence quark masses are collected
 into the subscript $C$.  We notice significant fluctuations
 of the short-distance window continuum error estimates since
 for the $t_0=0.1$~fm, $t_1=0.2$~fm window the 48c-64c and 64c-96c continuum
 extrapolations are in fortuitous agreement.  This indicates limits of the reliability of the corresponding error estimate. 
 The subscript $\frac{m_l}{m_s}$ denotes the uncertainty propagated from $\lambda_0$,  $m_l$ denotes the uncertainty from the light sea-quark mistuning,
 and $\frac{m_v}{m_s}$ denotes the light quark mass extrapolation uncertainty.
 The subscript $w_0/a$ denotes the scale setting uncertainty
 from the corresponding values in
 Table~\ref{table:ensembledata},
 and the subscript $w_0$ from the value in the caption
 of Table~\ref{table:ensembledata}.
 \label{table:windowphatrxunimp}
}
\end{table*}

\begin{table*}
\begin{tabular}{ccc|r}
$t_0$/fm & $t_1$/fm & $\Delta$/fm & $a^{\rm ud,conn.,isospin,W}_\mu  \times 10^{10}$ \\\hline
Total &  &  & $ 632(27)_{\rm S}(00)_{Z_V^{48}}(00)_{Z_V^{64}}(02)_{Z_V^{96}}(00)_{\rm C_3}(00)_{\rm C}(03)_{m_l}(00)_{\frac{m_l}{m_s}}(01)_{\frac{m_v}{m_s}}(00)_{\frac{w_0}{a^{48}}}(00)_{\frac{w_0}{a^{64}}}(01)_{\frac{w_0}{a^{96}}}(06)_{w_0} $ \\
\hline
0.0 & 0.1 & 0.15 & $ 4.60(00)_{\rm S}(00)_{Z_V^{48}}(01)_{Z_V^{64}}(00)_{Z_V^{96}}(03)_{\rm C_3}(28)_{\rm C}(00)_{m_l}(00)_{\frac{m_l}{m_s}}(14)_{\frac{m_v}{m_s}}(00)_{\frac{w_0}{a^{48}}}(00)_{\frac{w_0}{a^{64}}}(00)_{\frac{w_0}{a^{96}}}(01)_{w_0} $ \\
0.1 & 0.2 & 0.15 & $ 9.93(00)_{\rm S}(01)_{Z_V^{48}}(01)_{Z_V^{64}}(00)_{Z_V^{96}}(06)_{\rm C_3}(46)_{\rm C}(00)_{m_l}(00)_{\frac{m_l}{m_s}}(23)_{\frac{m_v}{m_s}}(00)_{\frac{w_0}{a^{48}}}(00)_{\frac{w_0}{a^{64}}}(00)_{\frac{w_0}{a^{96}}}(00)_{w_0} $ \\
0.2 & 0.3 & 0.15 & $ 15.4(0.0)_{\rm S}(0.0)_{Z_V^{48}}(0.0)_{Z_V^{64}}(0.0)_{Z_V^{96}}(0.1)_{\rm C_3}(1.1)_{\rm C}(0.0)_{m_l}(0.0)_{\frac{m_l}{m_s}}(0.6)_{\frac{m_v}{m_s}}(0.0)_{\frac{w_0}{a^{48}}}(0.0)_{\frac{w_0}{a^{64}}}(0.0)_{\frac{w_0}{a^{96}}}(0.0)_{w_0} $ \\
0.3 & 0.4 & 0.15 & $ 20.2(0.0)_{\rm S}(0.0)_{Z_V^{48}}(0.0)_{Z_V^{64}}(0.0)_{Z_V^{96}}(0.1)_{\rm C_3}(1.0)_{\rm C}(0.0)_{m_l}(0.0)_{\frac{m_l}{m_s}}(0.5)_{\frac{m_v}{m_s}}(0.0)_{\frac{w_0}{a^{48}}}(0.0)_{\frac{w_0}{a^{64}}}(0.0)_{\frac{w_0}{a^{96}}}(0.0)_{w_0} $ \\
0.4 & 0.5 & 0.15 & $ 24.71(04)_{\rm S}(02)_{Z_V^{48}}(05)_{Z_V^{64}}(01)_{Z_V^{96}}(02)_{\rm C_3}(21)_{\rm C}(01)_{m_l}(00)_{\frac{m_l}{m_s}}(09)_{\frac{m_v}{m_s}}(00)_{\frac{w_0}{a^{48}}}(00)_{\frac{w_0}{a^{64}}}(00)_{\frac{w_0}{a^{96}}}(01)_{w_0} $ \\
0.5 & 0.6 & 0.15 & $ 29.42(06)_{\rm S}(02)_{Z_V^{48}}(06)_{Z_V^{64}}(01)_{Z_V^{96}}(02)_{\rm C_3}(22)_{\rm C}(03)_{m_l}(00)_{\frac{m_l}{m_s}}(11)_{\frac{m_v}{m_s}}(00)_{\frac{w_0}{a^{48}}}(00)_{\frac{w_0}{a^{64}}}(00)_{\frac{w_0}{a^{96}}}(01)_{w_0} $ \\
0.6 & 0.7 & 0.15 & $ 33.87(10)_{\rm S}(03)_{Z_V^{48}}(06)_{Z_V^{64}}(01)_{Z_V^{96}}(02)_{\rm C_3}(22)_{\rm C}(04)_{m_l}(00)_{\frac{m_l}{m_s}}(10)_{\frac{m_v}{m_s}}(00)_{\frac{w_0}{a^{48}}}(01)_{\frac{w_0}{a^{64}}}(00)_{\frac{w_0}{a^{96}}}(04)_{w_0} $ \\
0.7 & 0.8 & 0.15 & $ 37.30(15)_{\rm S}(03)_{Z_V^{48}}(07)_{Z_V^{64}}(01)_{Z_V^{96}}(01)_{\rm C_3}(13)_{\rm C}(06)_{m_l}(00)_{\frac{m_l}{m_s}}(04)_{\frac{m_v}{m_s}}(01)_{\frac{w_0}{a^{48}}}(01)_{\frac{w_0}{a^{64}}}(00)_{\frac{w_0}{a^{96}}}(08)_{w_0} $ \\
0.8 & 0.9 & 0.15 & $ 39.52(21)_{\rm S}(03)_{Z_V^{48}}(07)_{Z_V^{64}}(01)_{Z_V^{96}}(01)_{\rm C_3}(01)_{\rm C}(08)_{m_l}(00)_{\frac{m_l}{m_s}}(04)_{\frac{m_v}{m_s}}(01)_{\frac{w_0}{a^{48}}}(02)_{\frac{w_0}{a^{64}}}(00)_{\frac{w_0}{a^{96}}}(13)_{w_0} $ \\
0.9 & 1.0 & 0.15 & $ 40.43(28)_{\rm S}(03)_{Z_V^{48}}(07)_{Z_V^{64}}(02)_{Z_V^{96}}(00)_{\rm C_3}(08)_{\rm C}(10)_{m_l}(00)_{\frac{m_l}{m_s}}(14)_{\frac{m_v}{m_s}}(02)_{\frac{w_0}{a^{48}}}(02)_{\frac{w_0}{a^{64}}}(01)_{\frac{w_0}{a^{96}}}(18)_{w_0} $ \\
1.0 & 1.1 & 0.15 & $ 40.56(46)_{\rm S}(04)_{Z_V^{48}}(10)_{Z_V^{64}}(03)_{Z_V^{96}}(01)_{\rm C_3}(19)_{\rm C}(14)_{m_l}(00)_{\frac{m_l}{m_s}}(12)_{\frac{m_v}{m_s}}(03)_{\frac{w_0}{a^{48}}}(05)_{\frac{w_0}{a^{64}}}(01)_{\frac{w_0}{a^{96}}}(23)_{w_0} $ \\
1.1 & 1.2 & 0.15 & $ 39.47(56)_{\rm S}(04)_{Z_V^{48}}(10)_{Z_V^{64}}(02)_{Z_V^{96}}(01)_{\rm C_3}(20)_{\rm C}(16)_{m_l}(01)_{\frac{m_l}{m_s}}(17)_{\frac{m_v}{m_s}}(04)_{\frac{w_0}{a^{48}}}(06)_{\frac{w_0}{a^{64}}}(01)_{\frac{w_0}{a^{96}}}(27)_{w_0} $ \\
1.2 & 1.3 & 0.15 & $ 37.54(67)_{\rm S}(03)_{Z_V^{48}}(09)_{Z_V^{64}}(01)_{Z_V^{96}}(00)_{\rm C_3}(18)_{\rm C}(18)_{m_l}(01)_{\frac{m_l}{m_s}}(22)_{\frac{m_v}{m_s}}(04)_{\frac{w_0}{a^{48}}}(06)_{\frac{w_0}{a^{64}}}(01)_{\frac{w_0}{a^{96}}}(32)_{w_0} $ \\
1.3 & 1.4 & 0.15 & $ 34.98(79)_{\rm S}(03)_{Z_V^{48}}(07)_{Z_V^{64}}(01)_{Z_V^{96}}(00)_{\rm C_3}(15)_{\rm C}(19)_{m_l}(01)_{\frac{m_l}{m_s}}(26)_{\frac{m_v}{m_s}}(04)_{\frac{w_0}{a^{48}}}(06)_{\frac{w_0}{a^{64}}}(01)_{\frac{w_0}{a^{96}}}(35)_{w_0} $ \\
1.4 & 1.5 & 0.15 & $ 31.98(91)_{\rm S}(02)_{Z_V^{48}}(06)_{Z_V^{64}}(00)_{Z_V^{96}}(00)_{\rm C_3}(14)_{\rm C}(19)_{m_l}(01)_{\frac{m_l}{m_s}}(27)_{\frac{m_v}{m_s}}(04)_{\frac{w_0}{a^{48}}}(06)_{\frac{w_0}{a^{64}}}(00)_{\frac{w_0}{a^{96}}}(38)_{w_0} $ \\
1.5 & 1.6 & 0.15 & $ 28.7(1.0)_{\rm S}(0.0)_{Z_V^{48}}(0.1)_{Z_V^{64}}(0.0)_{Z_V^{96}}(0.0)_{\rm C_3}(0.1)_{\rm C}(0.2)_{m_l}(0.0)_{\frac{m_l}{m_s}}(0.2)_{\frac{m_v}{m_s}}(0.0)_{\frac{w_0}{a^{48}}}(0.1)_{\frac{w_0}{a^{64}}}(0.0)_{\frac{w_0}{a^{96}}}(0.4)_{w_0} $ \\
1.6 & 1.7 & 0.15 & $ 24.64(82)_{\rm S}(01)_{Z_V^{48}}(02)_{Z_V^{64}}(03)_{Z_V^{96}}(13)_{\rm C_3}(00)_{\rm C}(16)_{m_l}(01)_{\frac{m_l}{m_s}}(43)_{\frac{m_v}{m_s}}(02)_{\frac{w_0}{a^{48}}}(02)_{\frac{w_0}{a^{64}}}(05)_{\frac{w_0}{a^{96}}}(39)_{w_0} $ \\
1.7 & 1.8 & 0.15 & $ 21.28(86)_{\rm S}(01)_{Z_V^{48}}(02)_{Z_V^{64}}(03)_{Z_V^{96}}(10)_{\rm C_3}(01)_{\rm C}(15)_{m_l}(00)_{\frac{m_l}{m_s}}(46)_{\frac{m_v}{m_s}}(02)_{\frac{w_0}{a^{48}}}(02)_{\frac{w_0}{a^{64}}}(05)_{\frac{w_0}{a^{96}}}(37)_{w_0} $ \\
1.8 & 1.9 & 0.15 & $ 18.22(88)_{\rm S}(01)_{Z_V^{48}}(01)_{Z_V^{64}}(03)_{Z_V^{96}}(04)_{\rm C_3}(00)_{\rm C}(13)_{m_l}(00)_{\frac{m_l}{m_s}}(51)_{\frac{m_v}{m_s}}(02)_{\frac{w_0}{a^{48}}}(02)_{\frac{w_0}{a^{64}}}(04)_{\frac{w_0}{a^{96}}}(34)_{w_0} $ \\
1.9 & 2.0 & 0.15 & $ 15.42(91)_{\rm S}(01)_{Z_V^{48}}(01)_{Z_V^{64}}(02)_{Z_V^{96}}(08)_{\rm C_3}(03)_{\rm C}(12)_{m_l}(00)_{\frac{m_l}{m_s}}(37)_{\frac{m_v}{m_s}}(02)_{\frac{w_0}{a^{48}}}(02)_{\frac{w_0}{a^{64}}}(04)_{\frac{w_0}{a^{96}}}(31)_{w_0} $ \\
\hline
0.0 & 0.2 & 0.15 & $ 14.53(00)_{\rm S}(01)_{Z_V^{48}}(02)_{Z_V^{64}}(01)_{Z_V^{96}}(02)_{\rm C_3}(19)_{\rm C}(00)_{m_l}(00)_{\frac{m_l}{m_s}}(09)_{\frac{m_v}{m_s}}(00)_{\frac{w_0}{a^{48}}}(00)_{\frac{w_0}{a^{64}}}(00)_{\frac{w_0}{a^{96}}}(01)_{w_0} $ \\
0.2 & 0.4 & 0.15 & $ 35.6(0.0)_{\rm S}(0.0)_{Z_V^{48}}(0.1)_{Z_V^{64}}(0.0)_{Z_V^{96}}(0.3)_{\rm C_3}(2.2)_{\rm C}(0.0)_{m_l}(0.0)_{\frac{m_l}{m_s}}(1.1)_{\frac{m_v}{m_s}}(0.0)_{\frac{w_0}{a^{48}}}(0.0)_{\frac{w_0}{a^{64}}}(0.0)_{\frac{w_0}{a^{96}}}(0.0)_{w_0} $ \\
0.4 & 0.6 & 0.15 & $ 54.12(10)_{\rm S}(05)_{Z_V^{48}}(11)_{Z_V^{64}}(01)_{Z_V^{96}}(00)_{\rm C_3}(01)_{\rm C}(04)_{m_l}(00)_{\frac{m_l}{m_s}}(02)_{\frac{m_v}{m_s}}(00)_{\frac{w_0}{a^{48}}}(00)_{\frac{w_0}{a^{64}}}(00)_{\frac{w_0}{a^{96}}}(03)_{w_0} $ \\
0.6 & 0.8 & 0.15 & $ 71.16(25)_{\rm S}(06)_{Z_V^{48}}(14)_{Z_V^{64}}(02)_{Z_V^{96}}(03)_{\rm C_3}(36)_{\rm C}(10)_{m_l}(00)_{\frac{m_l}{m_s}}(14)_{\frac{m_v}{m_s}}(01)_{\frac{w_0}{a^{48}}}(02)_{\frac{w_0}{a^{64}}}(00)_{\frac{w_0}{a^{96}}}(11)_{w_0} $ \\
0.8 & 1.0 & 0.15 & $ 79.96(49)_{\rm S}(06)_{Z_V^{48}}(14)_{Z_V^{64}}(03)_{Z_V^{96}}(01)_{\rm C_3}(07)_{\rm C}(18)_{m_l}(01)_{\frac{m_l}{m_s}}(18)_{\frac{m_v}{m_s}}(03)_{\frac{w_0}{a^{48}}}(04)_{\frac{w_0}{a^{64}}}(01)_{\frac{w_0}{a^{96}}}(31)_{w_0} $ \\
\hline
0.3 & 1.0 & 0.15 & $ 225.47(83)_{\rm S}(18)_{Z_V^{48}}(42)_{Z_V^{64}}(07)_{Z_V^{96}}(07)_{\rm C_3}(74)_{\rm C}(34)_{m_l}(01)_{\frac{m_l}{m_s}}(49)_{\frac{m_v}{m_s}}(05)_{\frac{w_0}{a^{48}}}(06)_{\frac{w_0}{a^{64}}}(02)_{\frac{w_0}{a^{96}}}(47)_{w_0} $ \\
0.3 & 1.3 & 0.15 & $ 343.4(2.7)_{\rm S}(0.4)_{Z_V^{48}}(0.9)_{Z_V^{64}}(0.3)_{Z_V^{96}}(0.2)_{\rm C_3}(1.6)_{\rm C}(0.8)_{m_l}(0.0)_{\frac{m_l}{m_s}}(1.1)_{\frac{m_v}{m_s}}(0.2)_{\frac{w_0}{a^{48}}}(0.3)_{\frac{w_0}{a^{64}}}(0.1)_{\frac{w_0}{a^{96}}}(1.3)_{w_0} $ \\
0.3 & 1.6 & 0.15 & $ 439.0(5.3)_{\rm S}(0.4)_{Z_V^{48}}(1.1)_{Z_V^{64}}(0.3)_{Z_V^{96}}(0.2)_{\rm C_3}(2.0)_{\rm C}(1.4)_{m_l}(0.0)_{\frac{m_l}{m_s}}(2.0)_{\frac{m_v}{m_s}}(0.3)_{\frac{w_0}{a^{48}}}(0.5)_{\frac{w_0}{a^{64}}}(0.1)_{\frac{w_0}{a^{96}}}(2.4)_{w_0} $ \\
\hline
0.4 & 1.0 & 0.15 & $ 205.25(82)_{\rm S}(17)_{Z_V^{48}}(39)_{Z_V^{64}}(06)_{Z_V^{96}}(04)_{\rm C_3}(29)_{\rm C}(33)_{m_l}(01)_{\frac{m_l}{m_s}}(01)_{\frac{m_v}{m_s}}(05)_{\frac{w_0}{a^{48}}}(06)_{\frac{w_0}{a^{64}}}(01)_{\frac{w_0}{a^{96}}}(45)_{w_0} $ \\
0.4 & 1.3 & 0.15 & $ 322.8(2.7)_{\rm S}(0.3)_{Z_V^{48}}(0.8)_{Z_V^{64}}(0.2)_{Z_V^{96}}(0.1)_{\rm C_3}(0.2)_{\rm C}(0.8)_{m_l}(0.0)_{\frac{m_l}{m_s}}(0.6)_{\frac{m_v}{m_s}}(0.2)_{\frac{w_0}{a^{48}}}(0.3)_{\frac{w_0}{a^{64}}}(0.1)_{\frac{w_0}{a^{96}}}(1.3)_{w_0} $ \\
0.4 & 1.6 & 0.15 & $ 418.5(5.3)_{\rm S}(0.4)_{Z_V^{48}}(1.0)_{Z_V^{64}}(0.2)_{Z_V^{96}}(0.0)_{\rm C_3}(0.7)_{\rm C}(1.4)_{m_l}(0.0)_{\frac{m_l}{m_s}}(1.4)_{\frac{m_v}{m_s}}(0.3)_{\frac{w_0}{a^{48}}}(0.5)_{\frac{w_0}{a^{64}}}(0.1)_{\frac{w_0}{a^{96}}}(2.4)_{w_0} $ \\
\hline
0.4 & 1.0 & 0.05 & $ 208.5(0.8)_{\rm S}(0.2)_{Z_V^{48}}(0.4)_{Z_V^{64}}(0.1)_{Z_V^{96}}(0.1)_{\rm C_3}(1.1)_{\rm C}(0.3)_{m_l}(0.0)_{\frac{m_l}{m_s}}(0.5)_{\frac{m_v}{m_s}}(0.0)_{\frac{w_0}{a^{48}}}(0.1)_{\frac{w_0}{a^{64}}}(0.0)_{\frac{w_0}{a^{96}}}(0.7)_{w_0} $ \\
0.4 & 1.0 & 0.1 & $ 207.60(79)_{\rm S}(17)_{Z_V^{48}}(40)_{Z_V^{64}}(06)_{Z_V^{96}}(09)_{\rm C_3}(84)_{\rm C}(32)_{m_l}(01)_{\frac{m_l}{m_s}}(29)_{\frac{m_v}{m_s}}(05)_{\frac{w_0}{a^{48}}}(06)_{\frac{w_0}{a^{64}}}(01)_{\frac{w_0}{a^{96}}}(43)_{w_0} $ \\
0.4 & 1.0 & 0.2 & $ 201.86(85)_{\rm S}(16)_{Z_V^{48}}(38)_{Z_V^{64}}(07)_{Z_V^{96}}(01)_{\rm C_3}(08)_{\rm C}(33)_{m_l}(01)_{\frac{m_l}{m_s}}(23)_{\frac{m_v}{m_s}}(05)_{\frac{w_0}{a^{48}}}(07)_{\frac{w_0}{a^{64}}}(02)_{\frac{w_0}{a^{96}}}(48)_{w_0} $ \\

\end{tabular}
\caption{
 Same as Table~\ref{table:windowphatrxunimp},
 but with the parity improvement.
 \label{table:windowphatrxipa}
}
\end{table*}

\begin{table*}
\begin{tabular}{ccc|r}
$t_0$/fm & $t_1$/fm & $\Delta$/fm & $a^{\rm ud,conn.,isospin,W}_\mu  \times 10^{10}$ \\\hline
Total &  &  & $ 628(26)_{\rm S}(00)_{Z_V^{48}}(00)_{Z_V^{64}}(02)_{Z_V^{96}}(02)_{\rm C_3}(00)_{\rm C}(03)_{m_l}(00)_{\frac{m_l}{m_s}}(02)_{\frac{m_v}{m_s}}(00)_{\frac{w_0}{a^{48}}}(00)_{\frac{w_0}{a^{64}}}(01)_{\frac{w_0}{a^{96}}}(06)_{w_0} $ \\
\hline
0.0 & 0.1 & 0.15 & $ 4.32(00)_{\rm S}(00)_{Z_V^{48}}(01)_{Z_V^{64}}(00)_{Z_V^{96}}(02)_{\rm C_3}(18)_{\rm C}(00)_{m_l}(00)_{\frac{m_l}{m_s}}(09)_{\frac{m_v}{m_s}}(00)_{\frac{w_0}{a^{48}}}(00)_{\frac{w_0}{a^{64}}}(00)_{\frac{w_0}{a^{96}}}(01)_{w_0} $ \\
0.1 & 0.2 & 0.15 & $ 9.29(00)_{\rm S}(01)_{Z_V^{48}}(02)_{Z_V^{64}}(00)_{Z_V^{96}}(05)_{\rm C_3}(40)_{\rm C}(00)_{m_l}(00)_{\frac{m_l}{m_s}}(20)_{\frac{m_v}{m_s}}(00)_{\frac{w_0}{a^{48}}}(00)_{\frac{w_0}{a^{64}}}(00)_{\frac{w_0}{a^{96}}}(00)_{w_0} $ \\
0.2 & 0.3 & 0.15 & $ 14.7(0.0)_{\rm S}(0.0)_{Z_V^{48}}(0.0)_{Z_V^{64}}(0.0)_{Z_V^{96}}(0.1)_{\rm C_3}(1.1)_{\rm C}(0.0)_{m_l}(0.0)_{\frac{m_l}{m_s}}(0.5)_{\frac{m_v}{m_s}}(0.0)_{\frac{w_0}{a^{48}}}(0.0)_{\frac{w_0}{a^{64}}}(0.0)_{\frac{w_0}{a^{96}}}(0.0)_{w_0} $ \\
0.3 & 0.4 & 0.15 & $ 18.71(02)_{\rm S}(02)_{Z_V^{48}}(04)_{Z_V^{64}}(01)_{Z_V^{96}}(03)_{\rm C_3}(23)_{\rm C}(00)_{m_l}(00)_{\frac{m_l}{m_s}}(13)_{\frac{m_v}{m_s}}(00)_{\frac{w_0}{a^{48}}}(00)_{\frac{w_0}{a^{64}}}(00)_{\frac{w_0}{a^{96}}}(01)_{w_0} $ \\
0.4 & 0.5 & 0.15 & $ 24.518(36)_{\rm S}(22)_{Z_V^{48}}(50)_{Z_V^{64}}(04)_{Z_V^{96}}(02)_{\rm C_3}(13)_{\rm C}(14)_{m_l}(00)_{\frac{m_l}{m_s}}(03)_{\frac{m_v}{m_s}}(01)_{\frac{w_0}{a^{48}}}(00)_{\frac{w_0}{a^{64}}}(00)_{\frac{w_0}{a^{96}}}(08)_{w_0} $ \\
0.5 & 0.6 & 0.15 & $ 29.36(06)_{\rm S}(02)_{Z_V^{48}}(06)_{Z_V^{64}}(01)_{Z_V^{96}}(03)_{\rm C_3}(27)_{\rm C}(02)_{m_l}(00)_{\frac{m_l}{m_s}}(14)_{\frac{m_v}{m_s}}(00)_{\frac{w_0}{a^{48}}}(00)_{\frac{w_0}{a^{64}}}(00)_{\frac{w_0}{a^{96}}}(03)_{w_0} $ \\
0.6 & 0.7 & 0.15 & $ 33.70(10)_{\rm S}(03)_{Z_V^{48}}(06)_{Z_V^{64}}(01)_{Z_V^{96}}(04)_{\rm C_3}(34)_{\rm C}(04)_{m_l}(00)_{\frac{m_l}{m_s}}(17)_{\frac{m_v}{m_s}}(01)_{\frac{w_0}{a^{48}}}(01)_{\frac{w_0}{a^{64}}}(00)_{\frac{w_0}{a^{96}}}(02)_{w_0} $ \\
0.7 & 0.8 & 0.15 & $ 37.54(15)_{\rm S}(03)_{Z_V^{48}}(07)_{Z_V^{64}}(01)_{Z_V^{96}}(00)_{\rm C_3}(03)_{\rm C}(06)_{m_l}(00)_{\frac{m_l}{m_s}}(04)_{\frac{m_v}{m_s}}(01)_{\frac{w_0}{a^{48}}}(01)_{\frac{w_0}{a^{64}}}(00)_{\frac{w_0}{a^{96}}}(08)_{w_0} $ \\
0.8 & 0.9 & 0.15 & $ 39.33(21)_{\rm S}(03)_{Z_V^{48}}(08)_{Z_V^{64}}(01)_{Z_V^{96}}(02)_{\rm C_3}(09)_{\rm C}(08)_{m_l}(00)_{\frac{m_l}{m_s}}(02)_{\frac{m_v}{m_s}}(01)_{\frac{w_0}{a^{48}}}(02)_{\frac{w_0}{a^{64}}}(00)_{\frac{w_0}{a^{96}}}(14)_{w_0} $ \\
0.9 & 1.0 & 0.15 & $ 40.47(27)_{\rm S}(03)_{Z_V^{48}}(07)_{Z_V^{64}}(02)_{Z_V^{96}}(01)_{\rm C_3}(12)_{\rm C}(10)_{m_l}(00)_{\frac{m_l}{m_s}}(17)_{\frac{m_v}{m_s}}(02)_{\frac{w_0}{a^{48}}}(02)_{\frac{w_0}{a^{64}}}(01)_{\frac{w_0}{a^{96}}}(16)_{w_0} $ \\
1.0 & 1.1 & 0.15 & $ 40.49(45)_{\rm S}(04)_{Z_V^{48}}(10)_{Z_V^{64}}(03)_{Z_V^{96}}(01)_{\rm C_3}(21)_{\rm C}(14)_{m_l}(00)_{\frac{m_l}{m_s}}(11)_{\frac{m_v}{m_s}}(03)_{\frac{w_0}{a^{48}}}(05)_{\frac{w_0}{a^{64}}}(02)_{\frac{w_0}{a^{96}}}(25)_{w_0} $ \\
1.1 & 1.2 & 0.15 & $ 39.35(55)_{\rm S}(04)_{Z_V^{48}}(09)_{Z_V^{64}}(02)_{Z_V^{96}}(00)_{\rm C_3}(17)_{\rm C}(16)_{m_l}(01)_{\frac{m_l}{m_s}}(14)_{\frac{m_v}{m_s}}(04)_{\frac{w_0}{a^{48}}}(06)_{\frac{w_0}{a^{64}}}(01)_{\frac{w_0}{a^{96}}}(26)_{w_0} $ \\
1.2 & 1.3 & 0.15 & $ 37.55(66)_{\rm S}(03)_{Z_V^{48}}(09)_{Z_V^{64}}(01)_{Z_V^{96}}(00)_{\rm C_3}(20)_{\rm C}(18)_{m_l}(01)_{\frac{m_l}{m_s}}(23)_{\frac{m_v}{m_s}}(04)_{\frac{w_0}{a^{48}}}(06)_{\frac{w_0}{a^{64}}}(01)_{\frac{w_0}{a^{96}}}(32)_{w_0} $ \\
1.3 & 1.4 & 0.15 & $ 34.89(77)_{\rm S}(03)_{Z_V^{48}}(07)_{Z_V^{64}}(01)_{Z_V^{96}}(00)_{\rm C_3}(13)_{\rm C}(19)_{m_l}(01)_{\frac{m_l}{m_s}}(24)_{\frac{m_v}{m_s}}(04)_{\frac{w_0}{a^{48}}}(06)_{\frac{w_0}{a^{64}}}(01)_{\frac{w_0}{a^{96}}}(35)_{w_0} $ \\
1.4 & 1.5 & 0.15 & $ 31.96(89)_{\rm S}(02)_{Z_V^{48}}(06)_{Z_V^{64}}(00)_{Z_V^{96}}(00)_{\rm C_3}(14)_{\rm C}(19)_{m_l}(01)_{\frac{m_l}{m_s}}(26)_{\frac{m_v}{m_s}}(04)_{\frac{w_0}{a^{48}}}(06)_{\frac{w_0}{a^{64}}}(00)_{\frac{w_0}{a^{96}}}(37)_{w_0} $ \\
1.5 & 1.6 & 0.15 & $ 28.7(1.0)_{\rm S}(0.0)_{Z_V^{48}}(0.1)_{Z_V^{64}}(0.0)_{Z_V^{96}}(0.0)_{\rm C_3}(0.1)_{\rm C}(0.2)_{m_l}(0.0)_{\frac{m_l}{m_s}}(0.2)_{\frac{m_v}{m_s}}(0.0)_{\frac{w_0}{a^{48}}}(0.1)_{\frac{w_0}{a^{64}}}(0.0)_{\frac{w_0}{a^{96}}}(0.4)_{w_0} $ \\
1.6 & 1.7 & 0.15 & $ 24.59(81)_{\rm S}(01)_{Z_V^{48}}(02)_{Z_V^{64}}(03)_{Z_V^{96}}(12)_{\rm C_3}(00)_{\rm C}(16)_{m_l}(01)_{\frac{m_l}{m_s}}(42)_{\frac{m_v}{m_s}}(02)_{\frac{w_0}{a^{48}}}(02)_{\frac{w_0}{a^{64}}}(05)_{\frac{w_0}{a^{96}}}(39)_{w_0} $ \\
1.7 & 1.8 & 0.15 & $ 21.21(85)_{\rm S}(01)_{Z_V^{48}}(02)_{Z_V^{64}}(03)_{Z_V^{96}}(11)_{\rm C_3}(01)_{\rm C}(15)_{m_l}(00)_{\frac{m_l}{m_s}}(43)_{\frac{m_v}{m_s}}(02)_{\frac{w_0}{a^{48}}}(02)_{\frac{w_0}{a^{64}}}(05)_{\frac{w_0}{a^{96}}}(37)_{w_0} $ \\
1.8 & 1.9 & 0.15 & $ 18.13(87)_{\rm S}(01)_{Z_V^{48}}(02)_{Z_V^{64}}(02)_{Z_V^{96}}(05)_{\rm C_3}(00)_{\rm C}(13)_{m_l}(00)_{\frac{m_l}{m_s}}(47)_{\frac{m_v}{m_s}}(02)_{\frac{w_0}{a^{48}}}(02)_{\frac{w_0}{a^{64}}}(04)_{\frac{w_0}{a^{96}}}(33)_{w_0} $ \\
1.9 & 2.0 & 0.15 & $ 15.37(89)_{\rm S}(01)_{Z_V^{48}}(01)_{Z_V^{64}}(02)_{Z_V^{96}}(06)_{\rm C_3}(03)_{\rm C}(11)_{m_l}(00)_{\frac{m_l}{m_s}}(32)_{\frac{m_v}{m_s}}(02)_{\frac{w_0}{a^{48}}}(02)_{\frac{w_0}{a^{64}}}(04)_{\frac{w_0}{a^{96}}}(30)_{w_0} $ \\
\hline
0.0 & 0.2 & 0.15 & $ 13.61(00)_{\rm S}(01)_{Z_V^{48}}(02)_{Z_V^{64}}(00)_{Z_V^{96}}(03)_{\rm C_3}(23)_{\rm C}(00)_{m_l}(00)_{\frac{m_l}{m_s}}(12)_{\frac{m_v}{m_s}}(00)_{\frac{w_0}{a^{48}}}(00)_{\frac{w_0}{a^{64}}}(00)_{\frac{w_0}{a^{96}}}(01)_{w_0} $ \\
0.2 & 0.4 & 0.15 & $ 33.41(03)_{\rm S}(03)_{Z_V^{48}}(06)_{Z_V^{64}}(01)_{Z_V^{96}}(10)_{\rm C_3}(84)_{\rm C}(01)_{m_l}(00)_{\frac{m_l}{m_s}}(40)_{\frac{m_v}{m_s}}(00)_{\frac{w_0}{a^{48}}}(00)_{\frac{w_0}{a^{64}}}(00)_{\frac{w_0}{a^{96}}}(02)_{w_0} $ \\
0.4 & 0.6 & 0.15 & $ 53.88(10)_{\rm S}(05)_{Z_V^{48}}(10)_{Z_V^{64}}(02)_{Z_V^{96}}(03)_{\rm C_3}(28)_{\rm C}(04)_{m_l}(00)_{\frac{m_l}{m_s}}(14)_{\frac{m_v}{m_s}}(00)_{\frac{w_0}{a^{48}}}(00)_{\frac{w_0}{a^{64}}}(00)_{\frac{w_0}{a^{96}}}(02)_{w_0} $ \\
0.6 & 0.8 & 0.15 & $ 71.23(24)_{\rm S}(06)_{Z_V^{48}}(14)_{Z_V^{64}}(02)_{Z_V^{96}}(03)_{\rm C_3}(32)_{\rm C}(10)_{m_l}(00)_{\frac{m_l}{m_s}}(11)_{\frac{m_v}{m_s}}(01)_{\frac{w_0}{a^{48}}}(02)_{\frac{w_0}{a^{64}}}(00)_{\frac{w_0}{a^{96}}}(10)_{w_0} $ \\
0.8 & 1.0 & 0.15 & $ 79.81(48)_{\rm S}(06)_{Z_V^{48}}(15)_{Z_V^{64}}(03)_{Z_V^{96}}(01)_{\rm C_3}(04)_{\rm C}(18)_{m_l}(01)_{\frac{m_l}{m_s}}(14)_{\frac{m_v}{m_s}}(03)_{\frac{w_0}{a^{48}}}(04)_{\frac{w_0}{a^{64}}}(01)_{\frac{w_0}{a^{96}}}(30)_{w_0} $ \\
\hline
0.3 & 1.0 & 0.15 & $ 223.63(82)_{\rm S}(19)_{Z_V^{48}}(43)_{Z_V^{64}}(06)_{Z_V^{96}}(08)_{\rm C_3}(80)_{\rm C}(33)_{m_l}(01)_{\frac{m_l}{m_s}}(23)_{\frac{m_v}{m_s}}(05)_{\frac{w_0}{a^{48}}}(06)_{\frac{w_0}{a^{64}}}(01)_{\frac{w_0}{a^{96}}}(41)_{w_0} $ \\
0.3 & 1.3 & 0.15 & $ 340.7(2.6)_{\rm S}(0.4)_{Z_V^{48}}(0.9)_{Z_V^{64}}(0.2)_{Z_V^{96}}(0.2)_{\rm C_3}(0.5)_{\rm C}(0.8)_{m_l}(0.0)_{\frac{m_l}{m_s}}(0.4)_{\frac{m_v}{m_s}}(0.2)_{\frac{w_0}{a^{48}}}(0.3)_{\frac{w_0}{a^{64}}}(0.1)_{\frac{w_0}{a^{96}}}(1.2)_{w_0} $ \\
0.3 & 1.6 & 0.15 & $ 436.3(5.1)_{\rm S}(0.4)_{Z_V^{48}}(1.1)_{Z_V^{64}}(0.2)_{Z_V^{96}}(0.1)_{\rm C_3}(0.0)_{\rm C}(1.4)_{m_l}(0.0)_{\frac{m_l}{m_s}}(1.0)_{\frac{m_v}{m_s}}(0.3)_{\frac{w_0}{a^{48}}}(0.4)_{\frac{w_0}{a^{64}}}(0.1)_{\frac{w_0}{a^{96}}}(2.3)_{w_0} $ \\
\hline
0.4 & 1.0 & 0.15 & $ 204.93(80)_{\rm S}(17)_{Z_V^{48}}(39)_{Z_V^{64}}(06)_{Z_V^{96}}(06)_{\rm C_3}(56)_{\rm C}(33)_{m_l}(01)_{\frac{m_l}{m_s}}(11)_{\frac{m_v}{m_s}}(05)_{\frac{w_0}{a^{48}}}(06)_{\frac{w_0}{a^{64}}}(01)_{\frac{w_0}{a^{96}}}(42)_{w_0} $ \\
0.4 & 1.3 & 0.15 & $ 322.2(2.6)_{\rm S}(0.3)_{Z_V^{48}}(0.8)_{Z_V^{64}}(0.2)_{Z_V^{96}}(0.1)_{\rm C_3}(0.2)_{\rm C}(0.8)_{m_l}(0.0)_{\frac{m_l}{m_s}}(0.5)_{\frac{m_v}{m_s}}(0.2)_{\frac{w_0}{a^{48}}}(0.3)_{\frac{w_0}{a^{64}}}(0.1)_{\frac{w_0}{a^{96}}}(1.2)_{w_0} $ \\
0.4 & 1.6 & 0.15 & $ 417.9(5.1)_{\rm S}(0.4)_{Z_V^{48}}(1.0)_{Z_V^{64}}(0.2)_{Z_V^{96}}(0.1)_{\rm C_3}(0.3)_{\rm C}(1.4)_{m_l}(0.0)_{\frac{m_l}{m_s}}(1.0)_{\frac{m_v}{m_s}}(0.3)_{\frac{w_0}{a^{48}}}(0.5)_{\frac{w_0}{a^{64}}}(0.1)_{\frac{w_0}{a^{96}}}(2.3)_{w_0} $ \\
\hline
0.4 & 1.0 & 0.05 & $ 215.8(0.8)_{\rm S}(0.2)_{Z_V^{48}}(0.4)_{Z_V^{64}}(0.1)_{Z_V^{96}}(0.6)_{\rm C_3}(5.6)_{\rm C}(0.3)_{m_l}(0.0)_{\frac{m_l}{m_s}}(3.0)_{\frac{m_v}{m_s}}(0.1)_{\frac{w_0}{a^{48}}}(0.1)_{\frac{w_0}{a^{64}}}(0.0)_{\frac{w_0}{a^{96}}}(0.4)_{w_0} $ \\
0.4 & 1.0 & 0.1 & $ 208.76(78)_{\rm S}(17)_{Z_V^{48}}(40)_{Z_V^{64}}(05)_{Z_V^{96}}(01)_{\rm C_3}(13)_{\rm C}(33)_{m_l}(01)_{\frac{m_l}{m_s}}(21)_{\frac{m_v}{m_s}}(05)_{\frac{w_0}{a^{48}}}(06)_{\frac{w_0}{a^{64}}}(01)_{\frac{w_0}{a^{96}}}(36)_{w_0} $ \\
0.4 & 1.0 & 0.2 & $ 201.10(84)_{\rm S}(17)_{Z_V^{48}}(38)_{Z_V^{64}}(06)_{Z_V^{96}}(06)_{\rm C_3}(48)_{\rm C}(33)_{m_l}(01)_{\frac{m_l}{m_s}}(03)_{\frac{m_v}{m_s}}(05)_{\frac{w_0}{a^{48}}}(07)_{\frac{w_0}{a^{64}}}(01)_{\frac{w_0}{a^{96}}}(46)_{w_0} $ \\

\end{tabular}
\caption{
 Same as Table~\ref{table:windowphatrxunimp},
 but with the $p$ prescription.
 \label{table:windowprxunimp}
}
\end{table*}

\begin{table*}
\begin{tabular}{ccc|r}
$t_0$/fm & $t_1$/fm & $\Delta$/fm & $a^{\rm ud,conn.,isospin,W}_\mu  \times 10^{10}$ \\\hline
Total &  &  & $ 634(27)_{\rm S}(00)_{Z_V^{48}}(00)_{Z_V^{64}}(01)_{Z_V^{96}}(02)_{\rm C_3}(01)_{\rm C}(03)_{m_l}(00)_{\frac{m_l}{m_s}}(01)_{\frac{m_v}{m_s}}(00)_{\frac{w_0}{a^{48}}}(00)_{\frac{w_0}{a^{64}}}(01)_{\frac{w_0}{a^{96}}}(06)_{w_0} $ \\
\hline
0.0 & 0.1 & 0.15 & $ 5.69(00)_{\rm S}(00)_{Z_V^{48}}(01)_{Z_V^{64}}(00)_{Z_V^{96}}(03)_{\rm C_3}(21)_{\rm C}(00)_{m_l}(00)_{\frac{m_l}{m_s}}(11)_{\frac{m_v}{m_s}}(00)_{\frac{w_0}{a^{48}}}(00)_{\frac{w_0}{a^{64}}}(00)_{\frac{w_0}{a^{96}}}(00)_{w_0} $ \\
0.1 & 0.2 & 0.15 & $ 11.0(0.0)_{\rm S}(0.0)_{Z_V^{48}}(0.0)_{Z_V^{64}}(0.0)_{Z_V^{96}}(0.1)_{\rm C_3}(1.1)_{\rm C}(0.0)_{m_l}(0.0)_{\frac{m_l}{m_s}}(0.6)_{\frac{m_v}{m_s}}(0.0)_{\frac{w_0}{a^{48}}}(0.0)_{\frac{w_0}{a^{64}}}(0.0)_{\frac{w_0}{a^{96}}}(0.0)_{w_0} $ \\
0.2 & 0.3 & 0.15 & $ 16.0(0.0)_{\rm S}(0.0)_{Z_V^{48}}(0.0)_{Z_V^{64}}(0.0)_{Z_V^{96}}(0.2)_{\rm C_3}(1.6)_{\rm C}(0.0)_{m_l}(0.0)_{\frac{m_l}{m_s}}(0.8)_{\frac{m_v}{m_s}}(0.0)_{\frac{w_0}{a^{48}}}(0.0)_{\frac{w_0}{a^{64}}}(0.0)_{\frac{w_0}{a^{96}}}(0.0)_{w_0} $ \\
0.3 & 0.4 & 0.15 & $ 20.3(0.0)_{\rm S}(0.0)_{Z_V^{48}}(0.0)_{Z_V^{64}}(0.0)_{Z_V^{96}}(0.1)_{\rm C_3}(1.1)_{\rm C}(0.0)_{m_l}(0.0)_{\frac{m_l}{m_s}}(0.6)_{\frac{m_v}{m_s}}(0.0)_{\frac{w_0}{a^{48}}}(0.0)_{\frac{w_0}{a^{64}}}(0.0)_{\frac{w_0}{a^{96}}}(0.0)_{w_0} $ \\
0.4 & 0.5 & 0.15 & $ 24.65(04)_{\rm S}(02)_{Z_V^{48}}(05)_{Z_V^{64}}(01)_{Z_V^{96}}(02)_{\rm C_3}(19)_{\rm C}(01)_{m_l}(00)_{\frac{m_l}{m_s}}(08)_{\frac{m_v}{m_s}}(00)_{\frac{w_0}{a^{48}}}(00)_{\frac{w_0}{a^{64}}}(00)_{\frac{w_0}{a^{96}}}(02)_{w_0} $ \\
0.5 & 0.6 & 0.15 & $ 29.36(06)_{\rm S}(03)_{Z_V^{48}}(06)_{Z_V^{64}}(01)_{Z_V^{96}}(03)_{\rm C_3}(27)_{\rm C}(03)_{m_l}(00)_{\frac{m_l}{m_s}}(14)_{\frac{m_v}{m_s}}(00)_{\frac{w_0}{a^{48}}}(00)_{\frac{w_0}{a^{64}}}(00)_{\frac{w_0}{a^{96}}}(01)_{w_0} $ \\
0.6 & 0.7 & 0.15 & $ 33.82(10)_{\rm S}(03)_{Z_V^{48}}(06)_{Z_V^{64}}(01)_{Z_V^{96}}(03)_{\rm C_3}(25)_{\rm C}(04)_{m_l}(00)_{\frac{m_l}{m_s}}(12)_{\frac{m_v}{m_s}}(00)_{\frac{w_0}{a^{48}}}(01)_{\frac{w_0}{a^{64}}}(00)_{\frac{w_0}{a^{96}}}(04)_{w_0} $ \\
0.7 & 0.8 & 0.15 & $ 37.28(15)_{\rm S}(03)_{Z_V^{48}}(07)_{Z_V^{64}}(01)_{Z_V^{96}}(01)_{\rm C_3}(15)_{\rm C}(06)_{m_l}(00)_{\frac{m_l}{m_s}}(05)_{\frac{m_v}{m_s}}(01)_{\frac{w_0}{a^{48}}}(01)_{\frac{w_0}{a^{64}}}(00)_{\frac{w_0}{a^{96}}}(08)_{w_0} $ \\
0.8 & 0.9 & 0.15 & $ 39.52(21)_{\rm S}(03)_{Z_V^{48}}(07)_{Z_V^{64}}(01)_{Z_V^{96}}(01)_{\rm C_3}(01)_{\rm C}(08)_{m_l}(00)_{\frac{m_l}{m_s}}(04)_{\frac{m_v}{m_s}}(01)_{\frac{w_0}{a^{48}}}(02)_{\frac{w_0}{a^{64}}}(00)_{\frac{w_0}{a^{96}}}(13)_{w_0} $ \\
0.9 & 1.0 & 0.15 & $ 40.44(28)_{\rm S}(03)_{Z_V^{48}}(07)_{Z_V^{64}}(02)_{Z_V^{96}}(00)_{\rm C_3}(08)_{\rm C}(10)_{m_l}(00)_{\frac{m_l}{m_s}}(14)_{\frac{m_v}{m_s}}(02)_{\frac{w_0}{a^{48}}}(02)_{\frac{w_0}{a^{64}}}(01)_{\frac{w_0}{a^{96}}}(18)_{w_0} $ \\
1.0 & 1.1 & 0.15 & $ 40.57(46)_{\rm S}(04)_{Z_V^{48}}(10)_{Z_V^{64}}(03)_{Z_V^{96}}(01)_{\rm C_3}(19)_{\rm C}(14)_{m_l}(00)_{\frac{m_l}{m_s}}(12)_{\frac{m_v}{m_s}}(03)_{\frac{w_0}{a^{48}}}(05)_{\frac{w_0}{a^{64}}}(01)_{\frac{w_0}{a^{96}}}(23)_{w_0} $ \\
1.1 & 1.2 & 0.15 & $ 39.48(57)_{\rm S}(04)_{Z_V^{48}}(10)_{Z_V^{64}}(02)_{Z_V^{96}}(01)_{\rm C_3}(21)_{\rm C}(16)_{m_l}(01)_{\frac{m_l}{m_s}}(17)_{\frac{m_v}{m_s}}(04)_{\frac{w_0}{a^{48}}}(06)_{\frac{w_0}{a^{64}}}(01)_{\frac{w_0}{a^{96}}}(27)_{w_0} $ \\
1.2 & 1.3 & 0.15 & $ 37.56(68)_{\rm S}(03)_{Z_V^{48}}(09)_{Z_V^{64}}(01)_{Z_V^{96}}(00)_{\rm C_3}(18)_{\rm C}(18)_{m_l}(01)_{\frac{m_l}{m_s}}(23)_{\frac{m_v}{m_s}}(04)_{\frac{w_0}{a^{48}}}(06)_{\frac{w_0}{a^{64}}}(01)_{\frac{w_0}{a^{96}}}(32)_{w_0} $ \\
1.3 & 1.4 & 0.15 & $ 35.00(80)_{\rm S}(03)_{Z_V^{48}}(07)_{Z_V^{64}}(01)_{Z_V^{96}}(00)_{\rm C_3}(16)_{\rm C}(19)_{m_l}(01)_{\frac{m_l}{m_s}}(27)_{\frac{m_v}{m_s}}(04)_{\frac{w_0}{a^{48}}}(06)_{\frac{w_0}{a^{64}}}(01)_{\frac{w_0}{a^{96}}}(35)_{w_0} $ \\
1.4 & 1.5 & 0.15 & $ 31.99(92)_{\rm S}(02)_{Z_V^{48}}(06)_{Z_V^{64}}(00)_{Z_V^{96}}(00)_{\rm C_3}(15)_{\rm C}(19)_{m_l}(01)_{\frac{m_l}{m_s}}(27)_{\frac{m_v}{m_s}}(04)_{\frac{w_0}{a^{48}}}(06)_{\frac{w_0}{a^{64}}}(00)_{\frac{w_0}{a^{96}}}(38)_{w_0} $ \\
1.5 & 1.6 & 0.15 & $ 28.7(1.0)_{\rm S}(0.0)_{Z_V^{48}}(0.1)_{Z_V^{64}}(0.0)_{Z_V^{96}}(0.0)_{\rm C_3}(0.1)_{\rm C}(0.2)_{m_l}(0.0)_{\frac{m_l}{m_s}}(0.2)_{\frac{m_v}{m_s}}(0.0)_{\frac{w_0}{a^{48}}}(0.1)_{\frac{w_0}{a^{64}}}(0.0)_{\frac{w_0}{a^{96}}}(0.4)_{w_0} $ \\
1.6 & 1.7 & 0.15 & $ 24.65(83)_{\rm S}(01)_{Z_V^{48}}(02)_{Z_V^{64}}(03)_{Z_V^{96}}(13)_{\rm C_3}(00)_{\rm C}(16)_{m_l}(01)_{\frac{m_l}{m_s}}(43)_{\frac{m_v}{m_s}}(02)_{\frac{w_0}{a^{48}}}(02)_{\frac{w_0}{a^{64}}}(05)_{\frac{w_0}{a^{96}}}(39)_{w_0} $ \\
1.7 & 1.8 & 0.15 & $ 21.20(86)_{\rm S}(01)_{Z_V^{48}}(02)_{Z_V^{64}}(03)_{Z_V^{96}}(10)_{\rm C_3}(01)_{\rm C}(14)_{m_l}(00)_{\frac{m_l}{m_s}}(37)_{\frac{m_v}{m_s}}(02)_{\frac{w_0}{a^{48}}}(02)_{\frac{w_0}{a^{64}}}(05)_{\frac{w_0}{a^{96}}}(37)_{w_0} $ \\
1.8 & 1.9 & 0.15 & $ 18.23(88)_{\rm S}(01)_{Z_V^{48}}(01)_{Z_V^{64}}(03)_{Z_V^{96}}(04)_{\rm C_3}(00)_{\rm C}(13)_{m_l}(00)_{\frac{m_l}{m_s}}(51)_{\frac{m_v}{m_s}}(02)_{\frac{w_0}{a^{48}}}(02)_{\frac{w_0}{a^{64}}}(04)_{\frac{w_0}{a^{96}}}(34)_{w_0} $ \\
1.9 & 2.0 & 0.15 & $ 15.57(91)_{\rm S}(01)_{Z_V^{48}}(01)_{Z_V^{64}}(02)_{Z_V^{96}}(08)_{\rm C_3}(03)_{\rm C}(12)_{m_l}(00)_{\frac{m_l}{m_s}}(61)_{\frac{m_v}{m_s}}(02)_{\frac{w_0}{a^{48}}}(02)_{\frac{w_0}{a^{64}}}(04)_{\frac{w_0}{a^{96}}}(30)_{w_0} $ \\
\hline
0.0 & 0.2 & 0.15 & $ 16.7(0.0)_{\rm S}(0.0)_{Z_V^{48}}(0.0)_{Z_V^{64}}(0.0)_{Z_V^{96}}(0.2)_{\rm C_3}(1.3)_{\rm C}(0.0)_{m_l}(0.0)_{\frac{m_l}{m_s}}(0.7)_{\frac{m_v}{m_s}}(0.0)_{\frac{w_0}{a^{48}}}(0.0)_{\frac{w_0}{a^{64}}}(0.0)_{\frac{w_0}{a^{96}}}(0.0)_{w_0} $ \\
0.2 & 0.4 & 0.15 & $ 36.3(0.0)_{\rm S}(0.0)_{Z_V^{48}}(0.1)_{Z_V^{64}}(0.0)_{Z_V^{96}}(0.3)_{\rm C_3}(2.7)_{\rm C}(0.0)_{m_l}(0.0)_{\frac{m_l}{m_s}}(1.4)_{\frac{m_v}{m_s}}(0.0)_{\frac{w_0}{a^{48}}}(0.0)_{\frac{w_0}{a^{64}}}(0.0)_{\frac{w_0}{a^{96}}}(0.0)_{w_0} $ \\
0.4 & 0.6 & 0.15 & $ 54.00(10)_{\rm S}(05)_{Z_V^{48}}(11)_{Z_V^{64}}(01)_{Z_V^{96}}(00)_{\rm C_3}(09)_{\rm C}(04)_{m_l}(00)_{\frac{m_l}{m_s}}(05)_{\frac{m_v}{m_s}}(00)_{\frac{w_0}{a^{48}}}(00)_{\frac{w_0}{a^{64}}}(00)_{\frac{w_0}{a^{96}}}(03)_{w_0} $ \\
0.6 & 0.8 & 0.15 & $ 71.11(25)_{\rm S}(06)_{Z_V^{48}}(14)_{Z_V^{64}}(02)_{Z_V^{96}}(04)_{\rm C_3}(40)_{\rm C}(10)_{m_l}(00)_{\frac{m_l}{m_s}}(17)_{\frac{m_v}{m_s}}(01)_{\frac{w_0}{a^{48}}}(02)_{\frac{w_0}{a^{64}}}(00)_{\frac{w_0}{a^{96}}}(11)_{w_0} $ \\
0.8 & 1.0 & 0.15 & $ 79.96(49)_{\rm S}(06)_{Z_V^{48}}(15)_{Z_V^{64}}(03)_{Z_V^{96}}(01)_{\rm C_3}(07)_{\rm C}(18)_{m_l}(01)_{\frac{m_l}{m_s}}(18)_{\frac{m_v}{m_s}}(03)_{\frac{w_0}{a^{48}}}(04)_{\frac{w_0}{a^{64}}}(01)_{\frac{w_0}{a^{96}}}(31)_{w_0} $ \\
\hline
0.3 & 1.0 & 0.15 & $ 225.40(85)_{\rm S}(19)_{Z_V^{48}}(43)_{Z_V^{64}}(08)_{Z_V^{96}}(07)_{\rm C_3}(73)_{\rm C}(34)_{m_l}(01)_{\frac{m_l}{m_s}}(49)_{\frac{m_v}{m_s}}(05)_{\frac{w_0}{a^{48}}}(06)_{\frac{w_0}{a^{64}}}(02)_{\frac{w_0}{a^{96}}}(47)_{w_0} $ \\
0.3 & 1.3 & 0.15 & $ 343.3(2.7)_{\rm S}(0.4)_{Z_V^{48}}(0.9)_{Z_V^{64}}(0.3)_{Z_V^{96}}(0.2)_{\rm C_3}(1.6)_{\rm C}(0.8)_{m_l}(0.0)_{\frac{m_l}{m_s}}(1.1)_{\frac{m_v}{m_s}}(0.2)_{\frac{w_0}{a^{48}}}(0.3)_{\frac{w_0}{a^{64}}}(0.1)_{\frac{w_0}{a^{96}}}(1.3)_{w_0} $ \\
0.3 & 1.6 & 0.15 & $ 439.5(5.3)_{\rm S}(0.4)_{Z_V^{48}}(1.1)_{Z_V^{64}}(0.3)_{Z_V^{96}}(0.2)_{\rm C_3}(2.1)_{\rm C}(1.4)_{m_l}(0.0)_{\frac{m_l}{m_s}}(1.4)_{\frac{m_v}{m_s}}(0.3)_{\frac{w_0}{a^{48}}}(0.5)_{\frac{w_0}{a^{64}}}(0.1)_{\frac{w_0}{a^{96}}}(2.4)_{w_0} $ \\
\hline
0.4 & 1.0 & 0.15 & $ 205.08(83)_{\rm S}(17)_{Z_V^{48}}(40)_{Z_V^{64}}(06)_{Z_V^{96}}(05)_{\rm C_3}(42)_{\rm C}(33)_{m_l}(01)_{\frac{m_l}{m_s}}(05)_{\frac{m_v}{m_s}}(05)_{\frac{w_0}{a^{48}}}(07)_{\frac{w_0}{a^{64}}}(01)_{\frac{w_0}{a^{96}}}(45)_{w_0} $ \\
0.4 & 1.3 & 0.15 & $ 321.6(2.7)_{\rm S}(0.3)_{Z_V^{48}}(0.8)_{Z_V^{64}}(0.2)_{Z_V^{96}}(0.1)_{\rm C_3}(0.0)_{\rm C}(0.8)_{m_l}(0.0)_{\frac{m_l}{m_s}}(1.0)_{\frac{m_v}{m_s}}(0.2)_{\frac{w_0}{a^{48}}}(0.3)_{\frac{w_0}{a^{64}}}(0.1)_{\frac{w_0}{a^{96}}}(1.3)_{w_0} $ \\
0.4 & 1.6 & 0.15 & $ 418.3(5.3)_{\rm S}(0.4)_{Z_V^{48}}(1.0)_{Z_V^{64}}(0.2)_{Z_V^{96}}(0.0)_{\rm C_3}(0.5)_{\rm C}(1.4)_{m_l}(0.0)_{\frac{m_l}{m_s}}(1.3)_{\frac{m_v}{m_s}}(0.3)_{\frac{w_0}{a^{48}}}(0.5)_{\frac{w_0}{a^{64}}}(0.1)_{\frac{w_0}{a^{96}}}(2.4)_{w_0} $ \\
\hline
0.4 & 1.0 & 0.05 & $ 208.4(0.8)_{\rm S}(0.2)_{Z_V^{48}}(0.4)_{Z_V^{64}}(0.1)_{Z_V^{96}}(0.1)_{\rm C_3}(1.2)_{\rm C}(0.3)_{m_l}(0.0)_{\frac{m_l}{m_s}}(0.6)_{\frac{m_v}{m_s}}(0.0)_{\frac{w_0}{a^{48}}}(0.1)_{\frac{w_0}{a^{64}}}(0.0)_{\frac{w_0}{a^{96}}}(0.7)_{w_0} $ \\
0.4 & 1.0 & 0.1 & $ 207.4(0.8)_{\rm S}(0.2)_{Z_V^{48}}(0.4)_{Z_V^{64}}(0.1)_{Z_V^{96}}(0.1)_{\rm C_3}(1.0)_{\rm C}(0.3)_{m_l}(0.0)_{\frac{m_l}{m_s}}(0.4)_{\frac{m_v}{m_s}}(0.0)_{\frac{w_0}{a^{48}}}(0.1)_{\frac{w_0}{a^{64}}}(0.0)_{\frac{w_0}{a^{96}}}(0.4)_{w_0} $ \\
0.4 & 1.0 & 0.2 & $ 201.83(86)_{\rm S}(17)_{Z_V^{48}}(38)_{Z_V^{64}}(07)_{Z_V^{96}}(01)_{\rm C_3}(05)_{\rm C}(33)_{m_l}(01)_{\frac{m_l}{m_s}}(22)_{\frac{m_v}{m_s}}(05)_{\frac{w_0}{a^{48}}}(07)_{\frac{w_0}{a^{64}}}(02)_{\frac{w_0}{a^{96}}}(48)_{w_0} $ \\

\end{tabular}
\caption{
 Same as Table~\ref{table:windowprxunimp},
 but with the parity improvement.
 \label{table:windowprxipa}
}
\end{table*}

\begin{table*}
\begin{tabular}{ccc|r}
$t_0$/fm & $t_1$/fm & $\Delta$/fm & $a^{\rm s,conn.isospin,W}_\mu  \times 10^{10}$ \\\hline
Total &  &  & $ 52.83(22)_{\rm S}(04)_{Z_V^{64}}(20)_{Z_V^{96}}(30)_{\rm C_3}(03)_{\frac{w_0}{a^{64}}}(13)_{\frac{w_0}{a^{96}}}(53)_{w_0} $ \\
\hline
0.0 & 0.1 & 0.15 & $ 0.81(00)_{\rm S}(00)_{Z_V^{64}}(00)_{Z_V^{96}}(12)_{\rm C_3}(00)_{\frac{w_0}{a^{64}}}(00)_{\frac{w_0}{a^{96}}}(00)_{w_0} $ \\
0.1 & 0.2 & 0.15 & $ 1.666(01)_{\rm S}(01)_{Z_V^{64}}(06)_{Z_V^{96}}(10)_{\rm C_3}(00)_{\frac{w_0}{a^{64}}}(00)_{\frac{w_0}{a^{96}}}(01)_{w_0} $ \\
0.2 & 0.3 & 0.15 & $ 2.57(00)_{\rm S}(00)_{Z_V^{64}}(01)_{Z_V^{96}}(16)_{\rm C_3}(00)_{\frac{w_0}{a^{64}}}(00)_{\frac{w_0}{a^{96}}}(00)_{w_0} $ \\
0.3 & 0.4 & 0.15 & $ 3.448(05)_{\rm S}(02)_{Z_V^{64}}(13)_{Z_V^{96}}(63)_{\rm C_3}(00)_{\frac{w_0}{a^{64}}}(01)_{\frac{w_0}{a^{96}}}(04)_{w_0} $ \\
0.4 & 0.5 & 0.15 & $ 4.170(07)_{\rm S}(03)_{Z_V^{64}}(16)_{Z_V^{96}}(08)_{\rm C_3}(00)_{\frac{w_0}{a^{64}}}(02)_{\frac{w_0}{a^{96}}}(08)_{w_0} $ \\
0.5 & 0.6 & 0.15 & $ 4.666(10)_{\rm S}(03)_{Z_V^{64}}(18)_{Z_V^{96}}(53)_{\rm C_3}(01)_{\frac{w_0}{a^{64}}}(04)_{\frac{w_0}{a^{96}}}(16)_{w_0} $ \\
0.6 & 0.7 & 0.15 & $ 4.866(13)_{\rm S}(04)_{Z_V^{64}}(18)_{Z_V^{96}}(67)_{\rm C_3}(01)_{\frac{w_0}{a^{64}}}(06)_{\frac{w_0}{a^{96}}}(24)_{w_0} $ \\
0.7 & 0.8 & 0.15 & $ 4.799(16)_{\rm S}(03)_{Z_V^{64}}(18)_{Z_V^{96}}(08)_{\rm C_3}(02)_{\frac{w_0}{a^{64}}}(08)_{\frac{w_0}{a^{96}}}(33)_{w_0} $ \\
0.8 & 0.9 & 0.15 & $ 4.505(17)_{\rm S}(03)_{Z_V^{64}}(17)_{Z_V^{96}}(03)_{\rm C_3}(02)_{\frac{w_0}{a^{64}}}(10)_{\frac{w_0}{a^{96}}}(39)_{w_0} $ \\
0.9 & 1.0 & 0.15 & $ 4.058(19)_{\rm S}(03)_{Z_V^{64}}(15)_{Z_V^{96}}(43)_{\rm C_3}(02)_{\frac{w_0}{a^{64}}}(11)_{\frac{w_0}{a^{96}}}(45)_{w_0} $ \\
1.0 & 1.1 & 0.15 & $ 3.527(19)_{\rm S}(03)_{Z_V^{64}}(13)_{Z_V^{96}}(57)_{\rm C_3}(02)_{\frac{w_0}{a^{64}}}(12)_{\frac{w_0}{a^{96}}}(46)_{w_0} $ \\
1.1 & 1.2 & 0.15 & $ 2.973(19)_{\rm S}(02)_{Z_V^{64}}(11)_{Z_V^{96}}(57)_{\rm C_3}(02)_{\frac{w_0}{a^{64}}}(12)_{\frac{w_0}{a^{96}}}(46)_{w_0} $ \\
1.2 & 1.3 & 0.15 & $ 2.441(18)_{\rm S}(02)_{Z_V^{64}}(09)_{Z_V^{96}}(62)_{\rm C_3}(02)_{\frac{w_0}{a^{64}}}(11)_{\frac{w_0}{a^{96}}}(44)_{w_0} $ \\
1.3 & 1.4 & 0.15 & $ 1.955(17)_{\rm S}(01)_{Z_V^{64}}(07)_{Z_V^{96}}(52)_{\rm C_3}(02)_{\frac{w_0}{a^{64}}}(10)_{\frac{w_0}{a^{96}}}(40)_{w_0} $ \\
1.4 & 1.5 & 0.15 & $ 1.534(15)_{\rm S}(01)_{Z_V^{64}}(06)_{Z_V^{96}}(47)_{\rm C_3}(02)_{\frac{w_0}{a^{64}}}(09)_{\frac{w_0}{a^{96}}}(35)_{w_0} $ \\
1.5 & 1.6 & 0.15 & $ 1.181(13)_{\rm S}(01)_{Z_V^{64}}(05)_{Z_V^{96}}(41)_{\rm C_3}(01)_{\frac{w_0}{a^{64}}}(08)_{\frac{w_0}{a^{96}}}(30)_{w_0} $ \\
1.6 & 1.7 & 0.15 & $ 0.894(12)_{\rm S}(01)_{Z_V^{64}}(03)_{Z_V^{96}}(35)_{\rm C_3}(01)_{\frac{w_0}{a^{64}}}(06)_{\frac{w_0}{a^{96}}}(25)_{w_0} $ \\
1.7 & 1.8 & 0.15 & $ 0.667(10)_{\rm S}(00)_{Z_V^{64}}(03)_{Z_V^{96}}(30)_{\rm C_3}(01)_{\frac{w_0}{a^{64}}}(05)_{\frac{w_0}{a^{96}}}(21)_{w_0} $ \\
1.8 & 1.9 & 0.15 & $ 0.491(08)_{\rm S}(00)_{Z_V^{64}}(02)_{Z_V^{96}}(25)_{\rm C_3}(01)_{\frac{w_0}{a^{64}}}(04)_{\frac{w_0}{a^{96}}}(17)_{w_0} $ \\
1.9 & 2.0 & 0.15 & $ 0.357(07)_{\rm S}(00)_{Z_V^{64}}(01)_{Z_V^{96}}(20)_{\rm C_3}(01)_{\frac{w_0}{a^{64}}}(03)_{\frac{w_0}{a^{96}}}(13)_{w_0} $ \\
\hline
0.0 & 0.2 & 0.15 & $ 2.48(00)_{\rm S}(00)_{Z_V^{64}}(01)_{Z_V^{96}}(11)_{\rm C_3}(00)_{\frac{w_0}{a^{64}}}(00)_{\frac{w_0}{a^{96}}}(00)_{w_0} $ \\
0.2 & 0.4 & 0.15 & $ 6.02(01)_{\rm S}(00)_{Z_V^{64}}(02)_{Z_V^{96}}(10)_{\rm C_3}(00)_{\frac{w_0}{a^{64}}}(00)_{\frac{w_0}{a^{96}}}(01)_{w_0} $ \\
0.4 & 0.6 & 0.15 & $ 8.837(18)_{\rm S}(06)_{Z_V^{64}}(34)_{Z_V^{96}}(61)_{\rm C_3}(01)_{\frac{w_0}{a^{64}}}(06)_{\frac{w_0}{a^{96}}}(24)_{w_0} $ \\
0.6 & 0.8 & 0.15 & $ 9.666(29)_{\rm S}(07)_{Z_V^{64}}(37)_{Z_V^{96}}(59)_{\rm C_3}(03)_{\frac{w_0}{a^{64}}}(14)_{\frac{w_0}{a^{96}}}(57)_{w_0} $ \\
0.8 & 1.0 & 0.15 & $ 8.562(36)_{\rm S}(06)_{Z_V^{64}}(33)_{Z_V^{96}}(45)_{\rm C_3}(04)_{\frac{w_0}{a^{64}}}(21)_{\frac{w_0}{a^{96}}}(84)_{w_0} $ \\
\hline
0.3 & 1.0 & 0.15 & $ 30.51(08)_{\rm S}(02)_{Z_V^{64}}(12)_{Z_V^{96}}(14)_{\rm C_3}(01)_{\frac{w_0}{a^{64}}}(04)_{\frac{w_0}{a^{96}}}(17)_{w_0} $ \\
0.3 & 1.3 & 0.15 & $ 39.45(13)_{\rm S}(03)_{Z_V^{64}}(15)_{Z_V^{96}}(04)_{\rm C_3}(01)_{\frac{w_0}{a^{64}}}(08)_{\frac{w_0}{a^{96}}}(31)_{w_0} $ \\
0.3 & 1.6 & 0.15 & $ 44.12(17)_{\rm S}(03)_{Z_V^{64}}(17)_{Z_V^{96}}(18)_{\rm C_3}(02)_{\frac{w_0}{a^{64}}}(10)_{\frac{w_0}{a^{96}}}(41)_{w_0} $ \\
\hline
0.4 & 1.0 & 0.15 & $ 27.06(08)_{\rm S}(02)_{Z_V^{64}}(10)_{Z_V^{96}}(08)_{\rm C_3}(01)_{\frac{w_0}{a^{64}}}(04)_{\frac{w_0}{a^{96}}}(16)_{w_0} $ \\
0.4 & 1.3 & 0.15 & $ 36.01(13)_{\rm S}(03)_{Z_V^{64}}(14)_{Z_V^{96}}(10)_{\rm C_3}(01)_{\frac{w_0}{a^{64}}}(08)_{\frac{w_0}{a^{96}}}(30)_{w_0} $ \\
0.4 & 1.6 & 0.15 & $ 40.68(17)_{\rm S}(03)_{Z_V^{64}}(15)_{Z_V^{96}}(24)_{\rm C_3}(02)_{\frac{w_0}{a^{64}}}(10)_{\frac{w_0}{a^{96}}}(41)_{w_0} $ \\
\hline
0.4 & 1.0 & 0.05 & $ 27.9(0.1)_{\rm S}(0.0)_{Z_V^{64}}(0.1)_{Z_V^{96}}(1.1)_{\rm C_3}(0.0)_{\frac{w_0}{a^{64}}}(0.0)_{\frac{w_0}{a^{96}}}(0.2)_{w_0} $ \\
0.4 & 1.0 & 0.1 & $ 27.70(08)_{\rm S}(02)_{Z_V^{64}}(11)_{Z_V^{96}}(05)_{\rm C_3}(01)_{\frac{w_0}{a^{64}}}(04)_{\frac{w_0}{a^{96}}}(17)_{w_0} $ \\
0.4 & 1.0 & 0.2 & $ 26.24(08)_{\rm S}(02)_{Z_V^{64}}(10)_{Z_V^{96}}(06)_{\rm C_3}(01)_{\frac{w_0}{a^{64}}}(04)_{\frac{w_0}{a^{96}}}(16)_{w_0} $ \\

\end{tabular}
\caption{
 Same as Table~\ref{table:windowphatrxunimp},
  but for the strange quark mass data.
\label{table:windowphatrxunimpstrange}
}
\end{table*}

\begin{table*}
\begin{tabular}{ccc|r}
$t_0$/fm & $t_1$/fm & $\Delta$/fm & $a^{\rm s,conn.isospin,W}_\mu  \times 10^{10}$\\\hline
Total &  &  & $ 53.08(22)_{\rm S}(04)_{Z_V^{64}}(20)_{Z_V^{96}}(58)_{\rm C_3}(03)_{\frac{w_0}{a^{64}}}(13)_{\frac{w_0}{a^{96}}}(53)_{w_0} $ \\
\hline
0.0 & 0.1 & 0.15 & $ 0.887(00)_{\rm S}(01)_{Z_V^{64}}(03)_{Z_V^{96}}(40)_{\rm C_3}(00)_{\frac{w_0}{a^{64}}}(00)_{\frac{w_0}{a^{96}}}(00)_{w_0} $ \\
0.1 & 0.2 & 0.15 & $ 1.728(01)_{\rm S}(01)_{Z_V^{64}}(06)_{Z_V^{96}}(86)_{\rm C_3}(00)_{\frac{w_0}{a^{64}}}(00)_{\frac{w_0}{a^{96}}}(02)_{w_0} $ \\
0.2 & 0.3 & 0.15 & $ 2.59(00)_{\rm S}(00)_{Z_V^{64}}(01)_{Z_V^{96}}(24)_{\rm C_3}(00)_{\frac{w_0}{a^{64}}}(00)_{\frac{w_0}{a^{96}}}(00)_{w_0} $ \\
0.3 & 0.4 & 0.15 & $ 3.451(05)_{\rm S}(03)_{Z_V^{64}}(13)_{Z_V^{96}}(47)_{\rm C_3}(00)_{\frac{w_0}{a^{64}}}(01)_{\frac{w_0}{a^{96}}}(04)_{w_0} $ \\
0.4 & 0.5 & 0.15 & $ 4.169(08)_{\rm S}(03)_{Z_V^{64}}(16)_{Z_V^{96}}(15)_{\rm C_3}(00)_{\frac{w_0}{a^{64}}}(02)_{\frac{w_0}{a^{96}}}(08)_{w_0} $ \\
0.5 & 0.6 & 0.15 & $ 4.665(11)_{\rm S}(03)_{Z_V^{64}}(18)_{Z_V^{96}}(59)_{\rm C_3}(01)_{\frac{w_0}{a^{64}}}(04)_{\frac{w_0}{a^{96}}}(16)_{w_0} $ \\
0.6 & 0.7 & 0.15 & $ 4.866(13)_{\rm S}(04)_{Z_V^{64}}(19)_{Z_V^{96}}(73)_{\rm C_3}(01)_{\frac{w_0}{a^{64}}}(06)_{\frac{w_0}{a^{96}}}(24)_{w_0} $ \\
0.7 & 0.8 & 0.15 & $ 4.799(16)_{\rm S}(04)_{Z_V^{64}}(18)_{Z_V^{96}}(06)_{\rm C_3}(02)_{\frac{w_0}{a^{64}}}(08)_{\frac{w_0}{a^{96}}}(33)_{w_0} $ \\
0.8 & 0.9 & 0.15 & $ 4.504(18)_{\rm S}(03)_{Z_V^{64}}(17)_{Z_V^{96}}(02)_{\rm C_3}(02)_{\frac{w_0}{a^{64}}}(10)_{\frac{w_0}{a^{96}}}(39)_{w_0} $ \\
0.9 & 1.0 & 0.15 & $ 4.058(19)_{\rm S}(03)_{Z_V^{64}}(16)_{Z_V^{96}}(43)_{\rm C_3}(02)_{\frac{w_0}{a^{64}}}(11)_{\frac{w_0}{a^{96}}}(45)_{w_0} $ \\
1.0 & 1.1 & 0.15 & $ 3.527(19)_{\rm S}(03)_{Z_V^{64}}(14)_{Z_V^{96}}(58)_{\rm C_3}(02)_{\frac{w_0}{a^{64}}}(12)_{\frac{w_0}{a^{96}}}(46)_{w_0} $ \\
1.1 & 1.2 & 0.15 & $ 2.973(19)_{\rm S}(02)_{Z_V^{64}}(11)_{Z_V^{96}}(57)_{\rm C_3}(02)_{\frac{w_0}{a^{64}}}(12)_{\frac{w_0}{a^{96}}}(46)_{w_0} $ \\
1.2 & 1.3 & 0.15 & $ 2.440(18)_{\rm S}(02)_{Z_V^{64}}(09)_{Z_V^{96}}(62)_{\rm C_3}(02)_{\frac{w_0}{a^{64}}}(11)_{\frac{w_0}{a^{96}}}(44)_{w_0} $ \\
1.3 & 1.4 & 0.15 & $ 1.955(17)_{\rm S}(01)_{Z_V^{64}}(08)_{Z_V^{96}}(52)_{\rm C_3}(02)_{\frac{w_0}{a^{64}}}(10)_{\frac{w_0}{a^{96}}}(40)_{w_0} $ \\
1.4 & 1.5 & 0.15 & $ 1.534(15)_{\rm S}(01)_{Z_V^{64}}(06)_{Z_V^{96}}(47)_{\rm C_3}(02)_{\frac{w_0}{a^{64}}}(09)_{\frac{w_0}{a^{96}}}(35)_{w_0} $ \\
1.5 & 1.6 & 0.15 & $ 1.181(13)_{\rm S}(01)_{Z_V^{64}}(05)_{Z_V^{96}}(41)_{\rm C_3}(01)_{\frac{w_0}{a^{64}}}(08)_{\frac{w_0}{a^{96}}}(30)_{w_0} $ \\
1.6 & 1.7 & 0.15 & $ 0.894(12)_{\rm S}(01)_{Z_V^{64}}(03)_{Z_V^{96}}(35)_{\rm C_3}(01)_{\frac{w_0}{a^{64}}}(06)_{\frac{w_0}{a^{96}}}(25)_{w_0} $ \\
1.7 & 1.8 & 0.15 & $ 0.667(10)_{\rm S}(00)_{Z_V^{64}}(03)_{Z_V^{96}}(31)_{\rm C_3}(01)_{\frac{w_0}{a^{64}}}(05)_{\frac{w_0}{a^{96}}}(21)_{w_0} $ \\
1.8 & 1.9 & 0.15 & $ 0.491(08)_{\rm S}(00)_{Z_V^{64}}(02)_{Z_V^{96}}(25)_{\rm C_3}(01)_{\frac{w_0}{a^{64}}}(04)_{\frac{w_0}{a^{96}}}(17)_{w_0} $ \\
1.9 & 2.0 & 0.15 & $ 0.357(07)_{\rm S}(00)_{Z_V^{64}}(01)_{Z_V^{96}}(21)_{\rm C_3}(01)_{\frac{w_0}{a^{64}}}(03)_{\frac{w_0}{a^{96}}}(13)_{w_0} $ \\
\hline
0.0 & 0.2 & 0.15 & $ 2.615(02)_{\rm S}(02)_{Z_V^{64}}(10)_{Z_V^{96}}(46)_{\rm C_3}(00)_{\frac{w_0}{a^{64}}}(00)_{\frac{w_0}{a^{96}}}(02)_{w_0} $ \\
0.2 & 0.4 & 0.15 & $ 6.05(01)_{\rm S}(00)_{Z_V^{64}}(02)_{Z_V^{96}}(19)_{\rm C_3}(00)_{\frac{w_0}{a^{64}}}(00)_{\frac{w_0}{a^{96}}}(01)_{w_0} $ \\
0.4 & 0.6 & 0.15 & $ 8.834(18)_{\rm S}(07)_{Z_V^{64}}(34)_{Z_V^{96}}(74)_{\rm C_3}(01)_{\frac{w_0}{a^{64}}}(06)_{\frac{w_0}{a^{96}}}(24)_{w_0} $ \\
0.6 & 0.8 & 0.15 & $ 9.665(29)_{\rm S}(07)_{Z_V^{64}}(37)_{Z_V^{96}}(67)_{\rm C_3}(03)_{\frac{w_0}{a^{64}}}(14)_{\frac{w_0}{a^{96}}}(57)_{w_0} $ \\
0.8 & 1.0 & 0.15 & $ 8.562(36)_{\rm S}(06)_{Z_V^{64}}(33)_{Z_V^{96}}(44)_{\rm C_3}(04)_{\frac{w_0}{a^{64}}}(21)_{\frac{w_0}{a^{96}}}(84)_{w_0} $ \\
\hline
0.3 & 1.0 & 0.15 & $ 30.51(09)_{\rm S}(02)_{Z_V^{64}}(12)_{Z_V^{96}}(14)_{\rm C_3}(01)_{\frac{w_0}{a^{64}}}(04)_{\frac{w_0}{a^{96}}}(17)_{w_0} $ \\
0.3 & 1.3 & 0.15 & $ 39.45(13)_{\rm S}(03)_{Z_V^{64}}(15)_{Z_V^{96}}(03)_{\rm C_3}(02)_{\frac{w_0}{a^{64}}}(08)_{\frac{w_0}{a^{96}}}(31)_{w_0} $ \\
0.3 & 1.6 & 0.15 & $ 44.12(17)_{\rm S}(03)_{Z_V^{64}}(17)_{Z_V^{96}}(17)_{\rm C_3}(02)_{\frac{w_0}{a^{64}}}(10)_{\frac{w_0}{a^{96}}}(41)_{w_0} $ \\
\hline
0.4 & 1.0 & 0.15 & $ 27.06(08)_{\rm S}(02)_{Z_V^{64}}(10)_{Z_V^{96}}(10)_{\rm C_3}(01)_{\frac{w_0}{a^{64}}}(04)_{\frac{w_0}{a^{96}}}(16)_{w_0} $ \\
0.4 & 1.3 & 0.15 & $ 36.00(13)_{\rm S}(03)_{Z_V^{64}}(14)_{Z_V^{96}}(08)_{\rm C_3}(01)_{\frac{w_0}{a^{64}}}(08)_{\frac{w_0}{a^{96}}}(30)_{w_0} $ \\
0.4 & 1.6 & 0.15 & $ 40.67(17)_{\rm S}(03)_{Z_V^{64}}(16)_{Z_V^{96}}(22)_{\rm C_3}(02)_{\frac{w_0}{a^{64}}}(10)_{\frac{w_0}{a^{96}}}(41)_{w_0} $ \\
\hline
0.4 & 1.0 & 0.05 & $ 27.9(0.1)_{\rm S}(0.0)_{Z_V^{64}}(0.1)_{Z_V^{96}}(1.1)_{\rm C_3}(0.0)_{\frac{w_0}{a^{64}}}(0.0)_{\frac{w_0}{a^{96}}}(0.2)_{w_0} $ \\
0.4 & 1.0 & 0.1 & $ 27.69(08)_{\rm S}(02)_{Z_V^{64}}(11)_{Z_V^{96}}(03)_{\rm C_3}(01)_{\frac{w_0}{a^{64}}}(04)_{\frac{w_0}{a^{96}}}(17)_{w_0} $ \\
0.4 & 1.0 & 0.2 & $ 26.24(08)_{\rm S}(02)_{Z_V^{64}}(10)_{Z_V^{96}}(07)_{\rm C_3}(01)_{\frac{w_0}{a^{64}}}(04)_{\frac{w_0}{a^{96}}}(16)_{w_0} $ \\

\end{tabular}
\caption{
 Same as Table~\ref{table:windowphatrxunimpstrange},
 but for the $p$ prescription.
 \label{table:windowprxunimpstrange}
}
\end{table*}

\bibliography{main}% Produces the bibliography via BibTeX.

\end{document}